%% file: dis.tex
\documentclass[12pt]{report}
\usepackage{times}
\usepackage{mathptm}
\usepackage{uidis}
\usepackage{natbib}
\usepackage{html}
\usepackage[xdvi]{graphicx}
\begin{document}
\pagenumbering{roman}
\title{Monte Carlo Methods: Application to Hydrogen Gas \\and Hard Spheres}
\author{Mark Douglas Dewing}
\advisor{David M. Ceperley}
\prevdegrees{ B.S., Michigan Technological University, 1993\\
              M.S., University of Illinois at Urbana-Champaign, 1995 }
\gradyear{2001}
\maketitle

\setcounter{page}{3}
\input{coms}   
\include{abstract}

\include{ack}

\tableofcontents
\listoftables
\listoffigures
\include{notation}
\clearpage
\pagenumbering{arabic}
\include{chap1}

\include{chap2}

\include{chap3}

\include{chap4}

\include{chap5}

\include{chap-hs}

\include{chap7}

\include{chap8}

\appendix
\include{appen1}

\include{appen2}

\include{appen3}

\singlespace
\bibliographystyle{uidis3}
\bibliography{dis}
\oneandhalfspace
\include{vita}
\end{document}

%% file: coms.tex

\newcommand{\be}{\begin{equation}}
\newcommand{\ee}{\end{equation}}
\newcommand{\bea}{\begin{eqnarray}}
\newcommand{\eea}{\end{eqnarray}}
\newcommand{\bra}[1]{\langle #1 |}
\newcommand{\ket}[1]{| #1 \rangle}
\newcommand{\braket}[3]{\langle #1 | #2 | #3 \rangle}
\newcommand{\overlap}[2]{\langle #1 | #2 \rangle}
\newcommand{\expect}[1]{\langle #1 \rangle}
\newcommand{\abs}[1]{\left| #1 \right|}
\newcommand{\del}{\nabla}

\newcommand{\iprod}[2]{\left\langle #1 \left| #2\right. \right\rangle}
\newcommand{\Det}[1]{\left| #1 \right|}
\newcommand{\pd}[2]{\frac{\partial #1}{\partial #2}}
\newcommand{\vc}[1]{{\bf #1}}
\newcommand{\spd}[3]{\frac{\partial^2 #1}{\partial #2 \partial #3}}

\newcommand{\calP}{\mbox{$\mathcal P$}}
\newcommand{\calO}{\mbox{$\mathcal O$}}
\newcommand{\calV}{\mbox{$\mathcal V$}}

\newcommand{\ra}{\rightarrow}
\newcommand{\half}{\frac{1}{2}}

\newcommand{\fcmatrix}[4]{
   \matrix{
         #1 & \ldots & #2 \cr
          \vdots &\ddots & \vdots \cr
         #3 & \ldots & #4 \cr
 }}

%% file: abstract.tex
\begin{abstract}

Quantum Monte Carlo (QMC) methods are among the most accurate
for computing ground state properties of quantum systems.
The two major types of QMC we use are Variational Monte Carlo (VMC), 
which evaluates
integrals arising from the variational principle, and Diffusion
Monte Carlo (DMC), which stochastically projects to the ground state from 
a trial wave function.
These methods are applied to a system of boson hard spheres to get 
exact, infinite system size results for the ground state at several densities.

The kinds of problems that can be simulated with Monte Carlo methods are
expanded through the development of
new algorithms for combining a
QMC simulation with a classical Monte Carlo simulation, which we call
Coupled Electronic-Ionic Monte Carlo (CEIMC).  The new CEIMC method
is applied to a system of molecular hydrogen 
at temperatures ranging from 2800K to 4500K and densities from 
0.25 to 0.46 g/cm$^3$.

VMC requires optimizing a parameterized wave function to find 
the minimum energy.
We examine several techniques for optimizing VMC wave functions, focusing on
the ability to optimize parameters appearing in the Slater determinant.  

Classical Monte Carlo simulations 
use an empirical interatomic potential
to compute equilibrium properties of various states of matter.
The CEIMC method replaces the empirical potential with a QMC calculation of the electronic energy.
This is similar in spirit to the Car-Parrinello technique,
which uses Density Functional Theory for the electrons
and molecular dynamics for the nuclei.

The challenges in constructing an efficient CEIMC simulation
center mostly around the noisy results generated
from the QMC computations of the electronic energy.
We introduce two complementary techniques, one for tolerating the noise and
the other for reducing it.
The penalty method modifies the Metropolis acceptance ratio 
to tolerate noise without introducing a bias in the simulation
of the nuclei.
For reducing the noise, we introduce 
the two-sided energy difference method, which uses correlated sampling 
to compute the energy change associated with a trial move
of the nuclear coordinates.  
Unlike the standard reweighting method,
it remains stable as the energy difference increases.

\end{abstract}

%% file: ack.tex
\chapter*{Acknowledgments}

First I would like to thank my advisor, David Ceperley, 
for supporting me in this research, for teaching these
Monte Carlo methods to me,
and for being available and patient
when answering questions.

I would also like to thank the graduate students and postdocs in the group
who have helped me and provided useful and interesting discussions.
And special thanks to Tadashi Ogitsu for the use of his DFT code.

I am grateful to my parents for their support during
my college and graduate school pursuits.
I enjoyed the refreshing summer visits to their farm.
Thanks to my brother Luke, whose living in Colorado was convenient for
ski trips.

During my time here, I have benefited greatly from friendships and 
relationships with
people in 
Graduate Intervarsity Christian Fellowship, 
Grace Community Church, and
Illini Life Christian Fellowship.
They have given me a great deal of strength and encouragement
when I needed it.

My work was supported by the computational facilities at NCSA, 
by a Graduate Research Trainee fellowship NSF Grant No. DGE93-54978, 
and by NSF Grant No. DMR 98-02373.

And finally, apologies to my cat, Lucy, for not giving her enough attention 
while finishing this work.

%% file: notation.tex
\chapter*{Guide to Notation}
\begin{list}{}{
   \setlength{\leftmargin}{1.75in}
   \setlength{\labelwidth}{1.25in}
   \setlength{\labelsep}{0.5in}
}

\item[CEIMC] Coupled Electronic-Ionic Monte Carlo

\item[DMC] Diffusion Monte Carlo

\item[DFT] Density Functional Theory

\item[GBRW] Gradient Biased Random Walk

\item[LDA] Local Density Approximation (in Density Functional Theory)

\item[PIMC] Path Integral Monte Carlo

\item[QMC] Quantum Monte Carlo

\item[SGA] Stochastic Gradient Approximation

\item[VMC] Variational Monte Carlo

\vskip .5cm

\item[$\alpha$] General variational parameter.

\item[$a$] Instantaneous acceptance probability in penalty method.

\item[$a_0$]  Bohr radius, unit of length in atomic units.   \\
          1 $a_0$ = $5.29 \times 10^{-9}$ m.

\item[$A(s \ra s')$] Acceptance probability in Metropolis method.

\item[$\beta$]  Inverse temperature, $\frac{1}{k_B T}$.

\item[$\delta$] QMC estimate of an energy  difference.

\item[$\Delta$] Exact energy difference.  Also the sampling box size
 in the Metropolis algorithm.

\item[$d$] Bond length.

\item[$D$] Slater determinant.

\item[$\zeta$] A variational parameter in H$_2$ orbitals.

\item[$f$] A trial wave function (also denoted $\psi_T$). Also the
        relative noise parameter.

\item[$\eta$] Additional noise rejection ratio.

\item[$E_L$] The local energy of a trial wave function.

\item[$\theta_v$] Vibrational temperature.

\item[$h$] Step size parameter in SGA and GBRW.

\item[$H$] A many-body Hamiltonian.

\item[Ha] Hartree, unit of energy in atomic units.  1 Ha = 27.21 eV.

\item[$G(R \ra R',\tau)$] Green's function propagator in DMC.

\item[$k_B$] Boltzmann factor.

\item[$K$] Kinetic energy.

\item[$\lambda$] $\hbar^2/2m$, where $m$ is the particle's mass.

\item[$n_0$] Condensate fraction.

\item[$N$] The number of particles in a system.

\item[$\pi(s)$] Probability distribution to be sampled in a Markov process.

\item[$P$] Pressure, or a probability distribution in the two-sided method.

\item[$\calP(s\rightarrow s')$] Transition probability in a Markov chain.

\item[$Q$] Normalization integral .

\item[$\rho$] Density.
 
\item[$\rho_1(r)$]  Single particle density matrix.

\item[$r_s$] $\left(\frac{3}{4 \pi n}\right)^{1/3}$, 
      where $n$ is the electron number density.

\item[$r_{ij}$]
The separation between particle $j$ and particle $i$.

\item[$R$] The coordinates of all the particles in a many-body system.

\item[$\sigma$] Noise level (variance or standard error). 
          Also the hard sphere diameter.

\item[$s$] State in configuration space.

\item[$\tau$] DMC time step .

\item[$T$] Temperature.

\item[$T(s\rightarrow s')$] Sampling distribution in Metropolis method.

\item[$u(r)$] Jastrow factor in a many-body wave function.

\item[$V$] Volume of the simulation cell, or potential energy.

\item[\calV] Alternate notation for potential energy 
             (for formulas that have both potential energy and volume).

\item[$w$] Weight factor in correlated sampling methods.

\item[$w_l$] Width of H$_2$ orbital.

\item[$\psi_T$] A trial wave function (also denoted $f$).

\item[$\phi_0$] The exact many body ground state wave function.

\item[$\phi$] A single particle orbital.


\end{list}

%% file: chap1.tex
\chapter{Introduction}

The first computer simulations of a condensed matter system used the 
simplest potential, the
hard sphere \citep{metropolis53}.    
As computers and simulations progressed, more sophisticated and realistic
potentials came into use. 
These potentials are parameterized and then fit to reproduce 
various experimental quantities.
Both Molecular Dynamics (MD) and Monte Carlo (MC) methods are
used to generate ensemble averages of many-particle systems.

These potentials originate from the microscopic structure
of matter, described in terms of electrons, nuclei, and
the Schr\"odinger equation.
But the many-body Schr\"odinger equation is too complicated
to solve directly, so some approximations are needed.   
The one electron approximation is a successful approach, where
a single electron interacts with an external potential (ie, the nuclei)
and with a mean field generated by all the other electrons.
This is done by Hartree-Fock (HF) or with
Density Functional Theory (DFT) \citep{parr89}.
DFT is in principle exact, but contains an unknown exchange and
correlation functional that must be approximated.  The most
common one is the Local Density Approximation (LDA).

These first principles calculations are used in fitting the
potentials, which are then used in an MC or MD computation.  
But the problem of transferability still remains.  Empirical
potentials are only valid in situations for which they have been designed  
and fitted.

In 1985, Car and Parrinello introduced their method, which replaced the
empirical potential with a DFT calculation done `on the fly'
\citep{car85}.
They did a molecular dynamics simulation of the nuclei of liquid silicon 
and then
computed the density functional energy of the electrons at every MD step.
To improve the efficiency of the computation of the DFT energy,
they introduced a new iterative method for solving the DFT equations.
It has been a very successful method, with
the original paper being cited over 2300 times since its publication.

Previously, the DFT equations had been solved by eigenvalue methods.
But eigenvalue problems can also be regarded as optimization problems,
where an energy functional is minimized.
Car and Parrinello used an idea similar to simulated annealing, 
but they used molecular dynamics to move through
parameter space, rather than Monte Carlo.
This had the effect of making the equations of motion similar
between the electronic problem and the nuclear problem, with
the only difference being in the relative masses. Since the electronic
problem was not real electron dynamics, the electron mass does not correspond
to any physical quantity, and is only a parameter controlling the
convergence of the electronic part of the simulation.
Since then, other iterative methods have been introduced, usually
based on the Conjugate-Gradient method
\citep{payne92}.

A brief review of applications of the Car-Parrinello method to 
liquid problems is given by \cite{sprik00}.
This review also mentions that LDA and some other functionals are not 
good enough to accurately simulate water (there are improved functionals that
are acceptable).
Another review of molecular dynamics by \cite{tuckerman00} includes
material on treating the nuclei classically  and also using path integrals
to treat the nuclei quantum mechanically, done by \cite{marx96}.

Quantum Monte Carlo (QMC) methods have developed as another
means for accurately solving the many body Schr\"odinger equation 
\citep{hammond94,anderson95,ceperley96}.
The success of QMC lies partly in the fact these methods
explicitly include correlation among the electrons, which
can not be done directly with the one electron methods.
Particularly with the Local Density Approximation (LDA), DFT
has known difficulties in handling electron correlation \citep{grossman95}.

In the spirit of the Car-Parrinello method,
we integrate a Classical Monte Carlo simulation of the nuclei with
a QMC simulation for the electrons.  
This we call Coupled Electronic-Ionic Monte Carlo (CEIMC).
There are some challenges in constructing an efficient method.

The first problem we encounter is that the results of a 
QMC simulation are noisy.  The QMC energy
has some uncertainty associated with it, and it could bias the 
classical part of the simulation.
We could run the QMC simulation until the noise is negligible, but that
is very time-consuming.  A better way is use the penalty method,
which modifies the usual MC formulas to be tolerant of noise.



The electrons are assumed to be in their ground state, both in the Car-Parrinello
method and in our CEIMC method.  There are two internal 
effects that could excite the electrons - coupling to nuclear motion
and thermal excitations.
In the first case, we make the Born-Oppenheimer
approximation, where the nuclei are so much more massive
than the electrons that
the electrons are assumed to respond to nuclear
motion instantaneously, and so stay in their ground state.  
We neglect any occupation of excited states of the electrons due
to coupling to nuclear motion.


In the case of thermal excitation, 
let us examine several relevant energy scales.
If we consider a gas of degenerate electrons at a density of $n=0.0298$
electrons per cubic Bohr 
(i.e. $r_s = \left(\frac{3}{4 \pi n}\right)^{1/3} = 2.0$),
the Fermi temperature is about 140,000K.
The gap between the ground state and the first excited state of 
a hydrogen molecule at equilibrium
bond distance is about 124,000K.
As long as our temperatures are well below this (and they are), and
we are not at too high pressures (pressure decreases the gap),
the thermal occupation of excited states can be neglected.



Hydrogen is the most abundant element in the universe, making an
understanding of its properties important, particularly for
astrophysical applications.  
Models of the interiors of the gas giant planets
depends on a knowledge of the equation of state of hydrogen
\citep{hubbard84,stevenson88}.
Hydrogen is also the simplest element, but it still displays  remarkable
variety in its properties and phase diagram.
It has several solid phases at low temperature, and the
crystal structure of one of them (phase III) is not fully known yet.
At high temperature and pressure the fluid becomes metallic,
but the exact nature of the transition is not known.






Computer simulation can also be used to obtain results on model systems.
We will examine the hard sphere Bose gas, a simple and important
model.  For this model, all the approximations we make are controllable,
and we will look at how to deal with those approximations and obtain
exact results for this model.


\section{Thesis Overview}

Chapter 2 is an introduction to the basic classical and quantum
Monte Carlo techniques we will be using.
Chapter 3 presents
an improved QMC method for computing the energy
difference between two systems.
Chapter 4 is an examination of parameter optimization, which is essential
in VMC.
We present various methods for minimizing the energy, and give
some comparisons between them.

Successful CEIMC simulations are based on the penalty method for
tolerating noise in the Metropolis method, which is detailed in
Chapter 5. Some additional details are discussed,
and an example of CEIMC applied to a single H$_2$ molecule is given.
The results of computations of the ground state energy of the boson hard sphere model are presented in Chapter 6.
In Chapter 7, the CEIMC simulation method is applied to fluid 
molecular hydrogen.
We present data for a few state points and perform some analysis
of the simulation itself.

%% file: chap2.tex
\chapter{Monte Carlo Methods}

Monte Carlo integration methods are very useful for evaluating the basic
integrals of statistical and quantum physics.
In a system with $N_p$ particles, these integrals have the form
\be
\expect{O} = \frac{\int dR\ \pi(R) O(R)}{\int dR\ \pi(R)}
\label{basic-integral}
\ee
where $R$ is a $3N_p$ dimensional vector, $\pi(R)$ is a probability 
distribution, and $O(R)$ is the observable or quantity of interest.
These integrals have two important characteristics:
high dimensionality
and the integrands are sharply peaked - only small parts of phase space 
contribute significantly to the integral.

The high dimensionality makes a grid based scheme impractical in two ways.
First, suppose we have a 300 dimensional integral (100 particle simulation),
and want 10 grid points in each dimension.  Even this crude integration 
requires function evaluations at $10^{300}$ grid points!
Second, consider the trapezoidal rule (as a concrete example) 
in $d$ dimensions.  The error using $N$ samples will go as $\calO(N^{-2/d})$.
As we will show,  the error in Monte Carlo integration goes as 
$\calO(N^{-1/2})$. 
The Monte Carlo error is independent of the dimensionality whereas
the grid based method depends on it strongly.
For these high dimensionality problems, Monte Carlo is only practical choice.



\section{Basic Monte Carlo Integration}
Consider an integral of the form 
\be \label{simpleI}
I = \int_0^1 f(x) dx.
\ee
To evaluate by Monte Carlo, compute $f(x)$ at $N$ points sampled uniformly
from $[0,1]$.  An approximation to $I$ is given by
\be \label{simple_sum}
 I \approx \bar f = \frac{1}{N}\sum_{i=1}^N f(x_i)
\ee
The estimate of the statistical error in $\bar f$ will be 
\be
\sigma_I = \sigma_f/\sqrt{N} 
\ee
where $\sigma_f^2$ is the variance, and is given by
\be
\sigma_f^2 = \frac{1}{(N-1)} \sum_{i=1}^N \left(f(x_i) - \bar f \right)^2
\label{variance}
\ee
Thus the error goes as 
$\calO(N^{-1/2})$.  

The error bounds can be improved by sampling more points, or by reducing
the variance, $\sigma^2_f$.  
The latter can be accomplished with importance sampling.
Consider some probability, $P(x)$, that is an approximation to $f(x)$.
Write Eq. (\ref{simpleI}) as
\be
I = \int_0^1 P(x) \frac{f(x)}{P(x)} dx.
\ee
The estimate of $I$ is obtained by sampling $N$ points from $P(x)$ and 
computing
\be \label{importance_sum}
I \approx \sum_{i=1}^N \frac{f(x_i)}{P(x_i)}
\ee
If $P$ is a good approximation to $f$, then the variance of the sum
in Eq. (\ref{importance_sum})  is much less than the variance of the
sum in Eq. (\ref{simple_sum}).


The fact that the integrands of interest are sharply peaked, as mentioned
previously, makes importance sampling a necessity.
The most useful type of importance sampling for these problems 
is the Metropolis method.


\section{Metropolis Sampling}

The Metropolis method \citep{metropolis53} uses a Markov process 
to generate samples from 
a normalized probability distribution, $\pi(R)/\int dR\ \pi(R)$.  
These samples are then used to estimate Eq. (\ref{basic-integral})
by
\be
\bar O = \frac{1}{N} \sum_i O(R_i)
\label{metropolis_sum}
\ee
For generality in the following section, 
we will denote the state of the simulation by $s$.

A Markov process takes a transition probability between states, 
$\calP(s \ra s')$, and constructs a series of state points $s_1,s_2,...$
(called a chain).
An important characteristic of a Markov process is that the choice
of the next state point in the chain depends only on the current state point, 
not any previous state points. 

The Metropolis method constructs a
transition probability such that generated state points are sampled
from the desired distribution.
For this to work,
the transition probability must satisfy ergodicity. 
This means the 
Markov chain must eventually be able to reach any state in the system.
A sufficient condition for satisfying ergodicity is detailed balance,
\be
 \pi(s) \calP(s \rightarrow s') = \pi(s') \calP(s' \rightarrow s).
  \label{db}
\ee


The generalized Metropolis method breaks the transition 
probability into the product of two pieces - an {\it a priori } sampling 
distribution
$T(s \rightarrow s')$ and an acceptance probability $A(s \rightarrow s')$.
The Metropolis choice for the acceptance probability is
\be \label{acc_prob}
A(s \rightarrow s') = \min\left[ 1, \frac{\pi(s') T(s'\rightarrow s)}
     {\pi(s) T(s \rightarrow s')} \right]
\ee

The procedure is to sample a trial state, $s'$, according to $T(s \ra s')$ 
and evaluate Eq. (\ref{acc_prob}). 
The acceptance probability is compared with a uniform random number on $[0,1]$.
If $A$ is greater than the random number,
the move is accepted, $s'$ becomes the new $s$ and is
used in the average in Eq (\ref{metropolis_sum}).  
Otherwise the move is rejected, $s$ is not changed, and is reused in the 
average.

The original Metropolis procedure chooses a trial position uniformly inside a 
box centered around the current point,
\be
s' = s + y
\ee
where $y$ is a uniform random number on $[-\Delta/2,\Delta/2]$, 
with $\Delta$ being an adjustable parameter.
In this case, $T$ is uniform and will cancel out of Eq. (\ref{acc_prob}).

An important measure of the Metropolis procedure is the acceptance ratio - the
ratio of accepted moves to the number of trial moves.
It can be adjusted by the choice of $\Delta$.
If the acceptance ratio is too small, state space will be explored very
slowly because very few moves are accepted.  If the acceptance ratio
is high, it is likely that the trial moves are too small and once again,
diffusion through state space will be very slow.
Balancing these considerations leads to the standard rule of thumb that 
the optimal acceptance ratio is around 50\%.

A better consideration is maximization of the efficiency, 
\be
\xi = \frac{1}{\sigma^2 T}
\label{mc_efficiency}
\ee
where $T$ is the computer time taken to get an error estimate of $\sigma$.


\section{Classical Monte Carlo}
The probability distribution we wish to sample is the Boltzmann distribution
\be
\pi(s) \propto  \exp[-V(s)/kT].
\ee

The first simulations of this type were done with the hard sphere potential
\citep{metropolis53,wood57}.
Later simulations used Lennard-Jones potentials, and then other
types of empirical potentials.

The Metropolis procedure samples only the normalized $\pi(s)$.
Averages over this distribution are readily computed, but 
quantities that depend on the value of the normalization are difficult
to compute.
In classical systems, this includes quantities such the entropy and
the free energy.
There are techniques, however, for computing the free 
energy difference between two systems.

\section{Variational Monte Carlo}
Variational Monte Carlo (VMC)  is based on 
evaluating the integral that arises from the variational principle.
The variational principle states that the energy from applying the Hamiltonian
to a trial wave function must be greater than or equal to the exact 
ground state energy.
Typically the wave function is parameterized and then optimized
with respect to those parameters to find the minimum energy (or minimum
variance of the energy).
Monte Carlo is needed because the wave function contains explicit 
two (or higher) particle correlations
and this results in a non-factoring high dimensional integral.

The energy is written as
\be \label{variationalE}
E=  \frac{\int dR \psi_T(R) H \psi_T(R) }{\int dR \psi_T(R)^2} =
  \frac{\int dR \abs{\psi_T(R)}^2 E_L(R)}{\int dR \abs{\psi_T(R)}^2}
\ee
where
$E_L = \frac{H \psi_T}{\psi_T}$,
and is called the local energy.
Other diagonal matrix elements can be 
evaluated in a similar fashion.  
Off diagonal elements can also be evaluated, but with more effort.
\cite{mcmillan65} introduced the use of Metropolis sampling for evaluating 
this integral.



A typical form of the variational wave function is a Jastrow factor 
(two body correlations)
multiplied by two Slater determinants of one body orbitals.
\be
\psi_T = \exp\left[-\sum_{i<j} u(r_{ij})\right]
{\mathrm{Det}}\left(S^\uparrow\right)
{\mathrm{Det}}\left(S^\downarrow\right)
\ee
The Jastrow factor, $u$, will contain electron-nucleus and electron-electron
correlations. Appendix B has details on the derivatives that enter into
the local energy.

 As two electrons or an electron and a nucleus get close, there is a 
singularity in the Coulomb potential.  That singularity needs to be canceled
by kinetic energy terms in the wave function.  This requirement is known as the 
cusp condition.  Details are given in Appendix C.

Techniques for the efficient handling of the determinants were 
developed by \cite{ceperley77}.
The VMC algorithm is implemented so that
only single electron trial moves are proposed.
This causes a change in only one column of the
Slater matrix.  
The new determinant and its derivatives can be computed in 
$\calO(N)$ operations, given the inverse of the old Slater matrix.
This inverse is computed once at the beginning of the simulation
and  then updated whenever a trial move is accepted.
The update takes $\calO(N^2)$ operations.
By comparison, computing the determinant directly takes $\calO(N^3)$ operations.
This technique creates a situation where
there is more work done for an acceptance than for a rejection.
A lower acceptance ratio will be faster, since fewer updates need to be
performed.
Details of the updating procedure and some other properties of determinants
are given in Appendix~ A.

Optimization of the parameters in the wave function is a large topic,
so we will defer the discussion until a later chapter.
However, we will make one observation here.
If $\psi_T$ is an eigenstate, the local energy becomes constant and
any MC estimate for the energy will have zero variance.  
This zero-variance
principle allows searching for optimum parameters by 
minimizing the variance rather than minimizing the energy.
In principle this is true for any eigenstate, not just the ground state.

\subsection{Two Level Sampling}
\label{vmc_twolevel}

A multilevel sampling approach can be used to increase the
efficiency of VMC \citep{dewing00}.
Multilevel sampling has been used extensively in path integral Monte Carlo
\citep{ceperley95}.
The general idea is to use a coarse approximation to the desired probability
function for an initial accept/reject step.  If it is accepted
at this first level, a more accurate approximation is used, and another
accept/reject step is made.  This continues until the move is rejected or until
the most detailed level has been reached.
This method increases the speed of the calculation 
because the entire 
probability function need not be computed every time.

Consider splitting the wave function into two factors - the single
body part ($D$) and the two body part ($e^{-U}$).
Treat the single body part as the first level, and the whole wave function
as the second level.

First, a trial move, $R'$, is proposed and accepted with probability
\be
A_1 = \min \left[1,\frac{D^2(R')}{D^2(R)}\right]
\ee
If it is accepted at this stage, the two body part is computed and the trial
move is accepted with probability
\be
A_2 = \min \left[1,\frac{\exp[-2U(R')]}{\exp[-2U(R)]} \right]
\ee
It can be verified by substitution that this satisfies detailed balance
in Eq. (\ref{db}).
After an acceptance at this second level, the inverse Slater matrices
are updated, as described previously.

We compared the efficiency between the 
standard sampling method and
the two level sampling method on two test systems: a single Li$_2$ molecule
in free space, and a collection of 32 H$_2$ molecules in a periodic box
of side 19.344 a.u. (r$_s$ = 3.0).  The wave functions are taken from 
\cite{reynolds82}, and will be described in Section \ref{wave_functions}.

The step size, $\Delta$, is the obvious
parameter to adjust in maximizing the efficiency.
But we can also vary the number of steps between
computations of $E_L$.  The Metropolis method produces correlated state
points
(see more on serial correlations in Section \ref{data-analysis}), 
so successive samples of $E_L$ do not contain much new information.
In these tests we sampled $E_L$ every five steps.

\begin{figure}
  \includegraphics[height=2.0in,trim=0 0 0 0]{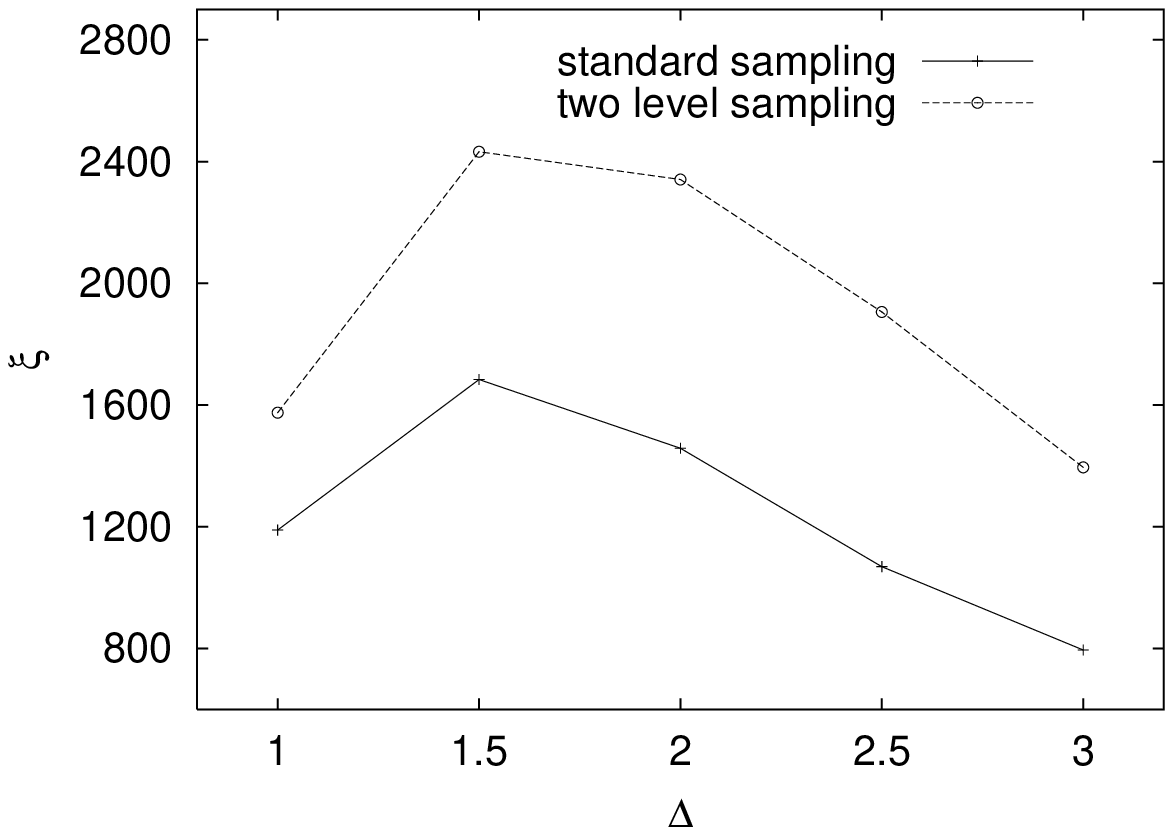}
  \hskip .2cm
  \includegraphics[height=2.0in,trim=0 0 0 0]{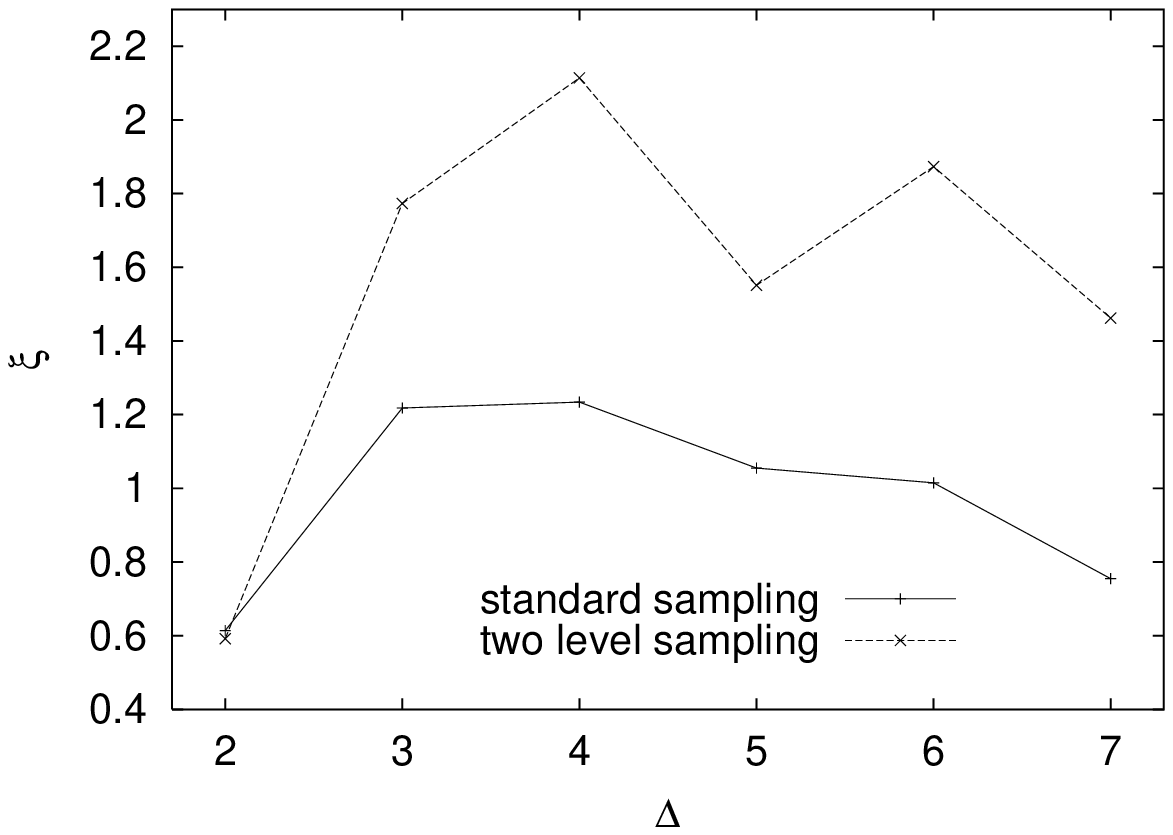}
  \caption{Efficiency of VMC. The graph on the left 
           is for Li$_2$.  The graph on the right is for 32 H$_2$ molecules.
   \label{vmc_efficiency}}
\end{figure}

Results for the different sampling methods with Li$_2$ are shown 
in Tables \ref{li2regular}  and 
\ref{li2twolevel}.  The Determinant Time and Jastrow Time columns
include only the time needed for computing the wave function ratio
in the Metropolis method, and not the time for computing the local energy.
The total time column does include the time for computing the local energy.
The efficiency is also shown on the left in Figure \ref{vmc_efficiency}.

For the two level method with Li$_2$, the second level acceptance ratio
is quite high, indicating the single body part is a good approximation
to the whole wave function.

Results for the collection of H$_2$ molecules are given in Tables 
\ref{h32regular} and \ref{h32twolevel}.  The efficiency is also
shown on the right graph in Figure \ref{vmc_efficiency}.

Comparing the maximum efficiency for each sampling method, two
level sampling is 39\% more efficient than standard sampling for Li$_2$,
and 72\% more efficient for the collection of H$_2$'s.

\begin{table}\caption{Timings for Li$_2$ molecule using
the standard sampling method. All times are in seconds on an SGI Origin 2000.}
\label{li2regular}
\begin{center}
\begin{tabular}{cccccc}
               &  Acceptance &   Determinant   & Jastrow  & Total  &\\
$\Delta$       &   Ratio     &     Time        &   Time   &  Time & $\xi$\\
 \hline
 1.0    & 0.610   & 48.3  & 340  & 516 & 1190 \\
 1.5    & 0.491   & 48.1  & 340  & 508 & 1680 \\
 2.0    & 0.407   & 48.2  & 340  & 503 & 1460 \\
 2.5    & 0.349   & 48.2  & 339  & 499 & 1070 \\
 3.0    & 0.307   & 48.2  & 339  & 496 &  800 \\
\end{tabular}
\end{center}
\end{table}

\begin{table}\caption{Timings for Li$_2$ molecule using
the two level sampling method. All times are in seconds on an SGI Origin 2000.}
\label {li2twolevel}
\begin{center}
\begin{tabular}{cccccc}
    &  First Level     &  Second Level  & Total Acc. &  \\
$\Delta$  & Acc. Ratio & Acc. Ratio & Ratio &Time &  $\xi$ \\
 \hline
1.0       & 0.674   & 0.899   & 0.606  & 400  & 1580 \\
1.5       & 0.543   & 0.894   & 0.485  & 347  & 2430 \\
2.0       & 0.447   & 0.897   & 0.401  & 304  & 2340 \\
2.5       & 0.379   & 0.902   & 0.342  & 276  & 1910 \\
3.0       & 0.331   & 0.906   & 0.300  & 256  & 1400 \\
\end{tabular}
\end{center}
\end{table}


\begin{table}\caption{Timings for the system of  32 H$_2$ molecules in a 
periodic box using
the standard sampling method. All times are in seconds on a Sun Ultra 5.}
\label{h32regular}
\begin{center}
\begin{tabular}{cccccc}
               &  Acceptance &   Determinant   & Jastrow  & Total  &\\
$\Delta$       &   Ratio     &     Time        &   Time   &  Time  & $\xi$ \\
  \hline
 2.0    & 0.606   & 167  & 1089  & 2015 & 0.61 \\
 3.0    & 0.455   & 167  & 1085  & 1891 & 1.22 \\
 4.0    & 0.338   & 166  & 1084  & 1794 & 1.23 \\
 5.0    & 0.250   & 166  & 1080  & 1722 & 1.06 \\
 6.0    & 0.185   & 164  & 1080  & 1668 & 1.02 \\
 7.0    & 0.139   & 162  & 1084  & 1629 & 0.76 \\
\end{tabular}
\end{center}
\end{table}

\begin{table}\caption{Timings for a system of 32 H$_2$ molecules in a 
periodic box using
the two level sampling method. All times are in seconds on a Sun Ultra 5.}
\label {h32twolevel}
\begin{center}
\begin{tabular}{cccccc}
    &  First Level     &  Second Level  & Total Acc. & Total & \\
$\Delta$  & Acc. Ratio & Acc. Ratio & Ratio &Time &  $\xi$ \\
 \hline
2.0       & 0.740   & 0.795   & 0.589  & 1804  & 0.59 \\
3.0       & 0.598   & 0.728   & 0.436  & 1421  & 1.77 \\
4.0       & 0.468   & 0.681   & 0.319  & 1185  & 2.11 \\
5.0       & 0.357   & 0.649   & 0.232  &  994  & 1.55 \\
6.0       & 0.370   & 0.627   & 0.169  &  849  & 1.87 \\
7.0       & 0.204   & 0.609   & 0.124  &  740  & 1.46 \\
\end{tabular}
\end{center}

\end{table}

\section{Diffusion Monte Carlo}

Diffusion Monte Carlo (DMC) is a method for computing the ground
state wave function.  It typically takes an order of magnitude more
computing time than VMC, and is most efficient when used in conjunction with 
a good VMC trial function. 

Formally, DMC and related methods work by 
converting the differential 
form of the Schr\"odinger equation into an integral equation and solving
that integral equation by stochastic methods. 
From another point of view, the Schr\"odinger equation in imaginary time and the 
diffusion equation are very similar, enabling one to use a 
random process to solve the imaginary time Schr\"odinger equation.
This similarity was recognized early and
was proposed as a computational scheme in the early days of computing
\citep{metropolis49}.
Unfortunately, without importance sampling, it is very inefficient 
computationally.

The ground state wave function can be obtained by the projection
\be
\phi_0  =  \lim_{\tau \ra \infty} e^{-\tau (H-E_0)} \psi_T
\ee
where $E_0$ is the ground state energy.
This can be seen by expanding $\psi_T$ in energy eigenstates,
\bea \nonumber
e^{-\tau(H-E_0)} \psi_T &=& e^{-\tau(H-E_0)} \sum_i \phi_i \\
 &=& \sum_n e^{-\tau(E_n-E_0)} \phi_n.
\eea
At large $\tau$, the contribution from the excited states will decay
exponentially, and only the ground state will remain.
To make a practical computation method, we write the projection in the
position basis as
\be
\psi(R',t+\tau) = \int dR \psi(R,t) G(R \rightarrow R'; \tau)
\label{green1}
\ee
where $G = \braket{R'}{e^{-\tau H}}{R}$ and $f(R,t)$ is the wave function
after some time $t$.  This equation is iterated to get to the large
time limit.
The fully interacting, many-body Green's function is too
hard to compute, so the various methods differ in how they approximate
the full projector.  In particular, DMC makes a short
time approximation, and the resulting pieces have natural interpretations
in terms of a diffusion process with branching.
The name Projector Monte Carlo or Green's Function Monte Carlo
is often applied to these methods.
Perhaps unfortunately, the name Green's Function Monte Carlo (GFMC) is
also applied to a specific technique that uses a spatial domain decomposition
for the Green's function.


For a more detailed presentation of DMC, with importance sampling, we
mostly follow \cite{reynolds82}.
  We start with the Schr\"odinger equation in imaginary time
\be
- \frac{\partial \phi(R,t)}{\partial t} = \left[ -\lambda \del^2 + V(R)
      - E_T \right] \phi(R,t)
\label{rawschrodinger}
\ee
where $\lambda = \hbar^2/2m$.
Importance sampling is added by 
multiplying
Eq. (\ref{rawschrodinger}) by a known trial function $\psi_T$.  The result,
written in terms of the ``mixed distribution''
$f(R,t) = \phi(R,t) \psi_T(R)$, is
\be
- \frac{\partial f(R,t)}{\partial t} =  -\lambda \del^2 f + (E_L(R)-E_T)f
    +\lambda \del \cdot (f F_Q(R))
\ee
where $E_L$ is the local energy, as in VMC,
and $F_Q = 2\del \psi_T/\psi_T$ (called the
quantum force).

Once again, the solution for $f$ 
in terms of a Green's function is
\be
f(R',t+\tau) = \int dR f(R,t) G(R \rightarrow R'; \tau)
\label{green}
\ee
For sufficiently short times, we can ignore the non-commutivity of 
the kinetic and potential terms in the Hamiltonian, 
$e^{-\tau H} \approx e^{-\tau T} e^{\tau V}$.
The explicit form for the short time Green's function in the position basis is
\begin{eqnarray}
G(R\rightarrow R';\tau) &=& (4\pi \lambda \tau)^{-3N/2}
      G_{\mathrm{branch}}(R\rightarrow R';\tau)
      G_{\mathrm{drift}}(R\rightarrow R';\tau) \\ 
 \label{shortgf}
G_{\mathrm{branch}}(R\rightarrow R';\tau) &=& \exp\left[ -\tau
\left\{ \bar E - E_T \right\} \right] 
 \label{dmc-branch}\\ 
G_{\mathrm{drift}}(R\rightarrow R';\tau) &=&  \exp\left[ -\left\{
R'-R-\lambda \tau F_Q(R) \right\}^2/4\lambda \tau \right]   
  \label{dmc-drift}
\end{eqnarray}
where $\bar E = \left[E_L(R) + E_L(R') \right]/2$.

The algorithm is started by generating a collection of configurations
(``walkers''), usually sampled from $\psi_T$.  
Equation (\ref{green}) proceeds by
applying a drifting random walk to each particle.  The new position
of the $i$th particle is given by
\be
\vec{r}'_i   =  \vec{r}_i + \lambda \tau  \vec{F}_Q(R) + \sqrt{2 \lambda \tau} 
    \ \vc{\chi}
\ee
where $\chi$ is  a normally distributed random variable with
zero mean and unit variance.
In a simple interpretation of Eq. (\ref{green}), this would always be the
new position.
But consider the case if $\psi_T$ becomes the true ground state, 
$\phi_0$. 
The branching term is
then constant and the algorithm becomes similar to VMC.
In this case we want to sample the correct distribution 
for any $\tau$.
This is done by adding a Metropolis rejection step, where
the trial move is accepted with probability
\be
A = \min \left[1,\frac{\psi_T(R')^2 G(R\ra R',\tau) }{\psi_T(R)^2 
       G(R'\ra R,\tau)} \right]
\ee
Each configuration is then weighted by
$G_{\mathrm{branch}}$.  
Because of rejections in the previous step, the time, $\tau$, in
Eq. (\ref{dmc-branch}) should be replaced by $\tau_{\mathrm{eff}}$, which is
\be
\tau_{\mathrm{eff}} = \frac{\expect{r^2_{\mathrm{accepted}}}}
                           {\expect{r^2_{\mathrm{total}}}} \tau
\ee
where $\expect{r^2_\mathrm{accepted}}$ is the mean square displacement 
of the accepted electron
moves and $\expect{r^2_\mathrm{total}}$ is the mean square displacement 
of all the proposed electron moves.

The weighting is done by adding or removing configurations from the collection
(branching).
This is done by computing the multiplicity
$M = \mathrm{int}(G_{\mathrm{branch}} + y)$,  where $y$ is a random
number on $[0,1]$.  This multiplicity is the number of copies of this
configuration that should be retained in the collection of walkers.
If it is zero, the configuration is deleted from the collection.  
If it is one, the configuration remains as is.  If it is greater than
one, additional copies of this configuration are added to the collection.



The number of walkers in the collection is kept roughly constant by 
adjusting $E_T$.
In particular, the trial energy is adjusted according to
\be
E_T = E_0 + \kappa \ln (P^*/P)
\ee
where $E_0$ is the best guess for the ground state energy,
$P$ is the current population, $P^*$ is the desired population,
and $\kappa$ is a feedback parameter.

The energy is computed by averaging the local energy over the
distribution of walkers.
Once the transients have decayed away, subsequent steps
are part of the ground state distribution.
The program is then run for however long is necessary to gather statistics
for the energy and other estimators.

There is a  problem with DMC for estimating quantities other
than the energy. The expectation value is not averaged over
the ground state, but over the mixed
distribution $\phi_0 \psi_T$.
This can be partly corrected by using the extrapolated estimator,
\be
 \braket{\phi_0}{A}{\phi_0} \approx 2 \braket{\phi_0}{A}{\psi_T} - 
       \braket{\psi_T}{A}{\psi_T}
\ee
.

Getting the correct estimator (also called a pure estimator) 
requires "forward walking", so named 
because the weight needed, $\phi_0/\psi_T$, is related to the asymptotic 
number of children of each walker \citep{liu74}.
 This can be implemented by
storing the value of the estimator and propagating it forward with
the walker for a given number of steps \citep{casulleras95}.


\subsection{Fermions}
  In all these methods, some quantity is treated as a probability, which 
requires
that it be positive.  In VMC this quantity is $\abs{\psi_T}^2$, which
is always positive. 
For DMC, we sample from $\psi_T \phi_0$, which can be negative
if the fermion nodes of $\psi_T$ are not the same as the nodes of $\phi_0$.  
The simplest cure is to fix the nodes of the ground state to be the same as 
$\psi_T$. 
This is known as the fixed-node approximation.
It is implemented in the DMC algorithm by rejecting moves that would
change the sign of the determinant of $\psi_T$.

\section{Statistical Errors}
\label{data-analysis}

The formula for the variance given in Eq. (\ref{variance}) assumes
that there are no serial correlations in the data.
However, the Metropolis sampling method produces correlated data, which
must be considered when estimating the statistical error.

Correlations in data are quantified by the autocorrelation function, 
defined for some estimator, $E$, as
\be
C(k) = \frac{1}{(N-k)} \sum_{i=1}^{N-k} (E_i - \bar E) (E_{i+k} - \bar E)
\ee
The autocorrelation time, $\kappa$, is computed as
\be
\kappa = 1 + \frac{2}{\sigma^2} \sum_{k=1}^{\mathrm{cutoff}} C(k)
\ee
This sum tends to be quite noisy.  As a heuristic strategy, we can approximate
$\kappa$ by the first place where the autocorrelation function drops
below 10\%.  The true variance of the mean is the simple variance 
of the individual data points multiplied by $\kappa$.

\begin{figure}
  \includegraphics[height=2.0in,trim=0 0 0 0]{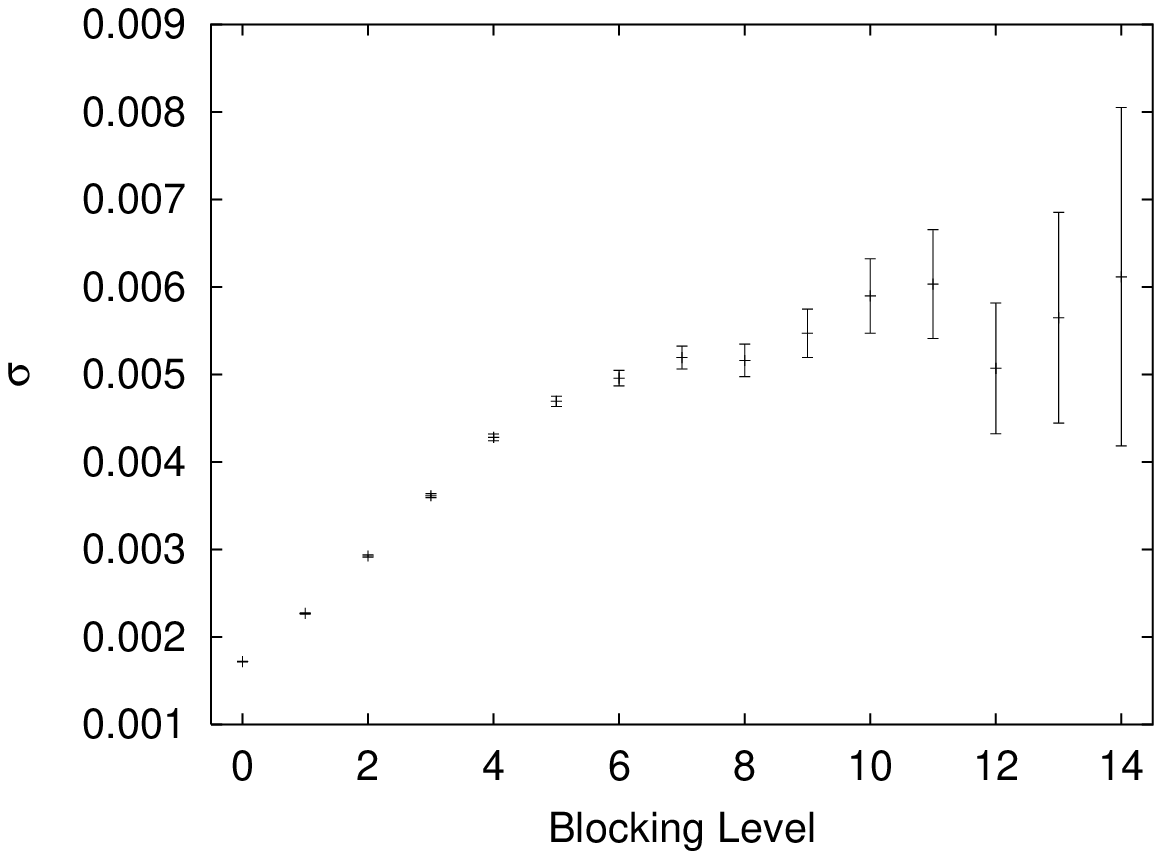}
  \hskip .2cm
  \includegraphics[height=2.0in,trim=0 0 0 0]{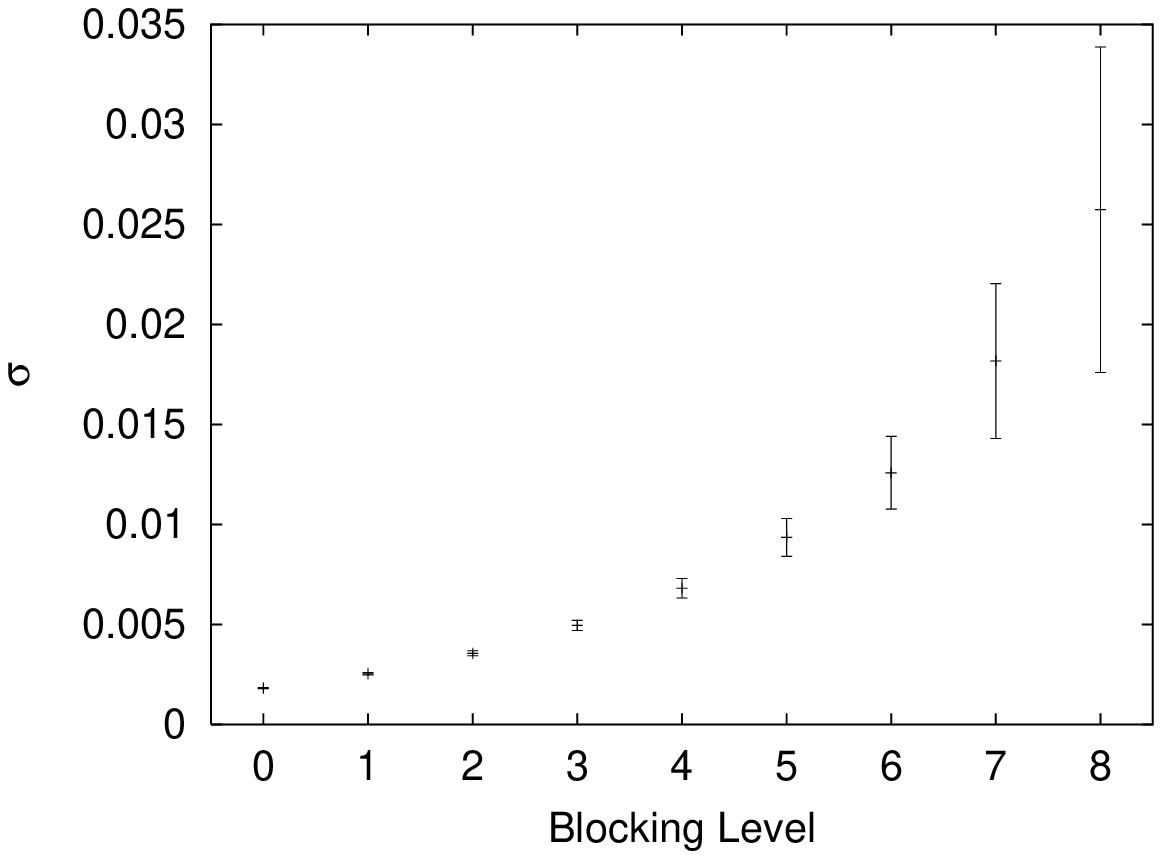}
  \caption{Examples of statistical data analysis using reblocking. 
           The error in the graph on the left 
           has converged,
          while the error in the graph on the right has not.
   \label{reb_example}}
\end{figure}

Another way to estimate the true error is by reblocking   
\citep{flyvbjerg89, nightingale99}.
At the second
level, take the average of every 2 data points.   Now compute the
variance of this set of data that has $N/2$ points.   Continue
this procedure recursively until the variance stops changing. 
\cite{nightingale99} gives a well-defined procedure for computing when
that occurs.
Figure \ref{reb_example} shows some example plots of error vs. reblocking
level.  On the left hand graph we see the expected plateau in the error
estimate.  On the right hand graph there is no plateau, indicating
that there is not enough data to reliably estimate the error.

\section{Wave Functions}
\label{wave_functions}

For our studies of molecular hydrogen,
we started with the wave function $\psi_{\mathrm{III}}$ from Reynolds 
\citep{reynolds82}.
The Jastrow factors are
\bea \nonumber
u_{ee} = -\sum_{i<j} \frac{a_e r_{ij}}{1 + b_{ee} r_{ij}} \\
u_{ne} = \sum_{i,\alpha} \frac{Z_\alpha a_n r_{i\alpha}}{1+b_{en} r_{i\alpha}}
\eea
where $Z$ is the nuclear charge and $b$ is the variational parameter.
The cusp conditions are satisfied by setting $a_n = 1$ and $a_e = 1/2$.
As noted in Appendix C, having the correct cusp condition for parallel spins
does not affect the energy much, so the same value for $a_e$ is used
for parallel and antiparallel electron spins.
The $b's$ from the two types of Jastrow factors are folded into
a single parameter, $\beta = a/b^2$.

The orbitals are floating Gaussians, with the form
\be
\phi_l(r) = \exp\left[\frac{-(r-c_l)^2}{w_l^2}\right]
\ee
where $c_l$ is the center of orbital, and $w_l$ is a
free parameter.  In molecular hydrogen, $c_l$ will be fixed at
the bond center.

The orbitals can be generalized to be anisotropic,
\be
\phi_l(r) = \exp\left[-(\vc{r}-\vc{c}_l)^T \cdot \vc{R}^T \Gamma \vc{R} 
              \cdot (\vc{r}-\vc{c}_l) \right]
\ee
where $\Gamma$ is a diagonal tensor and $\vc{R}$ is a rotation
matrix.  There are two parameters - the  width along the bond direction
(rotated so as to be the z-axis), and the width perpendicular to the bond
direction.  The elements of $\Gamma$ are defined to be 
$( 1/w_{xy}^2, 1/w_{xy}^2, 1/w_z^2)$.

Finally, additional energy reduction was found for the isolated H$_2$ molecule
by multiplying the orbital by $(1+\zeta \abs{\vc{r}-\vc{c}_l})$, where $\zeta$
is a variational parameter.

\section{Periodic Boundary Conditions}

The effects of an infinite system can be approximated by imposing periodic 
boundary conditions on a finite system.   
Every particle in the system then has an infinite number
of images.  Inter-particle distances are calculated 
using the minimum image convention, which uses only the distance to the closest
image.

Care needs to be taken with the wave function when using the minimum image
convention.  As the inter-particle distance crosses over from one image 
to another there
can be a discontinuity in the derivative of the wave function, leading to
a delta function spike in the energy.  If this is not accounted for,
the VMC energy can become lower than the true ground state because this 
delta function term in the energy 
has been neglected.  
Additionally, the Gaussian orbitals can lower their energy by 
having a width comparable to or larger than the box size.
Then sections of the orbital with large kinetic energy are outside
$L/2$, and do not get counted in the integral.
  This can be fixed by summing over images, or by insuring
the wave functions have the correct behavior at $\pm L/2$.
We use the latter solution.
  
The orbitals are multiplied by a cutoff function that ensures its value
and first derivative are zero at the box edge.  The function we use is
\be
f_c(r) = 1- \exp\left[-\gamma_c (r-r_m)^2\right]
\ee
where $r_m$ is fixed at $L/2$ and $\gamma_c$ is a variational parameter.

The Jastrow factors are constructed so that they obey the correct cusp 
conditions
as $r\rightarrow 0$ and so that the first and second derivatives are zero at 
$r_m \leq L/2$.   The simplest function that satisfies these conditions is 
a cubic polynomial.  Let $y=r/r_m$.  Then
\be
u(y) = a_1 y + a_2 y^2 + a_3 y^3, 
\ee
where $a_1 = {\rm (\hbox{cusp value})}*r_m$, $a_2 = -a_1$, and $a_3 = a_1/3$.
Variational freedom is gained by varying $r_m$, and by adding a general
function multiplied by $y^2 (y-1)^3$ to preserve the boundary conditions.
We choose a sum of Chebyshev polynomials
as the general function \citep{williamson96}.
The full Jastrow factor is then
\be
u(y) = a_1( y - y^2 + \frac{1}{3}y^3) + y^2(y-1)^3 \sum_i b_i T_i(2y-1),
\ee
where $r_m$ and the $b_i$ are variational parameters.
We use five Chebyshev polynomials for the electron-electron part  and
another five for the electron-nuclear part.
We optimized one set of $r_m$ and $b_i$ parameters for all electron-electron
pairs in any particular system, 
and another set of parameters for all electron-nuclear pairs.

\begin{figure}
 \begin{center}
  \includegraphics[height=3.0in,trim=0 0 0 0]{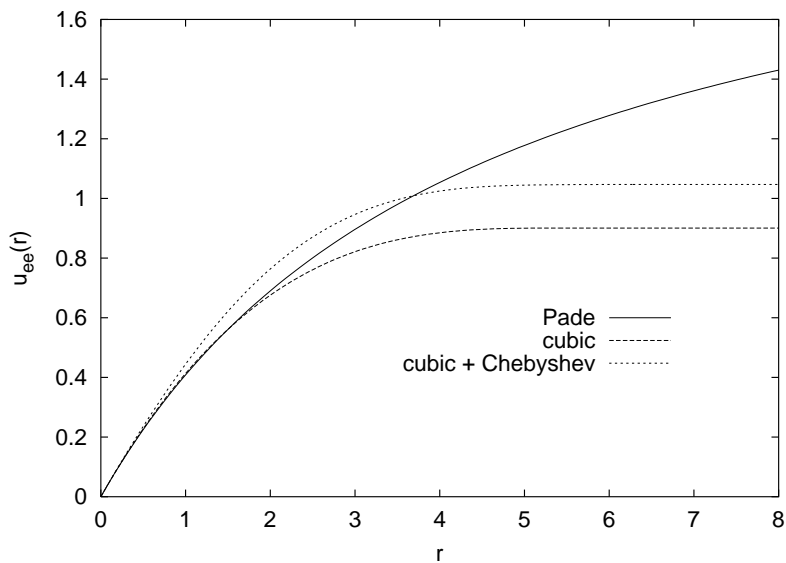}
  \caption{Optimized electron-electron Jastrow factor for different forms.
   \label{u_example}}
 \end{center}
\end{figure}

Comparisons of the energy and variance of various combinations 
of forms for the orbital and Jastrow
factors are shown in Table \ref{h2-energy}.  The variational parameters
are given in Tables \ref{pade-parameters} and \ref{cubic-parameters}.
A comparison of the electron-electron Jastrow factors is shown in Figure
\ref{u_example}.    Their short range behavior is similar, but the long
range behavior differs between the types of Jastrow factors.

The quality of wave functions is often measured by the percent of 
correlation energy recovered. For H$_2$, the HF (no correlation) 
energy is $-1.1336$ Hartrees 
and the exact (full correlation) energy is $-1.17447$ Ha.
\cite{sun89} compared a number of forms for electron correlation functions.
Their best value recovered $80\%$ of the correlation energy.
Using one of these forms, but with better optimization, 
\cite{huang90} recovered $84\%$ of the correlation energy.
\cite{snajdr99} obtained $93\%$ of the correlation energy using
a Linear Combination of Atomic Orbitals (LCAO) 
form with 1s, 2s and 2p orbitals, and using the Jastrow factors
of \cite{schmidt90}.

The variance is higher with those wave functions involving the cubic
polynomial, even though the energy is lower.
I believe this is mostly likely because the cubic polynomial does not have the 
correct $1/r$ behavior at large $r$, but the simple Pad\'e form does.  This
long range behavior contributes little to the average of the energy, but
it contributes more significantly to the variance.

\begin{table}
\caption{ Comparison of energies and variances for various forms 
           for orbitals and Jastrow factors for a single H$_2$ molecule.
  \label{h2-energy}}
\begin{center}
\begin{tabular}{ccc|ccc}
$\Psi$ &Orbital  & Jastrow  &   Energy & Variance & \% CE \\ \hline
A &  Isotropic  & simple Pad\'e            & $-1.1598(4)$ & 0.046 & 64.0(9) \\
B & Anisotropic & simple Pad\'e           & $-1.1643(2)$ & 0.040 & 75.0(6) \\
C & Anisotropic + $\zeta$ & simple Pad\'e      & $-1.1653(2)$ & 0.033 & 77.5(5) \\
D & Anisotropic + $\zeta$ & cubic            & $-1.1688(2)$ & 0.039 & 86.1(6) \\
E & Anisotropic + $\zeta$ & cubic+Chebyshev & $-1.1702(2)$ & 0.046 & 89.6(5) \\
\end{tabular}
\end{center}
\end{table}




\begin{table}
\caption{Values of variational parameters for H$_2$.
 \label{pade-parameters}}
\begin{center}
\begin{tabular} {c|l|l}
$\Psi$  & Orbital parameters  &  Jastrow parameters    \\ \hline
A &  $w = 2.74$  & $\beta = 9.913$   \\  
B & $w_{xy} = 2.514$, $w_{z} = 2.977$   & 
      $\beta = 10.002$ \\
C & $w_{xy} = 2.416$, $w_{z} = 2.833$, $\zeta = 0.0445$  
    & $\beta = 9.958$ \\
D & $w_{xy} = 2.357$, $w_{z} = 2.628$, $\zeta = 0.248$ 
  & e-e $r_m = 5.404$ \\
 \multicolumn{2}{c|}{} &  e-n $r_m = 5.376$ \\
\end{tabular} 
\end{center}
\end{table}
\begin{table}

%


\caption{Values of variational parameters for wave function E
 \label{cubic-parameters}}
\begin{center}
\begin{tabular} {c|r@{ $=$ }lr@{ $=$ }lr@{ $=$ }l}
Component  &  \multicolumn{3}{c}{Parameters} \\\hline
Orbital     & $w_{xy}$ & 2.299 & $w_{z}$ & 2.515&  $\zeta$ & 0.301   \\
Electron-electron Jastrow   &
   $r_m$ & 6.281  & $b_0$ & -1.012 & $b_1$ &  0.193  \\
 & $b_2$ & 0.619  & $b_3$ & 0.025 & $b_4$ & 0.138  \\
 Electron-nuclear Jastrow  &
   $r_m$ & 5.329  & $b_0$ & -2.084 & $b_1$ &  0.153  \\
 & $b_2$ & 0.952  & $b_3$ &  1.217 & $b_4$ &  1.027  \\
\end{tabular} 
\end{center}
\end{table}

%% file: chap3.tex
\chapter{Energy Difference Methods}

Very often it is the difference in energy between two systems that
is of interest, and not the absolute energy of a single system.
For a quantity such as the binding energy,
we want the difference between the energy of the 
molecule and the energy of the free atoms.
In our CEIMC simulations, we want the change in energy from moving a
few nuclei.
In VMC optimization, we want to know the change in energy from 
modifying some of the wave function parameters.


Correlated sampling methods can provide a more efficient approach to 
computing these energy differences.  But the widely used reweighting
method has some drawbacks. We will introduce a new method that
alleviates some of the drawbacks of reweighting while retaining its advantages.


\section{Direct Difference}
The simplest, and most straightforward way of computing the difference in energy
between two systems is to perform independent computations of the energy for 
each system.
Then the energy difference and error estimate are simply
\bea
\Delta E &=& E_1 - E_2 \\
\sigma (\Delta E) = &=& \sqrt{\sigma_1^2 + \sigma_2^2} 
\eea
This method is simple and robust, but has the drawback that 
the error is related to the error in computing a single system.
If the systems are very similar, either in variational parameters
or in nuclear positions, the energy difference is likely to be small
and difficult to resolve, since $\sigma_1$ and $\sigma_2$ are determined
by the entire system.
Similarities between the systems can be exploited with correlated
sampling.

\section{Reweighting}

 Reweighting is the simplest correlated sampling method.
\bea \label{rew} \nonumber
\Delta E &=& E_1 - E_2  \\ \nonumber
         &=&  \frac{\int dR \ \psi_1^2 \ E_{L1}}{\int dR \ \psi_1^2}- 
              \frac{\int dR \ \psi_2^2 \ E_{L2}}{\int dR \ \psi_2^2} \\
         &=&  \frac{\int dR \ \psi_1^2 \ E_{L1}}{\int dR \ \psi_1^2}-
           \frac {\int dR\ \psi_1^2 \left(\frac{\psi_2^2}{\psi_1^2}\right)\
              E_{L2} }
              {\int dR\ \psi_1^2\ \left(\frac{\psi_2^2}{\psi_1^2}\right) }
\eea

An estimate of $\Delta E$ for a finite simulation is
\be \label{rew_sum}
\Delta E \approx \frac{1}{N}\sum_{R_i \in \psi^2_1}
     \left[E_{L1}(R_i) - \frac{ w(R_i) E_{L2}(R_i)}
    { \sum_i w(R_i)} \right]
\ee
 where $w = \psi_2^2/\psi_1^2$.  The same set of sample points
is used for evaluating both terms.

Reweighting works well when $\psi_1$ and $\psi_2$ are not too different, and
thus have large overlap.  As the overlap between them decreases, reweighting
gets worse due to large fluctuations in the weights.
This effect can be quantified by computing
the effective number of points appearing in the sum in Eq. (\ref{rew_sum}), 
which is
\be
N_{\rm eff} = \frac{\sum_i w_i^2}{(\sum_i w_i)^2}
\ee
Eventually, one or a few large weights will come to dominate the sum, and
the effective number of points will be very small, and the variance in
$\Delta E$ will be very large.


Particularly pernicious is the case when the nodes differ between the
two systems.  The denominator of the weight can easily be very small,
causing a very large weight value.
This is encountered when using reweighting to optimize orbital parameters
in VMC
\citep{barnett97}.  

In Eq. (\ref{rew_sum}) we derived reweighting by drawing points from $\psi_1$ 
and computing the properties of both systems from them. 
It could also be derived by drawing points from $\psi_2$ as well.
We can compute the energy difference both ways and take the average. 
This gives us
the symmetrized reweighting method,
\bea \nonumber 
\Delta E =&& \frac{1}{2N}\sum_{R_i \in \psi^2_1}
     \left[E_{L1}(R_i) - \frac{ w_x(R_i) E_{L2}(R_i)}
    { \frac{1}{N}\sum_i w_x(R_i)} \right] \\
   &+&  \frac{1}{2N} \sum_{R_i \in \psi^2_2}
     \left[\frac{ w_y(R_i) E_{L1}(R_i)}{ \frac{1}{N}\sum_i w_y(R_i) }
     - E_{L2}(R_i)\right] 
\label{sym_rew_sum}
\eea
where $w_x = \psi_2^2/\psi_1^2$ and $w_y = \psi_1^2/\psi_2^2$.

\section{Bennett's Method for Free Energy Differences}

First let us digress to discuss computation of the normalization integral.
It was mentioned earlier that the Metropolis sampling method makes
it difficult to extract information about the normalization integral,
which the partition function in the classical case.  
\cite{bennett76} demonstrated a method for finding the free energy
difference between two systems.  We will describe his method in terms
of a ratio of normalizations.

We can compute the ratio of two normalizations, $Q_1$ and $Q_2$, 
in a fashion very similar to reweighting.
\bea
 Q_1/Q_2 &=& \int dR\ \psi^2_1(R)  \bigg/ \int dR\ \psi^2_2(R) \\
  &=& \int dR\ \psi^2_2(R) \frac{\psi^2_1(R)}{\psi^2_2(R)}  \bigg/ 
        \int dR\ \psi^2_2(R) \\
  &\approx& \sum_{R_i \in \psi_2} \frac{\psi^2_1(R_i)}{\psi^2_2(R_i)}
     \label{one-sidedQ}
\eea
This is a one-sided estimate, because it only uses samples from system two
to compute properties of system one.
Note that this sum is the same as the sum of the weights used in reweighting
 in Eq. (\ref{rew_sum}).


Bennett improved on this one-sided estimate,
starting with an identity written as
\be
  Q_1/Q_2 = \frac{ Q_1 \int dR\ \psi^2_2\  W \psi^2_1}
                 { Q_2 \int dR\ \psi^2_1\  W \psi^2_2} 
 \label{bennett-identity}
\ee
where $W$ is an arbitrary weight function.  He found the optimum $W$ by
minimizing the variance of the free energy difference,
\be
 W \propto  \frac{1}{Q_1 \psi_2^2 + Q_2 \psi_1^2}
 \label{w-val}
\ee

Let $Q = Q_1/Q_2$. Inserting Eq. (\ref{w-val}) into 
Eq. (\ref{bennett-identity}), we get
\be
1 =   \frac{\int dR\ \frac{\psi_2^2}{Q_2}  \psi_1^2 / \left(\psi_2^2 + 
        \psi_1^2/Q \right)}
          {\int dR\ \frac{\psi_1^2}{Q_1} \psi_2^2 / \left(Q \psi_2^2 + 
         \psi_1^2 \right)}
\ee
Let $x$ represent the configurations sampled from $\psi_1$ and $y$ 
the configurations sampled from $\psi_2$.
The finite sample version of this equation is
\be
\label{defineQ}
\sum_{i} 
    \frac{ Q \psi_2^2(x_i)}{ \psi_1^2(x_i) + Q \psi_2^2(x_i) } 
 = 
\sum_{i} 
    \frac{ \psi_1^2(y_i)}{ \psi_1^2(y_i) + Q \psi_2^2(y_i)}  
\ee

The value of $Q$ can be found by a simple iteration
\be
Q_{n+1} =  Q_n \left[\frac{
     \sum_{y} \frac{ \psi_1^2(y_i)}{ \psi_1^2(y_i) + Q_n \psi_2^2(y_i)}
    }{
  \sum_{x} \frac{ Q \psi_2^2(x_i)}{ \psi_1^2(x_i) + Q_n \psi_2^2(x_i) }
    }\right]
\ee
The iteration is started with $Q_0 = 1$  and stopped when the correction
factor in brackets is sufficiently close to one.  
Typically convergence takes less that ten iterations, 
but if $Q$ is much larger or smaller than one it can take more iterations.


We have written these formulas assuming that the number of sample points 
from each system is the same. 
Bennett derived them for case with differing numbers of samples in each sum, 
and found the best variance was usually
very near an equal ratio of computer time spent on each system.
In our case the systems are of equal complexity, so this means using
equal numbers of points is optimal, or very nearly so.

By properly combining information from both systems, we can get a much better
(lower variance) estimate of the ratio of their normalizations 
than if we had used information from only a single system (one-sided sampling).

\section{Two-Sided Sampling}

We can apply this notion to computing the energy difference between
two quantum systems.
Consider sampling from some distribution that contains information
about both $\psi_1$ and $\psi_2$.
The simplest such distribution is
\be
P = \frac{1}{2}\left[ \frac{\psi_1^2}{Q_1} + \frac{\psi_2^2}{Q_2}\right]
\ee

The energy difference can be written as
\bea \nonumber
\Delta E &=&  \frac{\int dR\ \psi_1^2 E_{L1}}{Q_1}  -
                        \frac{\int dR\ \psi_2^2 E_{L1}}{Q_2}  \\ 
           &=& \int dR\ P \left (\frac{\psi_1^2 E_{L1}}{Q_1 P} \right)-
            \int dR\ P \left( \frac{\psi_2^2 E_{L2}}{Q_2 P} \right) 
\eea

In the finite case, we have
\be
\Delta E \approx \frac{1}{2N} \sum_{R_i \in x,y}
                           \frac{\psi_1^2(R_i) E_{L1}(R_i)}{Q_1 P} -
                            \frac{\psi_2^2(R_i) E_{L2}(R_i)}{Q_2 P}
  \label{defineDE}
\ee
It is important to note that the sum covers samples taken from both  
$\psi_1$ and $\psi_2$.
The sum includes both ``direct'' terms (eg. $\psi_1$ and $E_{L1}$ evaluated 
at configurations
sampled from $\psi_1$) and ``cross'' terms (eg. $\psi_1$ and $E_{L2}$ 
evaluated at configurations sampled from $\psi_2$).

The denominator of the first term in Eq. (\ref{defineDE}) is
\be
Q_1 P = \frac{1}{2} \left[ \psi_1^2 + \frac{Q_1}{Q_2} \psi_2^2 \right]
\ee
The ratio  $Q = Q_1/Q_2$ is computed by the Bennett method.
The denominator of the second term can be computed similarly.


One major feature of the two-sided method is that it reproduces reweighting
in the large overlap regime, and the direct method in the low overlap
regime.  In the intermediate regime, it smoothly joins the two limits.


  To show this, first consider the case where the two systems are very different
and the wave functions have low overlap. Here $\psi^2_1(y_i)$ and $\psi^2_2(x_i)$
will be small.
Expand Eq. (\ref{defineDE}) into its four terms
\bea \nonumber
\Delta E  =
&\underbrace{
 \frac{1}{2N} \sum_{x} \frac{\psi_1^2(x_i) E_{L1}(x_i)}
      {\half\left[ \psi_1^2(x_i) + Q \psi_2^2(x_i)\right]}
 - \frac{1}{2N} \sum_{y} \frac{\psi_2^2(y_i) E_{L2}(y_i)}
      {\half\left[ \psi_1^2(y_i)/Q + \psi_2^2(y_i)\right]}
}_{\mathrm{direct}}   \\ 
+&\underbrace{
  \frac{1}{2N} \sum_{y}  \frac{\psi_1^2(y_i) E_{L1}(y_i)}
      {\half\left[ \psi_1^2(y_i) + Q \psi_2^2(y_i)\right]}
 -\frac{1}{2N} \sum_{x}  \frac{\psi_2^2(x_i) E_{L2}(x_i)}
      {\half\left[ \psi_1^2(x_i)/Q + \psi_2^2(x_i)\right]}
}_{\mathrm{cross}} 
\eea
Each denominator will have one large term  ($\psi_1^2(x_i)$ or  $\psi_2^2(y_i)$)
and one small term ($\psi^2_1(y_i)$ or $\psi^2_2(x_i)$).
The value of Q is moderate compared to the wave functions, 
so it will not affect the relative sizes of these terms.
Always having one large term in the denominator means there will
never be any excessively large contributions to the sum 
resulting from division by a 
small value, as happens in reweighting.
The cross terms have a small value ($\psi_1^2(y)$ or $\psi^2_2(x)$) in
the numerator, and so vanish.
The large terms in the denominators in the direct terms cancel the $\psi^2$'s in
the numerator, and we are left with
\be
\Delta E \approx \frac{1}{N}\sum_x E_{L1}(x_i) - \frac{1}{N}\sum_y E_{L2}(y_i)
\ee
which is just the direct method.


Now for the case where the systems are very similar and have large overlap.
Recall from Eq. (\ref{one-sidedQ}) that we can write one-sided estimates for 
Q as
\be
Q = \frac{1}{N}\sum_y w_y(y_i)
  =  1 \bigg/ \frac{1}{N}\sum_x w_x(x_i)
\ee
where $w_y = \psi^2_1(y_i)/\psi^2_2(y_i)$ and 
  $w_x = \psi^2_2(x_i)/\psi^2_1(x_i)$.
Write  the four terms of Eq. (\ref{defineDE}) in a different order
\bea \nonumber
\Delta E  =&&
 \frac{1}{N} \sum_{x} \frac{E_{L1}(x_i)}{1 + Q w_x(x_i)}
 -\frac{Q}{N} \sum_{x}  \frac{ w_x(x_i) E_{L2}(x_i) }
        { 1 + Q w_x(x_i)} \\
 &+& \frac{1}{Q N}\sum_{y}\frac{ w_y(y_1) E_{L1}(y_i)}
{1 + w_y(y_i)/ Q }
 - \frac{1}{N} \sum_{y} \frac{ E_{L2}(y_i)}{1 + w_y(y_i)/Q }
\label{large-limit}
\eea
Approximate the denominator of each term by two, replace the leading $Q$'s
with the appropriate one-sided approximation, and we get
\bea \nonumber
\Delta E \approx&& \frac{1}{2N} \sum_x \left[E_{L1}(x_i) - 
   \frac{w_x(x_i) E_{L2}(x_i) }
   {\frac{1}{N} \sum_x w_x(x_i) } \right] \\
  &+& \frac{1}{2N} \sum_y \left[ \frac{w_y(y_i) E_{L1}(y_i)}
  {\frac{1}{N} \sum_x w_y(y_i)} 
      - E_{L2}(y_i)\right]
\eea
which is the symmetric version of reweighting given in Eq. 
(\ref{sym_rew_sum}).

Due to computational considerations, it is useful to divide 
Eqns (\ref{defineQ}) and (\ref{defineDE}) by $\psi_1^2$ or $\psi_2^2$ as
appropriate, and work with the resulting ratios $w_x = \psi_2^2/\psi_1^2$
and $w_y = \psi_1^2/\psi_2^2$, as was done in Eq (\ref{large-limit}).
The values of the wave functions can easily over or under flow
double precision variables.  It is best to use the log of the
wave function, take differences, and then exponentiate.
Furthermore, an arbitrary normalization of the wave functions makes no
physical difference, but can result in very large or small numbers, even
after taking the difference of the logarithms.
This problem is ameliorated by subtracting the average value of the log
of the wave function from the individual values.
Sometimes this is not enough, however, and the value of the energy difference
exceeds the range representable in a double precision variable,
indicated by NaN (Not a Number).
In this case, the overlap is clearly very small and the two-sided method should
give the same results as the direct method.  The program checks for 
the energy difference being NaN, and if so, it substitutes the direct
method result (the data collected for the two-sided method is a superset of
that needed for the direct method).  Having done this, the subroutine
computing the two-sided energy difference will always return a reasonable
answer, an important consideration for a core routine in a program.

\section{Examples}

\begin{figure} 
 \begin{center}
  \includegraphics[height=4.5in,angle=270,trim=0 0 0 0]{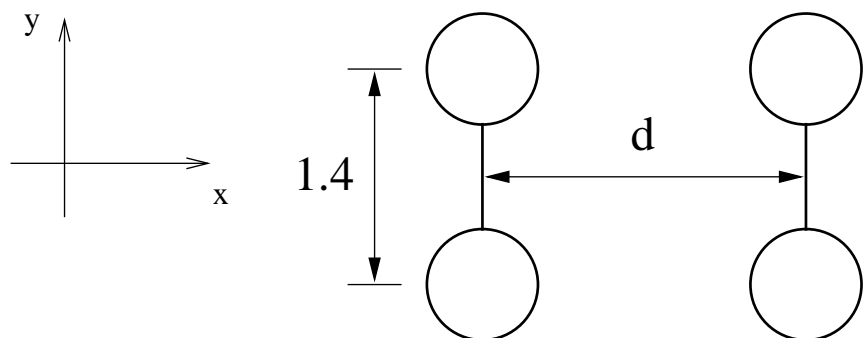}
  \caption{Two H$_2$ molecules in a parallel configuration
   \label{h2parallel} }
 \end{center}
\end{figure}

The first example is of two H$_2$ molecules in a parallel orientation
as shown in Figure \ref{h2parallel}.
The bond lengths are at equilibrium, 1.4 Bohr,  and the starting separation
between the molecules is $d=2.5$ Bohr.

\begin{figure}
 \begin{center}
  \includegraphics[height=2.0in,width=2.5in,trim=0 0 0 0]{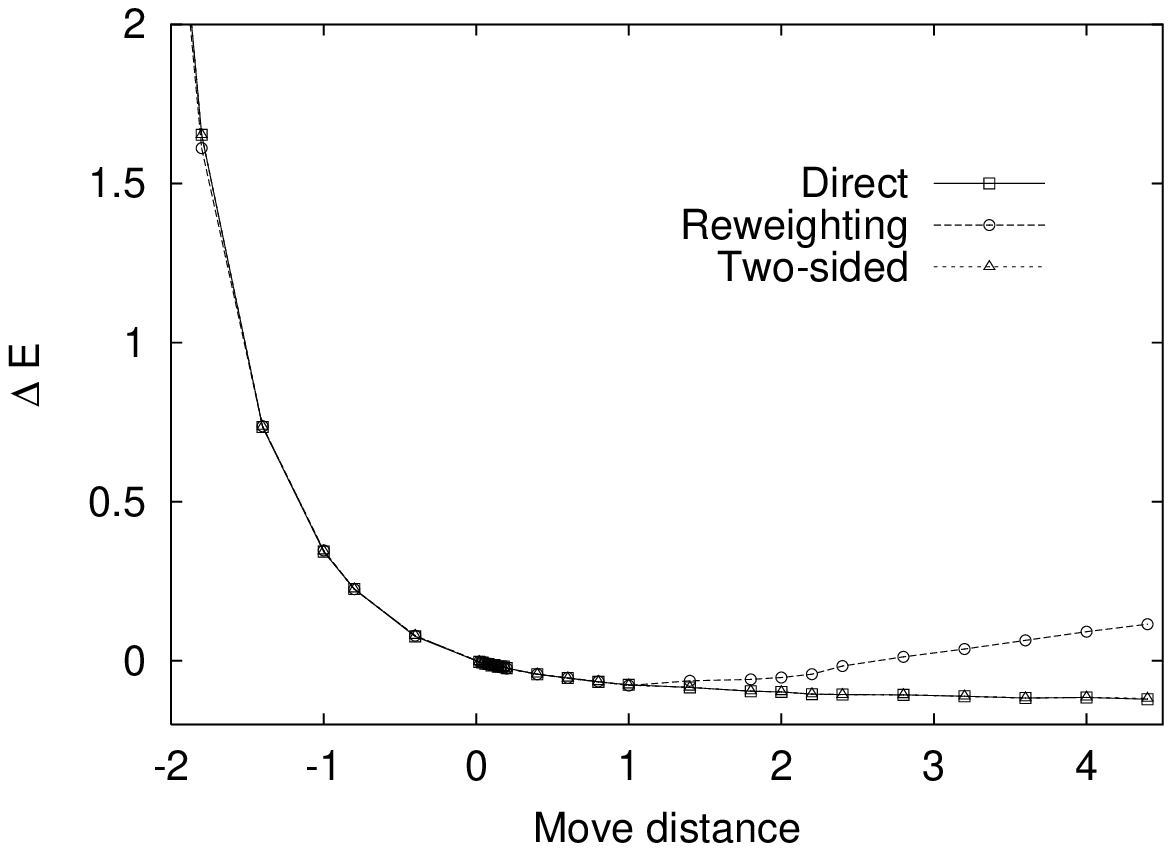}
  \hskip .5cm
  \includegraphics[height=2.0in,width=2.5in,trim=0 0 0 0]{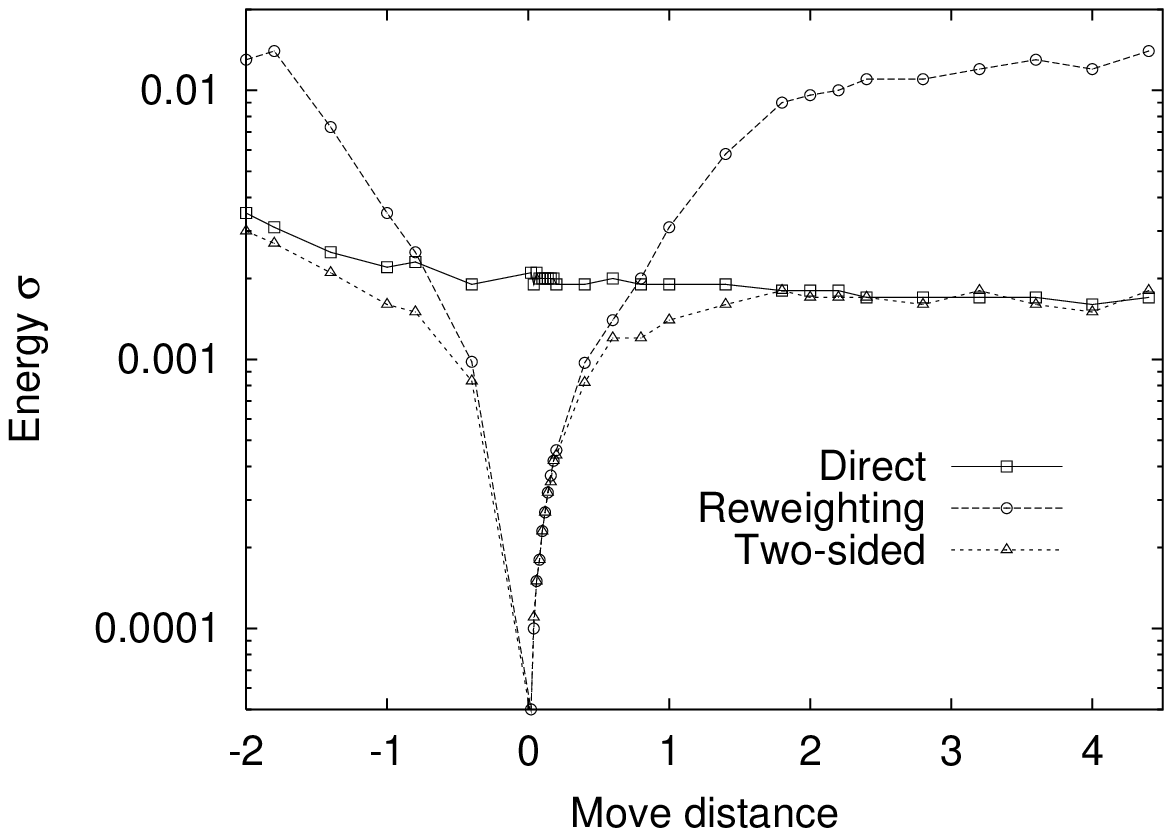}
  \caption{Energy difference (left) and the estimated statistical error 
 (on logscale) (right) for two 
        H$_2$ molecules in a parallel configuration, starting from d=2.5 Bohr.
   \label{h2endiff} }
 \end{center}
\end{figure}

The energy difference between that configuration and configurations
with other inter-molecular distances was computed using the direct
method, the two-sided method, and reweighting.
The resulting energy differences are shown on the left in 
Figure \ref{h2endiff}.   
Note that reweighting gets the wrong answer at large separations.
This is most likely due to a finite sample size bias.
More important is the error in that energy difference, shown on the right in 
Figure \ref{h2endiff}.
Note that both reweighting and the two-sided method have errors that
drop to zero as the overlap increases.
This graph also clearly shows the properties of the two-sided method 
mentioned previously, behaving like
reweighting at small changes in the separation (large wave function overlap), 
and smoothly crossing over to the direct method for at large changes in the 
separation (small wave function overlap).

\begin{figure} 
 \begin{center}
  \includegraphics[height=3.0in,trim=0 0 0 0]{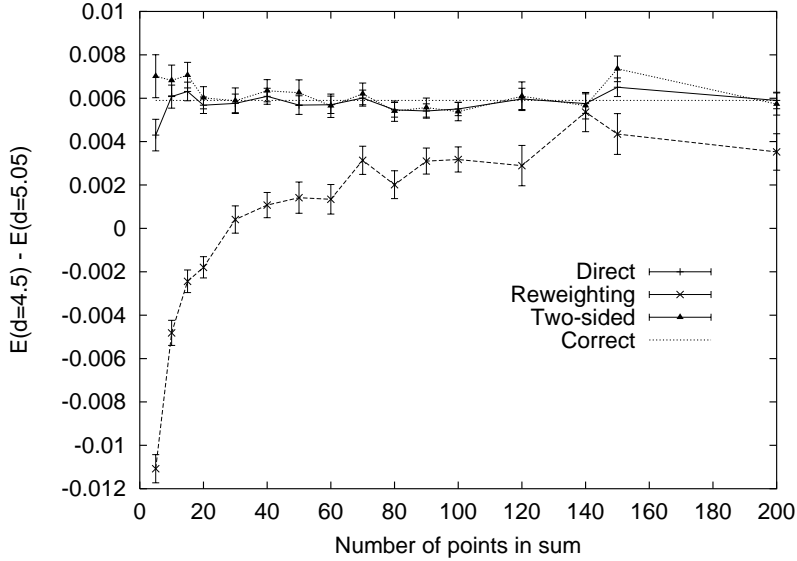}
  \caption{Finite sample size bias in the energy difference of Li$_2$.
   \label{li2bias}}
 \end{center}
\end{figure}

Reweighting and the two-sided method may give biased results
because  there are a finite number of sample points in the sums 
in Eqns. (\ref{sym_rew_sum}) and (\ref{defineDE}).
To test for this, a sum of a given length is repeated many times
and the average energy difference for that length computed.
The test for a bias was performed on a Li$_2$ molecule.
The energy difference was computed between
a bond length of 4.5 Bohr and the equilibrium bond length of 5.05 Bohr.
Figure \ref{li2bias} shows the results for different numbers of points
in the sum.    Reweighting shows a much larger finite sample size bias
than the two-sided method, which has almost none.

\subsection{Diffusion Monte Carlo}

Using the two-sided method (or reweighting, for that matter) with
DMC is slightly more complicated. 
The reweighting transformation applied to the basic DMC iteration gives
\bea
 f_1(R';t+\tau)  &=& \int dR\ f(R;t) G_1(R\ra R';\tau) \\
    &=& \int dR\ f(R';t) G_2(R\ra R';\tau) \frac{G_1(R\ra R';\tau)}
          {G_2(R\ra R';\tau)}
\eea
The weight $w=G_1/G_2$ must be computed over several iterations.
The final weight used in the correlated sampling formulas is a product of 
the weights of every iteration.

The weight factor is not quite right, due to the rejection step.
Since the rejection ratio for DMC is very small ($<1\%$), 
ignoring the issue should not introduce a large error.
\cite{umrigar00} give a more sophisticated method for dealing with rejections.

The fixed-node condition also has to be obeyed, and configurations
that cross a node while projecting have their weight set to zero.

A version of reweighting was implemented by \cite{wells85} 
as the differential Green's function Monte Carlo method (actually DMC).
He used the response to an external field to determine the dipole 
moment of LiH.  The same trial function
was used, so the drift term was the same between both systems. Only the
branching term was different; that entered as a weight factor.
In our case, the trial function and the nuclear positions may be different
between the two systems.

\begin{figure} 
 \begin{center}
  \includegraphics[height=3.0in,trim=0 0 0 0]{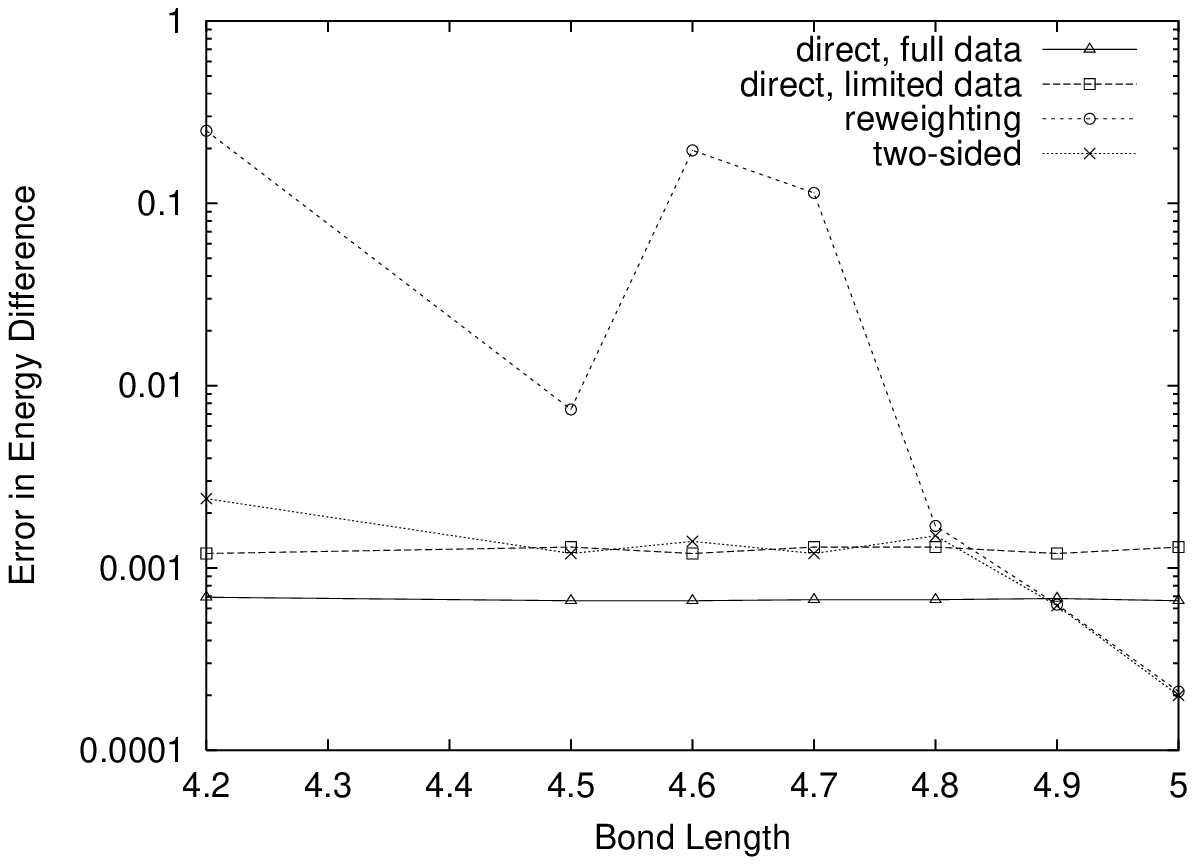} \\
  \includegraphics[height=3.0in,trim=0 0 0 0]{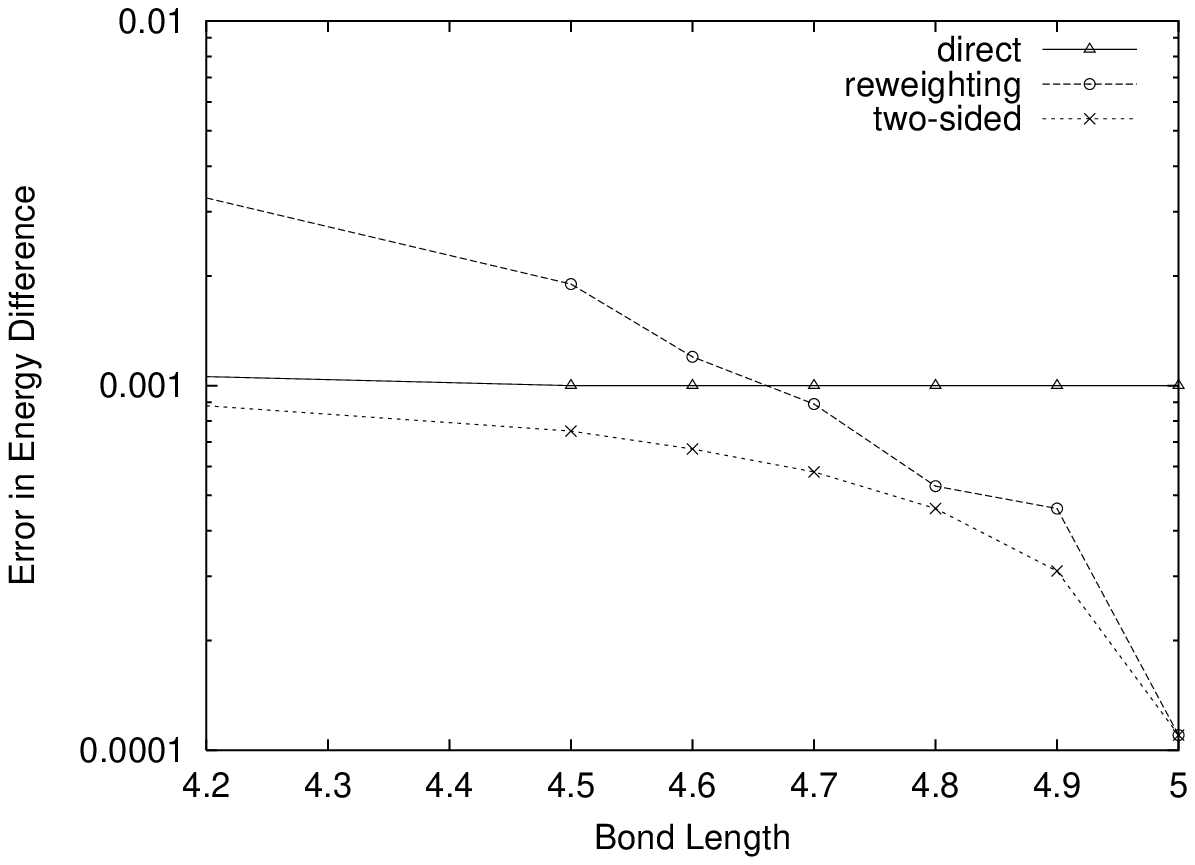} \\
  \caption{Error in energy difference of Li$_2$ using DMC (top) and 
           VMC (bottom)
   \label{qmcli2}}
 \end{center}
\end{figure}

The top of Figure \ref{qmcli2} shows the difference in DMC energies 
using the various methods.
The energy difference was computed starting
from the equilibrium bond length of 5.05 Bohr.
Partly because of the need to project for several DMC steps, the 
two-sided method has a fairly small range where it does better than
the direct method  (compared with the range for VMC).
For comparison, the VMC results are shown at the bottom of Figure \ref{qmcli2}.
There are two lines in the DMC graph for the direct method.  
The implementation had
a limitation where only one projection to accumulate the weights
would occur at a time.  We used 30 steps in the projection, and consequently
could only get the weights once every 30 steps.  
The limited data line
is computed from data collected once every 30 steps (the same
amount of data available to the correlated methods) and is the
line the two-side method joins on to.
The full data line used all the data available in the simulation
and so has a lower statistical error.

\subsection{Binding Energy}

To compute the binding energy, let
the non-interacting
system be $\psi_2$, and the fully interacting system be $\psi_1$.
The nuclear positions are the same for both systems, and the appropriate 
interaction terms are set to zero for the non-interacting system.

A pair of H$_2$ molecules in a parallel configuration was used as the test
system. The binding energy we are interested in is that of the interacting
molecules minus separate H$_2$ molecules (and not the separate atoms).
\be
E_B = E\left( (H_2)_2 \right) - 2 E (H_2)
\ee

There is a problem in that the electrons in the fully interacting system
can switch molecules and have no effect on the computation, but these 
configurations are very unlikely in the 
non-interacting system.   This leads to an artificially small overlap.
The solution in this symmetric case is to restrict the 
domain of integration.  
The electron coordinates are ordered along the x-axis 
so that $x_1 < x_2$ and $x_3 < x_4$.

To see that this restriction is exact, consider the integral
\be
\int_{-\infty}^\infty dx_1 \int_{-\infty}^\infty dx_2 \ 
        f(\abs{x_1-x_2}) g(x_1,x_2)
\ee
where $f$ corresponds to the electron-electron Jastrow factor and
$g$ is symmetric under the interchange of $x_1$ and $x_2$ and 
corresponds to the electron-nucleus
Jastrow factor and the square of the Slater determinant.
Change variables to $R = (x_1 + x_2)/2$ and $r=x_1-x_2$.
Now we have
\be
\int_{-\infty}^\infty dR \int_{-\infty}^\infty dr \ 
f(\abs{r}) g(R+r/2,R-r/2)
\ee
The integral over $r$ is even, and we only need to 
integrate over half of the interval ($r<0$ or $r>0$), which corresponds
to the restrictions $x_1 < x_2$ or $x_1 > x_2$.

\begin{figure} 
 \begin{center}
  \includegraphics[height=3.0in,trim=0 0 0 0]{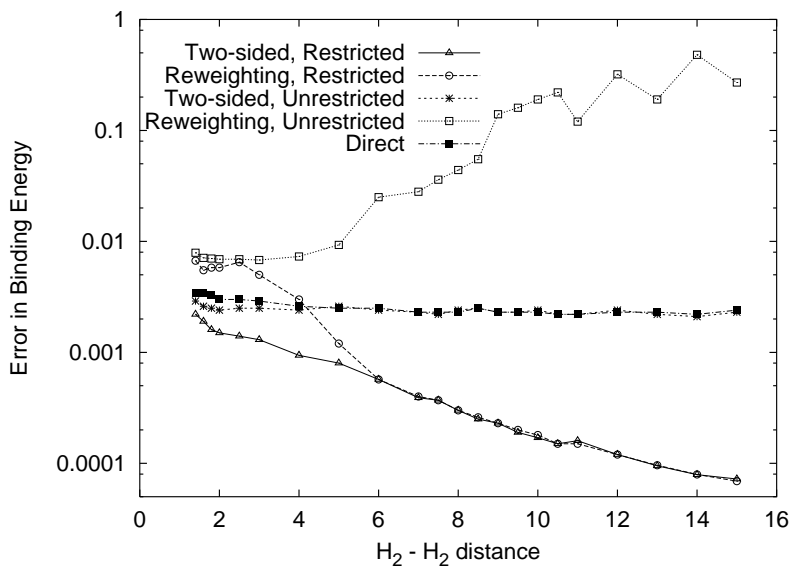}
  \caption{Error in VMC binding energy of H$_2$-H$_2$ system
   \label{swap}}
 \end{center}
\end{figure}

Figure \ref{swap} shows the error in the VMC binding energy for various 
intermolecular distances.  
Without restricting the domain of integration,
reweighting performs poorly, and the two-sided method reproduces the
results of the direct method.  With the restricted domain, the correlated 
methods perform quite well.

%% file: chap4.tex
\chapter{Wave Function Optimization}

Variational Monte Carlo (VMC) depends crucially on the optimization
of parameters in the wave function to find the minimum energy.
The general problem of function optimization is a well-studied area.
For a general introduction to various optimization
techniques, see  \cite{press92}.
For more in-depth work, consult
\cite{polak97}, \cite{dennis83}, or \cite{ortega70}.

The main difficultly in applying these techniques to optimizing 
VMC wave functions is noise - we
only get stochastic estimates for function values or gradients.
\cite{glynn86} describes several strategies for optimization in the presence
of noise.  We will divide these into three categories.

The first strategy is to convert the problem into a nearby smooth, 
non-noisy problem, and solve that problem instead.  
Fixed sample reweighting takes this approach by sampling some
set of configurations and optimizing with just these configurations.

The second approach is to reduce the noise to negligible levels, and 
proceed  with regular optimization techniques.   This is possible with
a Newton method, where the first and second derivatives of the function
are computed, and the number of iterations needed for convergence hopefully 
is small.

The third approach is to use a method tailored to handle noise.  
The Stochastic Gradient Approximation (SGA) is such a method. 
Also somewhat in this category, we will examine a method that is essentially a 
biased random
walk, and the moves are accepted or rejected based on the whether or not
the energy decreases.  

These approaches will be compared on several problems of different
sizes to see how they scale.
We will use a single H$_2$ molecule, and collections of 8,16, and 32
H$_2$ molecules in a box as trial problems.

\section{Energy vs. Variance Minimization}
There is a choice of objective functions - either the energy or the variance
of the energy can be minimized.
Under certain circumstances, variance minimization is more stable
that energy minimization.  For the reweighting method, this is definitely
true, but it may not be the case for the other optimization techniques.
It is generally held that variance optimization would produce better
values for observables other than energy \citep{williamson96}, 
but this may not always be the case \citep{snajdr99,snajdr00}.
The argument is that the variance is more sensitive to parts of the
wave function that do not contribute to the energy. As we have
seen in Chapter 2, having incorrect long range behavior in
the H$_2$ wave function does not affect the energy much, but does
cause the variance to rise.  In other words, variance minimization should yield
a ``smoother'' wave function, which should then have better non-energy
properties.


\section{Fixed Sample Reweighting}
Fixed sample reweighting with minimization of the variance was popularized
by \cite{umrigar88}, and has been used extensively since then.
The current state of the art is described by Kent \citep{kent99,kent-thesis99}.

The core of the method is the single sided reweighting method described 
in Chapter 3.
A number of configurations are sampled from a distribution with variational
parameters $\alpha_0$.  The energy at an arbitrary value of the variational
parameter, $\alpha$, is computed by
\be
\label{rew1}
E(\alpha) = \sum_i w(R_i;\alpha) E_L(R_i;\alpha)/ \sum_i w(R_i;\alpha)
\ee
where $w(R_i;\alpha) = \psi^2(R_i;\alpha)/\psi^2(R_i;\alpha_0)$.
Alternatively, one could compute the variance by
\be
\label{rew-var}
A(\alpha) = \sum_i w(R_i;\alpha) (E_L(R_i;\alpha)- E_T)^2/\sum_i w(R_i;\alpha)
\ee 
where $E_T$ can either be the weighted average energy (\ref{rew1}) or
it could be a guess at the desired energy.

The weights in these expressions can get very
large when the variational parameters move far from the sampled value
$\alpha_0$, and especially when the parameters that affect the nodes
are adjusted.
Then just a few configurations will dominate the sum, and the
energy estimator can often give meaningless low values.  The variance
estimator, however, will remain more stable in this situation.
For either estimator, the best fix is to regenerate the configurations 
being used when the parameters move too far away from $\alpha_0$.
This can be used in conjunction with the enhancements described below.

A second advantage of the variance estimator is that the weights can
be modified without changing the location of the minimum \citep{kent99}.  
The same is not true for the energy.  The problem of a few large weights
can be solved by limiting them to a maximum value \citep{filippi96}, 
or more simply
by just setting them all to one \citep{schmidt90}.  
\cite{barnett97} tame the fluctuating weights by sampling 
from a positive definite guiding function.
In this cases, $E_T$ should be
set to a best guess, or slightly below, because the energy
estimator will not be reliable.

 Further increases in stability can be gained by limiting outliers
in Eq. (\ref{rew-var}).
Large outlying values have a disproportionately large effect
on the variance, but their contribution is not that meaningful.
\cite{kent99} gives a procedure choosing a cutoff that will reduce
its effect as the number of samples increases.  We used a simpler approach,
removing from the sum any values greater than 5 standard deviations
from the average.

Another efficiency improvement can be exploited when the Jastrow $U$ is linear
in the variational parameters.  Then the variational parameters can be
factored out of the sum over interparticle distances, and the 
value of that sum can be stored.  
Fixed sample reweighting has been applied mostly to optimizing Jastrow
factors, and not parameters in the Slater determinant, so
this results in a dramatic time savings.
In our case we have both Jastrow and determinantal parameters, and
the code spends about 40\% of its time computing the Jastrow factor.
This percentage will decrease as the system size increases, since
the Jastrow computation is $\calO(N^2)$ but the determinantal part
requires matrix work that is $\calO(N^3)$. 
We did not implement this improvement, so bear
in mind when perusing the results that the reweighting time could
be reduced, probably by 30\%.

An additional advantage of reweighting is that, since it is solving
a smooth problem, an off-the-shelf minimizer can be used.
We used the DSMNF general minimizer from the PORT library, which
uses only the function values and does not need any derivatives.
Routines to minimize sums of squares are also available, but
we did not try them.
The fixed sample reweighting algorithm is then: 
Generate a set of configurations
and minimize the variance with this set.  
Generate a new set of configurations using the new variational parameters
and find the minimum variance again.
Repeat for several steps to ensure convergence.


\section{Newton Method}

The Newton method makes use of the first and second derivatives.
We can approximate a function near its minimum as a quadratic surface
\be
f(\vc{x}) \approx  f(\vc{x}_0) + (\vc{x}-\vc{x}_0)^T \cdot \vc{b}
  + (\vc{x}-\vc{x}_0)^T \cdot \vc{A} \cdot (\vc{x}-\vc{x}_0)
\ee
where $b_i = \pd{f}{x}$ and 
$A_{ij} = \pd{^2 f}{x_i \partial x_j}$ is the Hessian matrix.
The location of the minimum is then given by 
\be\label{newton_it}
\vc{x}_0 = \vc{x} - \vc{A}^{-1} \cdot \vc{b}
\ee
Since we are likely to start in a region where $f$ is not
quadratic, this step is iterated several times.

This procedure, along with analytic evaluation of the derivatives,
was applied to VMC energy minimization by \cite{lin00}.
Analytic derivatives of the local energy 
with respect to determinantal parameters are given by 
\cite{bueckert92}, but these were used in the context of
a reweighting minimization.

Recall the VMC energy is computed by 
\be
E = \expect{E_L} 
   = \int dR\ \psi^2(\alpha) E_L(\alpha) \left/ \int dR\ \psi^2(\alpha)\right.
\ee
We want the derivatives of E with respect to various variational
parameters, $\alpha$.   These could be computed with finite differences
and reweighting, but it is better to do some analytical work on this
expression first.  

\cite{lin00} 
use some Green's relations to eliminate explicit derivatives
of the local energy \citep{ceperley77},
and derive the following expression for the gradient,
\be
\pd{E}{\alpha_m} = 2\left[ 
    \expect{E_L\ \psi_{\ln,m}'}
  -  \expect{E_L} \expect{ \psi_{\ln,m}'} \right]
\ee
where 
\be
\psi_{\ln,m}' =  \pd{\ln \psi}{\alpha_m} = \frac{1}{\psi}\pd{\psi}{\alpha_m}.
\ee
They also give the expression for the Hessian,
\bea \nonumber
\spd{E}{\alpha_m}{\alpha_n} &=&  2 \left\{ 
     \expect{E_L\ \psi_{\ln,m,n}''} - \expect{E_L} \expect{\psi_{\ln,m,n}''} 
       \right.\\ \nonumber
     &&+ 2 \left[ \expect{ E_L\ \psi_{\ln,m}' \psi_{\ln,n}'}
             -   \expect{ E_L} \expect{\psi_{\ln,m}' \psi_{\ln,n}'}
        \right] \\ \nonumber
     &&-  \expect{\psi_{\ln,m}'} \pd{E}{\alpha_n} 
     - \expect{\psi_{\ln,n}'} \pd{E}{\alpha_m}   \\
    &&  \left.  + \expect{\psi_{\ln,m}' E'_{L,n}}
 \right\}
\eea
where 
\be
E_{L,n}' = \pd{E_L}{\alpha_n}
\ee
and 
\be
\psi_{\ln,m,n}'' = \spd{\ln \psi}{\alpha_m}{\alpha_n}.
\ee

Computing the first derivatives of the wave function
analytically is relatively easy.
Computing the second derivatives with respect to parameters 
in the Jastrow factor is also easy analytically.   
For parameters that appear in the determinant, however, second derivatives
are more difficult.
For this reason we compute most of the first derivatives analytically, and
use these to compute the second derivatives with a simple 
finite difference scheme.
The first derivative of the local energy was computed 
with finite differences.  The derivative of the orbital cutoff parameter,
which is the same for all the orbitals, was also computed with finite
differences.

An advantage of the Newton approach over the gradient-only approaches is
that it has information about how big of step should be taken, whereas
the step size is a parameter that must be tuned in the 
gradient-only methods.
The drawback, though, is a greater sensitivity to noise. 
The gradient and Hessian must be sufficiently accurate, or the Newton
iteration will get wildly wrong results.  
More precisely, it is the non-linear process of taking the inverse in
Eq. (\ref{newton_it}) that causes the problem.
Furthermore, this
sensitivity to noise increases with the number of parameters.

Another problem is parameter degeneracy, or near degeneracy.  This
will make the Hessian singular, or nearly so.  Even if it not exactly
singular, being nearly singular is the equivalent of dividing
by a small number, which will also greatly magnify the effects of noise.
The usual solution is use of the Singular
Value Decomposition (SVD).  
See \cite{press92} or \cite{kincaid91} for a description of the algorithm.
A more detailed look at "regularization"  (of which the SVD is one method) 
is in \cite{hansen98}. 
The SVD starts by decomposing a matrix as
\be
\vc{A} =  \vc{P} \vc{D} \vc{Q}
\ee
where $\vc{P}$ and $\vc{Q}$ are unitary matrices and 
$\vc{D}$ is a diagonal matrix. The elements of $\vc{D}$ are the eigenvalues
of $\vc{A}^T \vc{A}$.  For our square, symmetric matrix, these are
are the squares of the eigenvalues
of $\vc{A}$.  
We can also take $\vc{P}$ 
and $\vc{Q}$ to be the eigenvectors of $\vc{A}$.
The utility of the SVD is seen when we write the inverse of $\vc{A}$ as
\be
\label{psuedoinverse}
\vc{A}^{-1} = \vc{Q}^T \vc{D}^{-1} \vc{P}^T
\ee
If $\vc{A}$ is singular, then at least one of its eigenvalues is
 zero.
In this case, zero eigenvalues also indicate parameter degeneracy,
so it's not really necessary to move in the directions corresponding
to the zero eigenvalues.  To avoid moving in these these directions,
and to stabilize the inverse, 
set $1/d_i$ in Eq. (\ref{psuedoinverse}) 
to zero when $d_i$ is smaller than some cutoff.


With the eigenvalue decomposition we have an additional technique - 
negative eigenvalues correspond to uphill directions and mean we
are at a saddle point or 
are far from a region where the quadratic approximation is good.
\footnote{Much of the complexity of current Newton and 
quasi-Newton optimization methods
is in deciding how to move the parameters when the quadratic
approximation is not good.  Typically it involves a line minimization
in the gradient direction or some sort of back tracking.}
The simplest way of handling this
is to ignore negative eigenvalues.
So we remove small positive and all negative eigenvalues when
solving Eq. (\ref{newton_it}).

\begin{figure}
 \begin{center}
  \includegraphics[height=3.0in,trim=0 0 0 0]{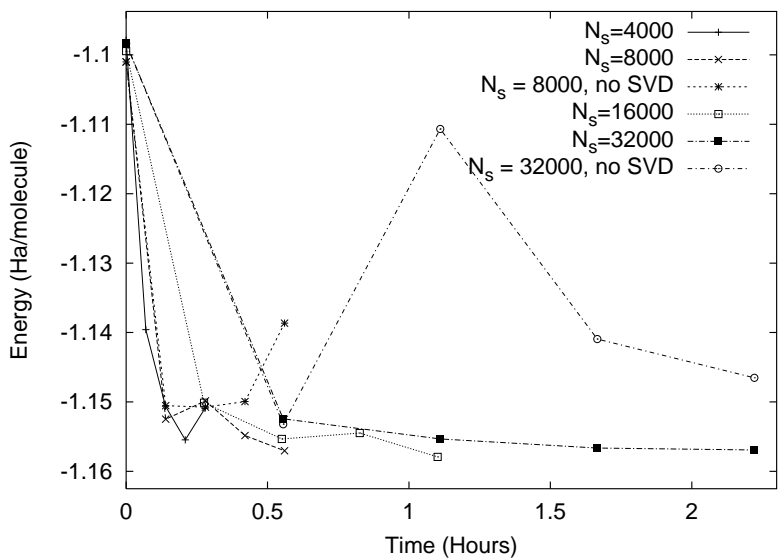}
  \caption{Examples using the Newton iteration with varying amounts of
noise. 
   \label{newton_example}}
 \end{center}
\end{figure}

Some examples of this Newton iteration with 8 H$_2$ molecules
are shown in Figure \ref{newton_example}.   
$N_s$ is the number of samples used in computing the gradient and Hessian.
Unless otherwise noted, the SVD method for solving 
Eq. (\ref{newton_it}) was used with removal of
eigenvalues less than $0.01$.
Other runs without using regularization are not shown
because they diverge very drastically.


\section{Stochastic Gradient Approximation}

The Stochastic Gradient Approximation (SGA)  was designed
by \cite{robbins51}
to handle optimization with noisy gradients.
It was first applied to VMC optimization by 
\cite{harju97}.

The SGA iteration can be written as
\begin{equation}
\label{sga}
\alpha_i = \alpha_{i-1} - h \gamma_i \nabla_\alpha E(\alpha_{i-1}).
\end{equation}
where $h$ is a step size parameter and $\gamma_i$ is some specially
chosen series.

There are some conditions on $\gamma_i$ that must be satisfied in order
for this iteration to converge. They are
\bea 
\gamma_i > 0  \label{sga-cond1}\\
 \sum_{i=1}^\infty \gamma_i = \infty \label{sga-cond2}\\
 \sum_{i=1}^\infty \gamma_i^2 < \infty \label{sga-cond3} 
\eea
The condition given by Eq. (\ref{sga-cond2}) allows the iteration to
reach anywhere in parameter space.  The condition in Eq. (\ref{sga-cond3}) 
is needed so the effects of noise will eventually be damped out.
An obvious choice for $\gamma_i$ is $1/i$.
For more discussion on these conditions and for some conditions on the
objective function, see \cite{young76} and \cite{tsypkin72}.


We can analyze the convergence in the limiting case of no noise in one
dimension.
First let us make a continuous version of Eq. (\ref{sga}) by letting
$\gamma(t) = dt/t$ and $d\alpha = \alpha(t)-\alpha(t-dt)$.  
Then in the $dt\ra 0$ limit, 
the SGA iteration  is
\be
\label{cont-sga}
\frac{d\alpha}{dt}  =  - \frac{h}{t} \del_\alpha f(\alpha(t))
\ee

Now let us assume that $f$ has a quadratic form,
$f(\alpha) = \half A \alpha^2 + B \alpha + f_0$, 
with a minimum at $\alpha = -B/A$.
Now Eq. (\ref{cont-sga}) is
\be
\frac{d\alpha}{dt}  =  - \frac{h}{t} \left[A \alpha + B  \right]
\ee
The solution is
\be
\alpha(t) = - B/A + a_0 t^{-hA}
\ee
where $a_0$ is a  constant of integration.
So we see that it will converge to the solution at $t \ra \infty$,
with a rate that is controlled by the curvature of the potential and
our choice of $h$.

Now consider generalizing  to the case where $\gamma(t) = dt/t^\delta$.
Our continuous equation is then
\be
\label{gen-cont-sga}
\frac{d\alpha}{dt}  =  - \frac{h}{t^\delta} \left[A \alpha + B  \right]
\ee
The solution is
\be
\alpha(t) = -B/A + a_0 \exp\left[-h A t^{1-\delta}/(1-\delta)\right]
\ee
We see that the smaller $\delta$ is, the faster the convergence.
If there were no noise, $\delta = 0$ would indeed be the best choice.

Now let us represent the noise in the gradient with an additive noise 
term, $\eta(t)$.
Then Eq. (\ref{gen-cont-sga}) is
\be
\label{cont-sga-noise}
\frac{d\alpha}{dt}  =  - \frac{h}{t^\delta} \left[A \alpha + B + \eta(t) \right]
\ee
Previously we considered case where noise was negligible. Now consider
the case where the noise dominates, so Eq. (\ref{cont-sga-noise}) becomes
\be
\frac{d\alpha}{dt}  =  - \frac{\eta(t)}{t^\delta} 
\ee
The solution is the integral
\be
\alpha(t) = - \int^T dt \frac{\eta(t)}{t^\delta}
\ee
To look at convergence, we need to compute the variance of $\alpha$ 
integrated over the noise.  
Take the noise to have a probability distribution $P(x,t)$ with zero mean
and variance $\sigma$.
The variance of $\alpha$ is then
\be
\sigma^2(\alpha) = - \int_{-\infty}^{\infty} dx  \int^T dt\ P(x,t) 
  \frac{x^2}{t^{2\delta}}
\ee
If we take $P(x,t)$ to have no dependence on $t$, the integrals factor and
we get
\be
\sigma^2(\alpha) = -\frac{\sigma^2 }{1-2\delta}\ \frac{1}{T^{2\delta-1}}
\ee

Here we see that larger values of $\delta$ lead to faster convergence of
the noise.  Since smaller values of $\delta$ lead to faster convergence
of the non-noisy problem, we need an intermediate value of $\delta$ 
to balance these effects.

One variation, suggested by \cite{nemirovsky83}, 
is to use $\delta = 1/2$ and use the
cumulative average of the variational parameters.  This value of $\delta$ 
violates the condition in Eq. (\ref{sga-cond3}), but this condition is there
to insure the noisy part converges.  Instead we use the cumulative averaging
process to remove the noise.

Another acceleration technique involves monitoring the sign of the gradient
\citep{tsypkin72}.
Far from the minimum the gradient will 
not often change sign between successive steps.  Close to the minimum,
the noise will eventually dominate, and the gradient will change sign
more often.  The acceleration procedure is to only update $\gamma_i$ when the
sign of the gradient changes.  This also has the advantage of adjusting
the convergence of each parameter separately.

  In practice, starting the series at $\gamma_1 = 1$ tends to make the first
steps have a dramatically larger effect on the parameters than subsequent
steps. Often, the first few steps would move the parameters very far
from the minimum, and then the iteration will take a long time to converge.
In this work we started the series at $i=10$ to minimize this effect. 

\begin{figure}
 \begin{center}
  \includegraphics[height=3.0in,trim=0 0 0 0]{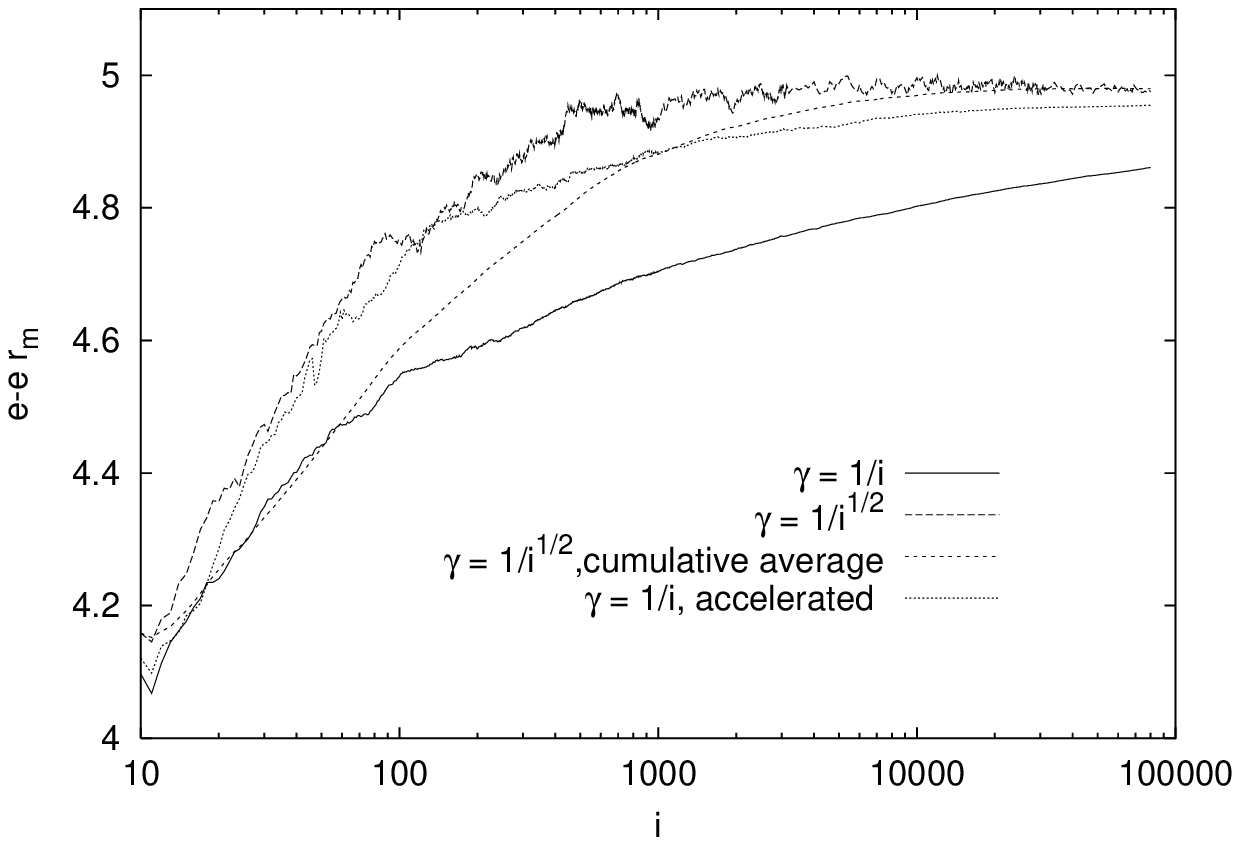}
  \includegraphics[height=3.0in,trim=0 0 0 0]{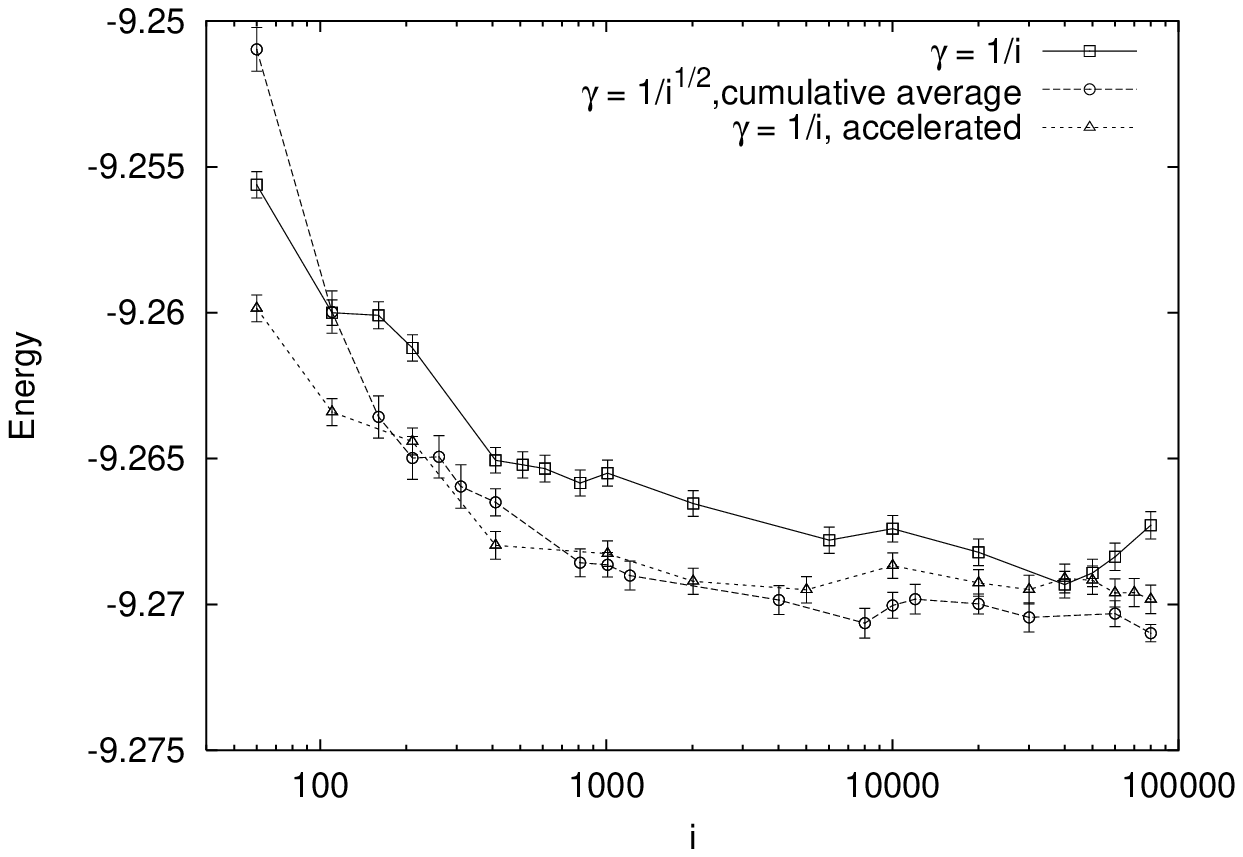}
  \caption{Examples of SGA. The graph on the top shows the convergence of
one variational parameter for several SGA algorithms.  The graph on the
bottom shows the resulting energy.
   \label{sga_example}}
 \end{center}
\end{figure}

We tried several of these SGA variants on the box of 8 H$_2$ molecules.
We used $h=3$ when $\gamma_i = 1/\sqrt{i}$  and $h=10$ when $\gamma_i = 1/i$.
This way the initial step sizes (give by $h\gamma_i$) were similar.
Figure \ref{sga_example} shows the convergence of one of the variational
parameters ($r_m$ for the electron-electron Jastrow).
The convergence of the energy is also shown.
We see that the two accelerated methods converge faster than the simple SGA.


\section{Gradient Biased Random Walk}

We introduce a new method that is made possible by the two-sided energy
difference method in Chapter 3.  Using this, 
it is relatively easy to determine whether a change of the trial parameters
lowers the energy or not.  This determination can be fitted onto a number
of methods, even a random walk.   
We evaluate the gradient and make a trial move in the gradient direction,
similar to the SGA.
Unlike the SGA, the move is accepted only if it lowers the energy.
Since the gradient is noisy,
we are effectively making a random walk that is biased in the gradient
direction, hence the name Gradient Biased Random Walk (GBRW).

The trial move is
\begin{equation}
\label{gbrw}
\alpha_T = \alpha_{i-1} - h \nabla_\alpha E(\alpha_{i-1}).
\end{equation}
where $h$ is randomly chosen from $[0,h_{\mathrm{max}}]$.
To provide some simple adaptivity, $h_{\mathrm{max}}$ is adjusted during the run.
If a trial move is rejected, $h_{\mathrm{max}}$
is decreased via multiplication
by some factor, usually 0.5 or 0.6.  If
a trial move is accepted, it is increased by multiplying
by the reciprocal of that same factor.


Currently the level of convergence of this method is controlled by 
how well the energy difference is computed.   
In other words, once the energy differences are of the same size
as the estimated error, it simply fluctuates.
There are several possibilities for making a convergent method.
The first is to take the cumulative average of the parameters, or
add a damping parameter as in the SGA. 
The second is to increase the number of samples to compute the
energy difference (and so decrease the noise) at each iteration.

\section{Comparison of methods}


We test the various optimization methods and compare their run times.
The test systems are an isolated H$_2$ molecule, and 8,16, and 32 molecules
in a box at $r_s = 3.0$, a fairly low density.
Each system has 12 parameters in the Jastrow factor, and  3 determinantal
parameters per molecule, plus one more for the box cutoff (which is the
same for all the orbitals).  Thus we have 15, 37, 61, and 109 variational
parameters, respectively.
For the starting parameters,we set the Jastrow cutoff to $r_m = 4.0$, 
the orbital widths
to $2.0$, the orbital box cutoff to $1.0$, and all the other parameters
to zero.

The Newton method used the regularization method with a cutoff of $0.01$
for $N=8, 16$ and a cutoff of $0.1$ for $N=32$.  No regularization was used
for $N=1$.
The SGA method used $\gamma_i = 1/\sqrt{i}$ and parameter averaging.
Reweighting used 16000 configurations for $N=1$ and 1000 configurations
for $N=8$ and $16$.
We did not attempt reweighting on the largest system.

\begin{figure}
 \begin{tabular}{cc}
  \includegraphics[height=2.0in,trim=0 0 0 0]{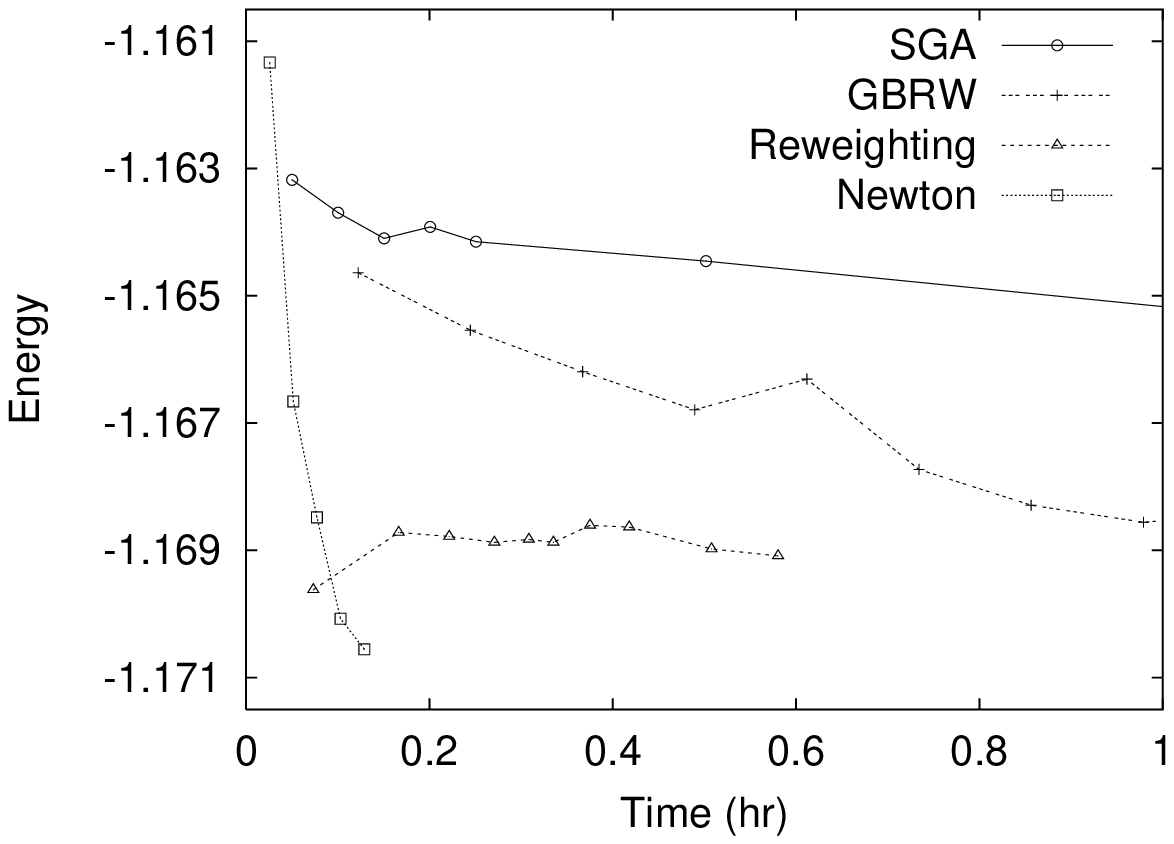} &
  \includegraphics[height=2.0in,trim=0 0 0 0]{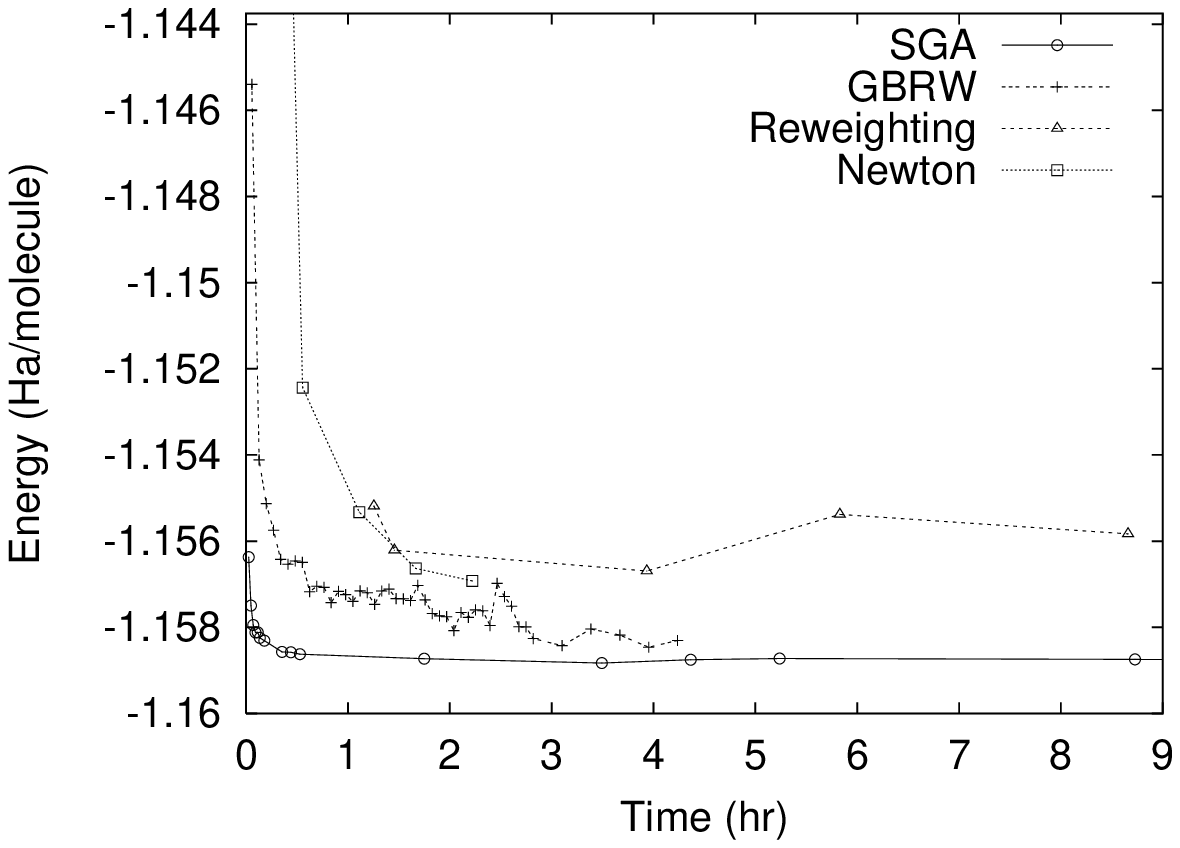} \\
  (a) & (b) \\
     & \\
  \includegraphics[height=2.0in,trim=0 0 0 0]{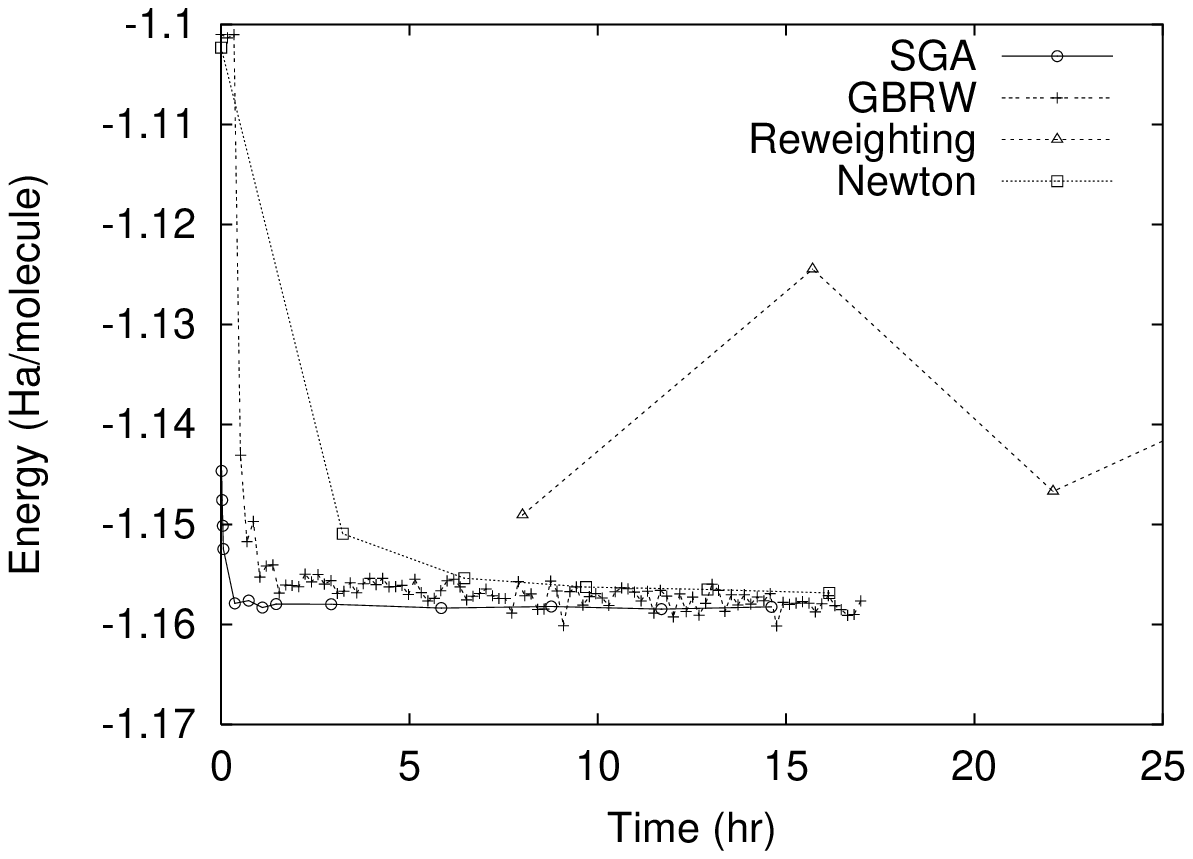} &
  \includegraphics[height=2.0in,trim=0 0 0 0]{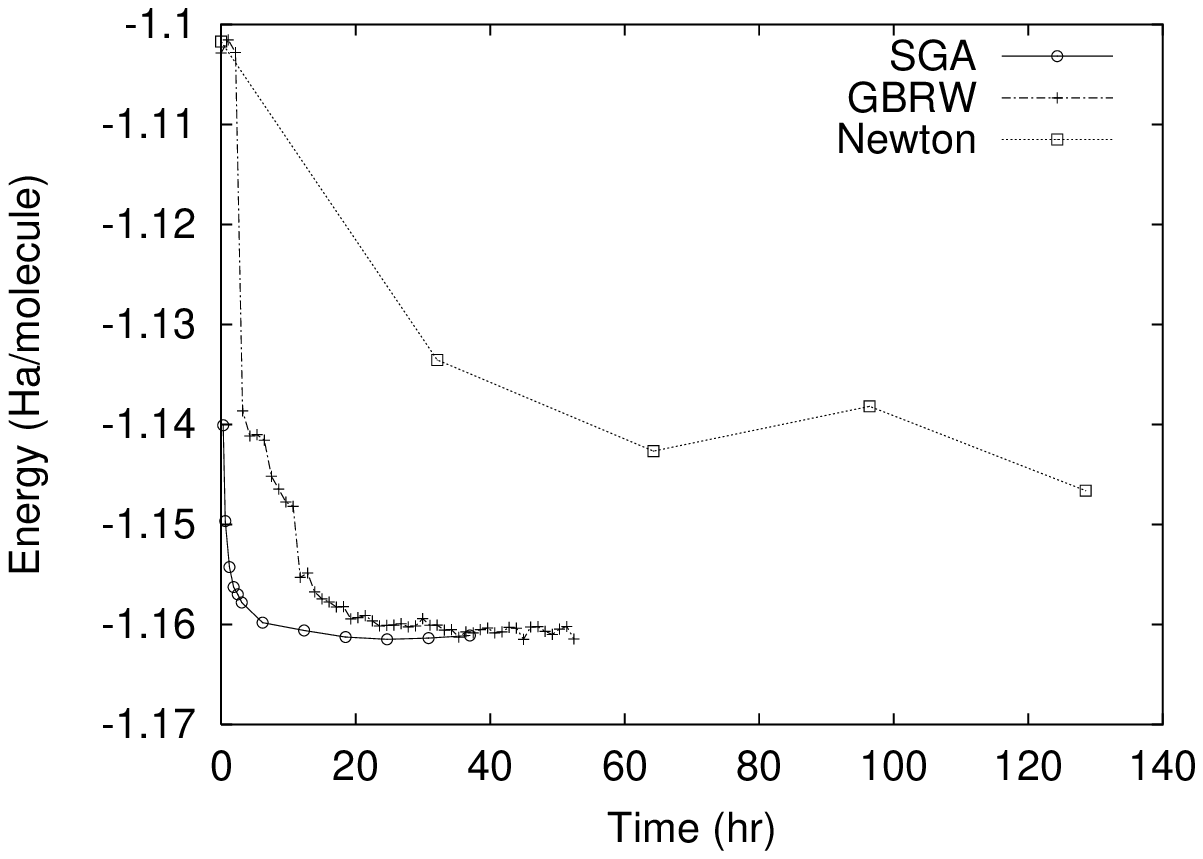} \\
  (c) & (d) \\
     & \\
 \end{tabular}
  \caption{Optimization methods applied to (a)\ Single H$_2$ \ (b)\ 8 H$_2$'s \
        (c)\ 16 H$_2$'s \ (d)\ 32 H$_2$'s \
   \label{opt_examples}}
\end{figure}

The best way to compare these methods would be
to run them all many times starting from different random number seeds.
The average of the resulting distribution would give the average quality of 
each method,
and the spread of the distribution would indicate the stability.
However, this is time-consuming and instead, as a first approximation
we present the results for a single run of each method in Figure 
\ref{opt_examples}.  The times are in hours on an AMD Duron 600 Mhz
(which is approximately 1/2 to 2/3 the speed of a 195 Mhz R10000 in an SGI 
Origin).



For the single molecule, it 
is clear that the Newton method is the best method.
The reweighting method also performs well, and the two gradient
methods take longer to converge.
As the system size increases, however, the gradient methods do better,
with the SGA method doing the best.

The Newton method in particular has difficulty with stability as the
system size increases.  It needs to be run long enough so the noise
is small enough that it does not affect the results.

The reweighting method performs surprisingly poorly on the larger systems.
Looking more closely at the results of reweighting for the $N=8$ case,
we get a total energy of $-9.244(2)$ Ha and a variance of $0.30$.  
From the SGA we get $E=-9.275(3)$ Ha and a variance of $0.42$.
From the GBRW we get $E=-9.268(2)$ Ha and a variance of $0.36$.
It appears that the problem is with the variance minimization
and we have a case where the minimum variance solution
is not the lowest energy solution.  
Although on the scale of the total energy, the difference between 
reweighting and the SGA 
is only $0.3\%$.  On the scale of the correlation energy 
in the isolated molecule, this difference is about $10\%$.


%




\section{Future Work}
We have compared a few basic methods for VMC parameter
optimization.  Many more improvements and modifications 
could be conceived and tried. 

Currently we ran these with set numbers of iterations and numbers of samples,
then looked at the results, and perhaps made adjustments and tried again.
What would be very helpful is some sort of adaptivity - adjusting
the number of samples or even the type of method as the
optimization proceeds in order to ensure convergence.

So far the gradient-only methods seem to have the advantage, but have
the disadvantage that they require a step size be set manually.
In order to generate a trial step size automatically,
a secant updating method could be tried, where successive
gradient evaluations are used to build up an approximate inverse Hessian
\citep{dennis83}.
These methods are often superior to using the actual Hessian \citep{press92} , 
but it
is not clear how the presence of noise will affect the algorithm.

Finally, it would be instructive to perform these comparisons on systems
containing atoms with higher atomic number.


%% file: chap5.tex
\chapter{Coupled Simulation Methods} 

  There are several issues we have to deal with when constructing an efficient
CEIMC simulation.   The first is noise from the QMC evaluation of the
energy.     We will discuss a modification to the Metropolis acceptance
ratio, called the penalty method, that will accommodate noise.
Next we will examine some of the details involved in a CEIMC simulation,
and finally give results for a single H$_2$ molecule.

\section{Penalty Method}

The Metropolis acceptance ratio, from Chapter 2, is 
$\min\left[1,\exp(-\Delta)\right]$,
where $\Delta = \beta[V(s') - V(s)]$.
The QMC simulation will yield a noisy estimate for $\Delta$,
which we denote as $\delta$ 

The exponential in the acceptance ratio
is nonlinear, so that $\expect{\exp(-\delta)} \neq \exp(\expect{-\delta})$.
The noise will introduce a bias into our acceptance ratio formula.
To avoid this bias
in our simulations, we can either run until the noise is negligible, or we can
try find a method that tolerates noise.



Typical energy differences for moves in our simulations are on the order
of $.01-.05$ Ha.  
If we want an error level of 10\% (statistical error of .001 Ha) 
\footnote{The usual error level considered chemical accuracy 
is 1 kcal/mol = .0016 Ha}
it would
take about 7 hours of computer time for a system of 16 H$_2$ molecules.
We need to perform hundreds of these steps as part of the classical simulation,
so clearly a method that could tolerate higher noise levels 
would be very beneficial.
The penalty method of \cite{ceperley99} does this, and our simulations
run with noise levels on the order of $.01$ Ha, which only takes about 4 minutes
of computer time.



In the penalty method, we start with detailed balance, written as
\be
A(s\ra s') = A(s' \ra s) \exp\left[-\Delta\right].
\ee
To deal with noise, we would like to satisfy detailed balance on average,
We introduce an instantaneous acceptance probability, $a(\delta)$, that
is a function of the estimated energy difference.
The average acceptance probability is the instantaneous one averaged
over the noise,
\be
A(s\ra s') = \int_{-\infty}^{\infty}  d\delta P(\delta; s\ra s') a(\delta)
\ee

The detailed balance equation we would like to satisfy is then
\be
\label{noisy_acc}
\int_{-\infty}^{\infty}  d\delta P(\delta; s\ra s') \left[ a(\delta) - 
e^{-\Delta}a(-\delta)\right] = 0
\ee

Suppose the noise is normally distributed with variance, $\sigma$.
Then
\be
P(\delta)  = (2 \sigma^2 \pi)^{-1/2} 
      \exp\left[-\frac{(\delta - \Delta)^2}{2\sigma^2} \right]
\ee
A simple solution to Eq. (\ref{noisy_acc}) is
\be
\label{penalty-acc}
a(\delta) = \min\left[1,\exp(-\delta-\frac{\sigma^2}{2})\right]
\ee
The extra $-\sigma^2/2$ term causes addition rejections of trial moves
due to noise.  For this reason it is called the penalty method.

To verify that the solution in Eq. (\ref{penalty-acc}) satisfies detailed
balance (\ref{noisy_acc}), let us compute the average acceptance probability
\bea \nonumber
A(\Delta) &=& \frac{1}{\sqrt{2\sigma^2\pi}} 
           \int_{-\infty}^{\infty}  d\delta \ 
           e^{-(\delta-\Delta)^2/2\sigma^2}
          \min\left[1,e^{-\delta-\sigma^2/2}\right] \\ \nonumber
 &=&  \frac{1}{\sqrt{2\sigma^2\pi}} \int_{-\infty}^{-\sigma^2/2}  d\delta \
           e^{-(\delta-\Delta)^2/2\sigma^2}
 + \frac{1}{\sqrt{2\sigma^2\pi}} \int_{-\sigma^2/2}^{\infty}  d\delta \
           e^{-(\delta-\Delta)^2/2\sigma^2} e^{-\delta-\sigma^2/2}  \\ \nonumber
 &=& \frac{1}{\sqrt{2\sigma^2\pi}} \int_{-\infty}^{-\sigma^2/2-\Delta} 
         d\delta' \ e^{-\delta'^2/2\sigma^2}
 + \frac{1}{\sqrt{2\sigma^2\pi}} \int_{\sigma^2/2-\Delta}^{\infty}  d\delta'' \
      e^{-\Delta} e^{-\delta''^2/2\sigma^2}  \\ \nonumber
 &=& 
   \half \mathrm{erfc}( (\sigma^2/2 + \Delta)/2\sigma^2)
  + \half e^{-\Delta} \mathrm{erfc}( (\sigma^2/2 - \Delta)/2\sigma^2)
\eea
where we have made the substitutions $\delta' = \delta - \Delta$ 
and $\delta'' = \delta' + \sigma^2$.
This expression for $A(\Delta)$ will satisfy detailed balance, 
$A(\Delta) = e^{-\Delta} A(-\Delta)$.

In practice, both the energy difference and the error are being estimated
from a finite set of data.  
Assume we have $n$ estimates for the energy difference, $y_1,...,y_n$.
Estimates for the mean and variance are given by 
\bea
\delta &=& \frac{1}{n} \sum_{i=1}^n y_i \\
\chi^2 &=& \frac{1}{n(n-1)}\sum_{i=1}^n (y_i-\delta)^2
\eea
and we have $\Delta = \expect{\delta}$ and $\sigma^2 = \expect{\chi^2}$.


The average acceptance ratio can be written as integral over $\delta$
and $\chi^2$.
The probability distribution for the estimated error is a 
chi-squared distribution.  
An asymptotic solution can be formed by expanding $a(\delta,\chi^2)$ and
performing the integrals to get the average acceptance ratio.
This is set equal to a power series for $\exp(-\sigma^2/2)$, 
and by matching powers of $\sigma$  we get the coefficients for the
original series for $a(\delta,\chi^2)$.  This series can by summed to
get a Bessel function, hence we call it the Bessel acceptance
formula.  It is convenient to expand the log of the Bessel acceptance formula
in powers of  $\chi^2/n$.
The Bessel acceptance formula is then
\be
a(\delta,\chi^2,n) =  \min\left[1,\exp(-\delta-u_B)\right]
\ee
where
\be
u_B  = \frac{\chi^2}{2} + \frac{\chi^4}{4(n+1)} + \frac{\chi^6}{3(n+1)(n+3)}
        + \frac{\chi^8 (5n+7)}{8(n+5)(n+3)(n+1)^2)} + \cdots
\ee
Note that as $n$ gets large, only the first term is important, 
which is just the regular penalty method.

%
%

\subsection{Other methods}
There is another method for handling noise, originally proposed by 
Kennedy, Kuti, and Bhanot
\citep{kennedy85,bhanot85},
that uses a power series expansion of $\exp[-\delta]$ to construct an
unbiased acceptance ratio.  
It has an advantage over the penalty method in that it does not assume
any particular distribution for the noise.
The method has a major drawback in that it depends
on the value of $\delta$ not becoming too large, and not just the
error estimate for $\delta$.  This could severely restrict the maximum
steps sizes for moving the nuclei in our simulations.
Methods for dealing with this restriction 
has recently been addressed by \cite{lin99} and \cite{bakeyev00}, but 
we did not explore these extensions. 



\subsection{Handling noisy data}

%

Using noisy data requires care in handling.
Particularly, inappropriate reuse of any single estimated value can lead 
to biased results.
For instance, in a classical simulation the energy difference would
be computed, and one of the two energies involved would be used
in accumulating the average energy.  See the top of 
Figure \ref{cmc-algorithm} for an outline of such a simulation.
However, this leads to a bias when noisy energies are involved.
This can be seen by considering a negative fluctuation in the energy of the
trial move.
This will make the energy difference smaller (or more negative)
and hence more likely to be accepted.  Thus the negative fluctuations
would be preferentially added to the accumulated average, and bias
the result downward.

This program outline could be corrected by computing a new value
for the energy in the average.   
However, there is another arrangement that is more amenable to the 
energy difference methods of Chapter 3.
The computation of the energy used in the average is the same quantity
needed for the old energy in the next iteration. So that computation
can be moved to the next iteration, as shown on the bottom 
in Figure \ref{cmc-algorithm}.


\begin{figure}
 \begin{tabbing}
  \framebox{Compute old Energy} \\
  loop \=over Classical steps  \\
       \> loop \=over Number of molecules \\
       \>     \> make trial move (translation and rotation of H$_2$ molecule) \\
       \>     \> \framebox{Compute trial Energy} \\
      \>     \> acceptance probability $=
               \min\left[ 1,\exp \left( -\beta \Delta E  -
                       (\beta\sigma)^2/2 \right) \right]$\\

          \> \>  accept/reject trial move\\
          \> \>  if accept, set old Energy = trial Energy\\
          \> \>
               \begin{picture}(190,15)(0,0)
                 \dashbox{5}(190,16){Use updated old Energy in average} 
                \end{picture}\\
       \> end loop \\
  end loop \\
 \end{tabbing}
 \begin{tabbing}
  loop \=over Classical steps  \\
       \> loop \=over Number of molecules \\
       \>     \> make trial move (translation and rotation of H$_2$ molecule) \\
       \>     \> \framebox{Compute old Energy}\framebox{Compute trial Energy} \\
      \>     \> acceptance probability $=
               \min\left[ 1,\exp \left( -\beta \Delta E - 
                       (\beta\sigma)^2/2 \right) \right]$\\
          \> \>  accept/reject trial move\\
          \> \>
             \begin{picture}(140,15)(0,0)
               \dashbox{5}(140,15){Use old Energy in average} \end{picture}\\
       \> end loop \\
  end loop \\
 \end{tabbing}
 \caption{CEIMC program outlines.  Boxes indicate quantum computations. 
        The dashed box indicates a quantity saved from a previous computation.
        The top algorithm is incorrect. The bottom algorithm is correct.
  \label{cmc-algorithm}}

\end{figure}


\begin{figure}
 \begin{center}
  \includegraphics[height=3.0in,trim=0 0 0 0]{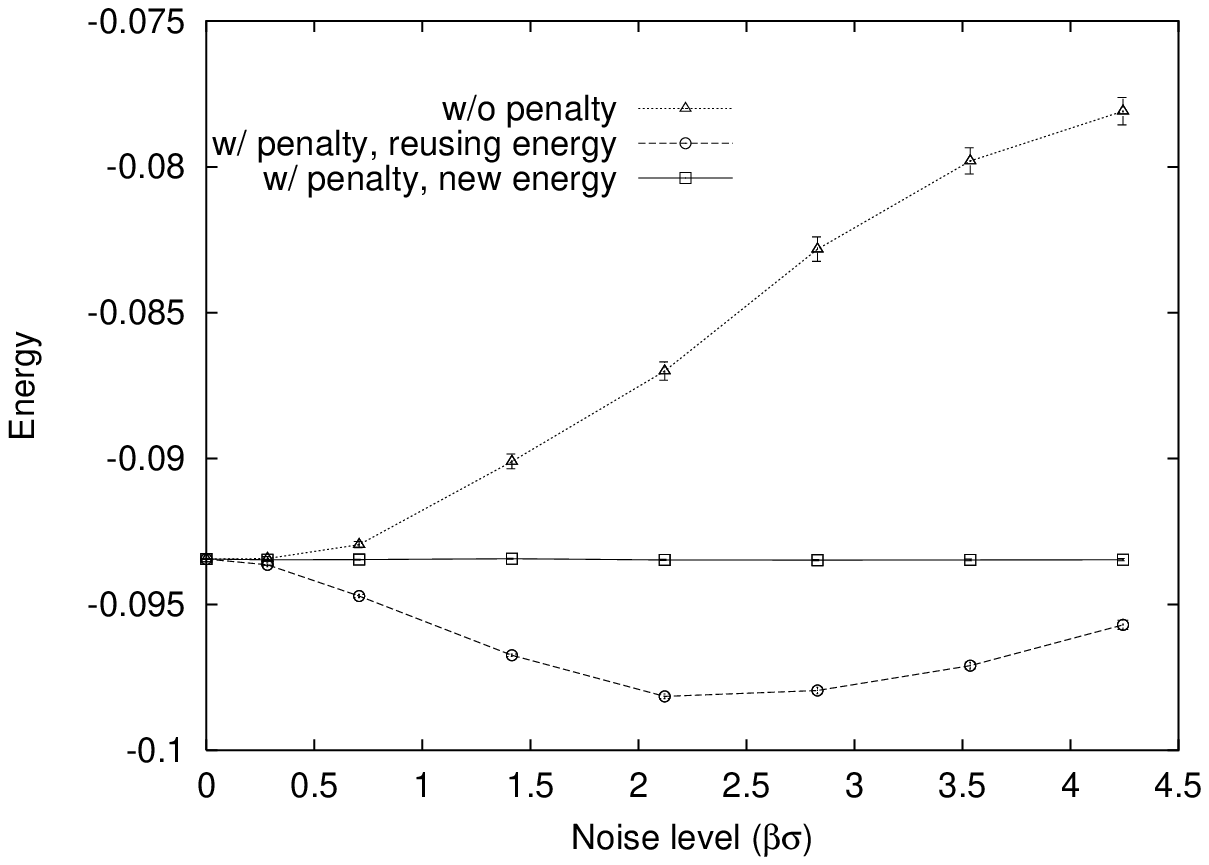}
  \caption{ Examples on a Lennard-Jones potential with synthetic noise.
   \label{lj-example}}
 \end{center}
\end{figure}

Several points are illustrated with a system of
two particle interacting via
a Lennard-Jones potential with $\epsilon = 0.1$ and $\sigma=1.5$
\footnote{Here $\sigma$ is the length scale for the LJ potential}.
The temperature of the system was  3160K ($\beta =100$).
Noise was simulated by adding a Gaussian random variable with known
variance to every energy computation.
Results for several algorithms versus noise level are shown in 
Figure \ref{lj-example}.  
The top curve shows the bias that results from having no penalty method.
The middle curve is the correct method, which we see is independent of the 
noise level.
The bottom curve demonstrates
the bias from reusing the energy involved in making the accept/reject decision.

The noise level of a system can be characterized by the relative noise
parameter, $f= (\beta \sigma)^2 t/ t_0$, where $t$ is the computer time
spent reducing the noise, and $t_0$ is the computer time spent on
other pursuits, such as optimizing the VMC wavefunction or equilibrating the
DMC runs.  A small $f$ means little time is being spent on reducing noise,
where a large $f$ means much time is being spent reducing noise.
The efficiency of a CEIMC simulation can be written in terms of this
parameter.
Our paper gives an example of a double well potential, and finds
the noise level that gives the maximum efficiency.  Generally it
falls around $\beta \sigma = 1-2$, with the optimial noise level increasing
as the relative noise parameter increases.
The one exception occurs when computing the first moment, which is
sensitive to crossing the barrier between the double wells.  
These crossings are assisted by an increased noise level, hence
the optimal noise level is much higher.

\section{Pre-rejection}

We can apply multi-level sampling ideas (see section \ref{vmc_twolevel}
for an application to VMC) to our CEIMC simulations as well.
The idea is to use an empirical potential to "pre-reject" moves that
would cause particles to overlap and be rejected anyway.

In this case, the trial move is proposed and accepted/rejected based on
a classical potential
\be
A_1 = \min \left[1,\frac{T(R\ra R')}{T(R' \ra R)} \exp(-\beta \Delta V_{cl})\right]
\ee
where $\Delta V_{cl} = V_{cl}(R') - V_{cl}(R)$ and $T$ is the
sampling probability for a move.
If it is accepted at this first level, the QMC energy difference is 
computed and accepted with probability
\be
A_2 = \min \left[1, \exp(-\beta \Delta V_{QMC} -u_B ) \exp(\beta \Delta V_{cl})\right]
\ee
where $u_B$ is noise penalty.

Compared to the cost of evaluating the QMC energy difference, 
computing the classical energy difference is free.
Reducing the number of QMC energy difference evaluations is valuable
in reducing the computer time required.

In Chapter 7, using the pre-rejection technique with a CEIMC-DMC simulation 
results in a first level (classical potential) acceptance
ratio of 0.43, and a second level (quantum potential) acceptance
ratio of  0.52.   The penalty
method rejects additional trial moves because of noise.  If these rejections
are counted as acceptances (ie, no penalty method or no noise), then
the second level acceptance ratio would be 0.71.
The classical potential is a fairly good representation for the 
DMC potential, and we can use that to reduce the number of DMC
energy difference evaluations needed.

\section{Trial Moves}

Molecular moves are separated into translation,
rotation and bond length changes.


The Silvera-Goldman potential is used as the empirical potential for
pre-rejecting translational moves.
Anisotropic potentials were tried for pre-rejecting rotational moves, but
they did not work very well.  It is not clear whether this was from the
the potentials being derived for isolated H$_2$-H$_2$ interaction,
or from inaccuracy in the trial wave function.

Bond stretching moves were pre-rejected using the, essentially exact, H$_2$ 
intramolecular potential of \cite{kolos64}.
The new bond length is sampled uniformly from a box of size $\Delta_b$ 
around the current position.  Because of phase space factors we need to
include a sampling probability of $T(R) = 1/R^2$ in the acceptance formula.



The trial move for classical Monte Carlo is usually presented as either
moving one particle at a time, or all of the particles at once.  
However, we can move other numbers of particles as well.
Table \ref{cm-eff} shows the efficiency for a classical system
with 32 H$_2$ molecules  for two densities and temperatures.
On the left is a low density system with $r_s = 3.0$ and at a
temperature of 5000K, and on the right 
higher density system with $r_s = 1.8$ and a temperature of 3000K.
For the lower density system,  the highest efficiency occurs when moving 
half the molecules at a time.   
Relatively high efficiency can also be found moving 2, 4 or 8 at a time as well.
For the higher density system, the most efficient regime shifts towards
smaller step sizes and fewer number of particles moved at a time.
\footnote{These results are not generally applicable to classical MC 
simulations, since much more efficient implementations are possible for systems
interacting with a two body potential.}


\begin{table}
\caption{Efficiency of classical Monte Carlo for moving several particles
at once.  The table on the left is for low density system at $r_s=3.0$ and
T=5000K.  The table on the right is for a high density system at $r_s=1.8$ and
T=3000K. The largest values of the efficiency are shown in boxes.
\label{cm-eff}
}
\begin{center}
\hbox{
\begin{tabular}{|c|cccc|}
\hline
  & \multicolumn{4}{c|}{$\Delta$} \\
\hline
N$_{\mathrm{m}}$  &  0.8  &  1.6  &  3.0  &  4.0 \\ \hline
1                       &  1.4  &  4.9  &  9.6  & 11.6 \\
2                       &  2.4  &  8.1  & 15.3  & \framebox{17.4} \\
4                       &  5.6  & 12.5  & \framebox{16.9}  & \framebox{17.7} \\
8                       &  6.9  & 14.8  & \framebox{18.0}  & 14.8 \\
16                      &  9.5  & \framebox{22.2}  &  7.4  & \\
32                      & 11.9  & 14.8  &  2.7  & \\ \hline
\end{tabular}

\hskip 2cm
\begin{tabular}{|c|ccccc|}
\hline
  & \multicolumn{5}{c|}{$\Delta$} \\
\hline
N$_{\mathrm{m}}$  &  0.4  &  0.8  &  1.2  &  1.6 & 2.0\\ \hline
1                 &  74  &  134  &  99  & \framebox{236} & \framebox{191}\\
2                 &  43  &  149  & \framebox{179}  & 121 & 114 \\
4                 &  118 &  \framebox{170}  & 141  & 66  & 23 \\
8                 &  \framebox{172}  & 128  & 24  & & \\
16                &  155  & 52  &   & &\\
32                &  39  &  &   & &\\ \hline
\end{tabular}
}
\end{center}

\end{table}

\section{Single H$_2$ molecule}

The CEIMC method was applied to a single H$_2$ molecular in free space,
at a temperature of 5000K.
Exact results are obtained by integrating the potential of \cite{kolos64}.
Results for the energy, pressure, and first and second moments of the 
bond length are given in Table \ref{ceimc-single}.
The Virial column is
computed by 
\be
\mathrm{Virial}  = \left[2\expect{K} + \expect{\calV}\right]
\ee
This is related by the virial theorem to the force on the nuclei,
which should be zero for an isolated molecule.
In Chapter 7, we will see this expression  used to compute the pressure.

\begin{figure}
 \begin{center}
  \includegraphics[height=3.0in,trim=0 0 0 0]{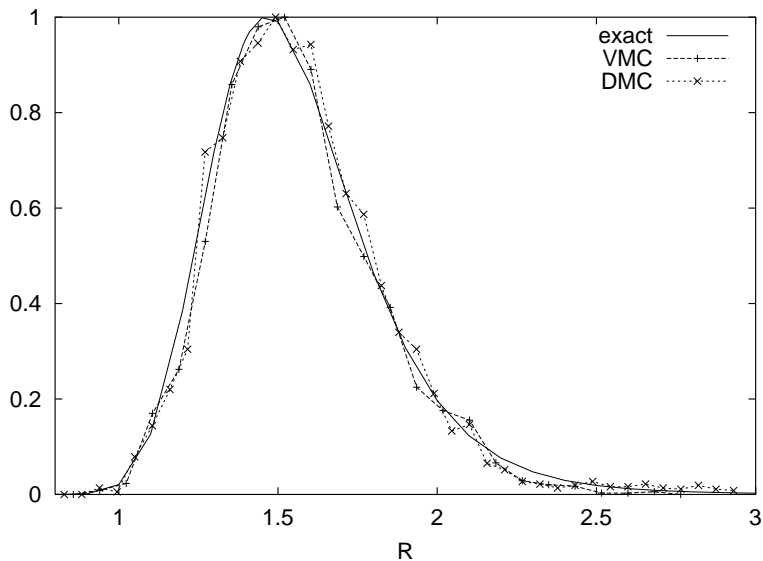}
  \caption{ H$_2$ bond length distribution.
   \label{h2-bl-dist}}
 \end{center}
\end{figure}

As we would expect, 
the VMC energy is higher than the exact energy.
The other quantities are close to their expected values.
Histograms of the bond length distribution are shown in Figure 
\ref{h2-bl-dist}. Here again, both VMC and DMC reproduce the
exact distribution well.



\begin{table}
   \caption{Results of CEIMC for isolated H$_2$ molecule at T=5000K.
     \label{ceimc-single}}
 \begin{center}
   \begin{tabular}{c|cccc}
     & Energy  & Virial  & $\expect{r}$ & $\sqrt{\expect{r^2}}$ \\ \hline
exact & -1.1630 & 0.0  & 1.57  & 1.60  \\
VMC   & -1.159(1) & -0.009(6)  & 1.56(2)  & 1.58(2) \\
DMC & -1.163(2) & -0.015(6)  & 1.58(2) & 1.60(2) \\
   \end{tabular}
 \end{center}
\end{table}



%% file: chap-hs.tex
\chapter{Hard Spheres}


A system of hard spheres is the simplest non-ideal many body system
to study. 
Statistical Monte Carlo techniques  and molecular dynamics were 
first applied to hard
spheres \citep{metropolis53,alder57}.
Additionally, some of the first applications of field theoretic methods 
to condensed matter systems were on hard spheres.
More recently,  the achievement of BEC in trapped atomic gases has renewed
interest in the theory of the hard sphere Bose gas \citep{dalfovo99}.

This chapter is a VMC and DMC study of a homogeneous, boson hard sphere
fluid.  Because they are bosons, there is no fixed node approximation in DMC,
and the result can be made essentially exact.  The approximations we
must control are finite size effects and timestep error in DMC.







  Applying perturbation theory to the Bose hard sphere gas yields the
low density expansion,
\begin{equation}
\label{low_density_exp}
E = 2 \pi \rho [ 1 + C_1 \sqrt{\rho} + C_2 \rho \ln \rho + C_3 \rho + ...]
\end{equation}
where $C_1 = 128/(15\sqrt{\pi})$ and $C_2 = 8(4\pi/3 - \sqrt{3})$.
Mean field theory gives the linear term.
Straight forward application of perturbation theory yields the
$C_1$ term, as was done by Huang, Yang, and Lee \citep{lee57,huang57}.
The next higher order of perturbation theory diverges, but this was solved
by including the depletion of the condensate by \cite{beliaev58}
to get $C_2$.
\cite{wu59} obtained the same results via a resummation technique.  
\cite{hugenholtz59} also obtained the logarithmic term.
They also obtained the functional form for the series, which includes
terms of the form $\rho^{n/2}$ and $\rho^{n/2}\log(\rho)$.

  Renormalization group techniques have recently been applied to examine this
divergence \citep{castellani97,braaten97a,braaten97}. In addition, Braaten and
Nieto have calculated $C_3$ \citep{braaten97a,braaten97}.
This term is also first term that depends on more that the two body $s$-wave 
scattering length.  It would require a solution to the three body scattering 
problem, which makes an explicit computation of that coupling constant 
difficult.



\cite{hansen71} used VMC on an 864 particle system to find the fluid-solid 
transition density.
\cite{kalos74} used the more accurate Green's Function Monte Carlo (GFMC)
to calculate the energy of the solid and liquid phase near freezing to
determine the freezing density. They used 256 particles.
In the liquid state they computed four points ranging in
density from 0.16 to 0.27.

 Recently \cite{giorgini99} did DMC calculations on the homogeneous Bose gas,
with various potentials, including the hard sphere potential.  
They used 500 particles and there was no mention of what DMC time step was used.
These calculations were also used to make a fitting to the extended form
of Eq. (\ref{low_density_exp}) (using terms up to $\rho^{5/2}$) 
by \cite{boronat00}.


There have been other attempts to get an equation of state by
using various fitting techniques to
combine the low density results and the GFMC results \citep{navarro87,keller96}.
As noted by \cite{keller96},
the earlier work had an error and used only half the energy of the actual GFMC 
results.
A new attempt at fitting the various
functional forms was not done, but a new value for $C_3$ was estimated.


The Hamiltonian for this system is
\begin{equation}
H = -\frac{1}{2} \sum_{i} \nabla_i ^2 + \sum_{i<j} v(|r_i-r_j|)
\end{equation}
where
\begin{equation}
v(r) = \cases{ \infty & $r < \sigma$ \cr
               0      & $r > \sigma$ \cr}
\end{equation}
We have set $\ \hbar=m=\sigma=1$ in all these calculations.


\section{Wave function}
An approximate wave function for the boson ground state that we use for a
trial function is
\begin{equation}
\psi = \prod_{i<j} f(r_{ij}) 
\end{equation}
 The individual functions are very similar to the ones used for hydrogen.  
The correlation function has a maximum range, $r_{max}$, 
beyond which $f$ is constant.
In order to be compatible with periodic boundary conditions, we require
$r_{max} \leq L/2$.
The ``cusp'' condition is
that the wave function must vanish linearly when two spheres get close,
$f(r \ra \sigma) \propto (r-\sigma)$.
(Unlike the electronic case, the slope is not fixed.)

A change of variables will simplify these expressions. Let $x=r-\sigma$,
$x_{max} = r_{max}  - \sigma$ and $y = x/x_{max}$.
Now $y$ lies in the range $[0,1]$.
In these variables, 
the boundary conditions on $f$ are
\bea \nonumber
f(y=0)&=&0  \\
f(y=1)&=&1  \\
f'(y=1)&=&f''(y=1)=0 \\
\eea
and the wave function is
\begin{equation}
f(y) = 3(y-y^2) + y^3 + y(y-1)^3 \sum_{i=0}^4 b_i T_i(2y-1)
\end {equation}
where $T_i$ are Chebyshev polynomials,
and $b_i$ are variational parameters.

The parameters for each density are given in Table \ref{varparm}.
They were obtained from optimization of the smallest system ($N=40$).
Then those same parameters were used for all system sizes at a 
particular density.

\begin{table}
\begin{center}
\caption{Variational parameters for hard sphere gas
\label{varparm}}
\begin{tabular}{c|cccccc}
 $\rho$ & $x_{max}$& $b_0$ & $b_1$ & $b_2$ & $b_3$ & $b_4$ \\ \hline
 .2     & 1.92   & -0.3209 & 0.25395 & 0.5624 & 0.0145 & -.05123 \\
 .1     & 2.68   & 0.9173 & 1.995 & 0.8147 & 0.2269  &0.0345 \\
 .05    & 2.68   & -0.33 & 0.674 & -0.12 & 0.056 & 0.0 \\
 .01    & 5.9152   & -1.95 & 0.86267 & -1.2982 & -0.08135 & -0.3152 \\
\end{tabular}
\end{center}
\end{table}

The cost of computing of the wave function and local energy 
is dominated by calculating the $N(N-1)/2$ interparticle
distances.  There are techniques for improving the scaling of computations
of short range 
interactions to achieve $\calO(N)$.  We used the cell method \citep{allen87}.
 The simulation box is divided into cubic cells and a list is made of all
the particles in each cell.  For simplicity, consider the case where
each cell is larger than the cutoff distance, $r_{max}$.
Then a particle in a cell will have a non-zero interaction only with 
the particles in the same cell and with particles in the neighboring
cells.  Particles in cells further away can be ignored.
There is an overhead in computing and maintaining these lists. 
We used the cell method on systems with 500 particles and larger.




There is a deficiency with the trial wave function that leads to undersampling
when three particles are in close proximity.  
In DMC, this leads to a large number of branching
walkers to compensate for the undersampling, which invariably causes problems
with maintaining a stable population.
One solution is to use a guiding function, which differs from the trial
wave function,  for the diffusion and branching.  
Then a weight, $\psi_T/\psi_G$, 
is associated with each sample point.
In this case the simplest guiding function is to use is $\psi_G=\psi_T^\alpha$.
We found that $\alpha = 0.9$ was sufficient to make the population
of walkers stable.

\begin{figure}
 \begin{tabular}{cc}
  \includegraphics[height=2.0in,trim=0 0 0 0]{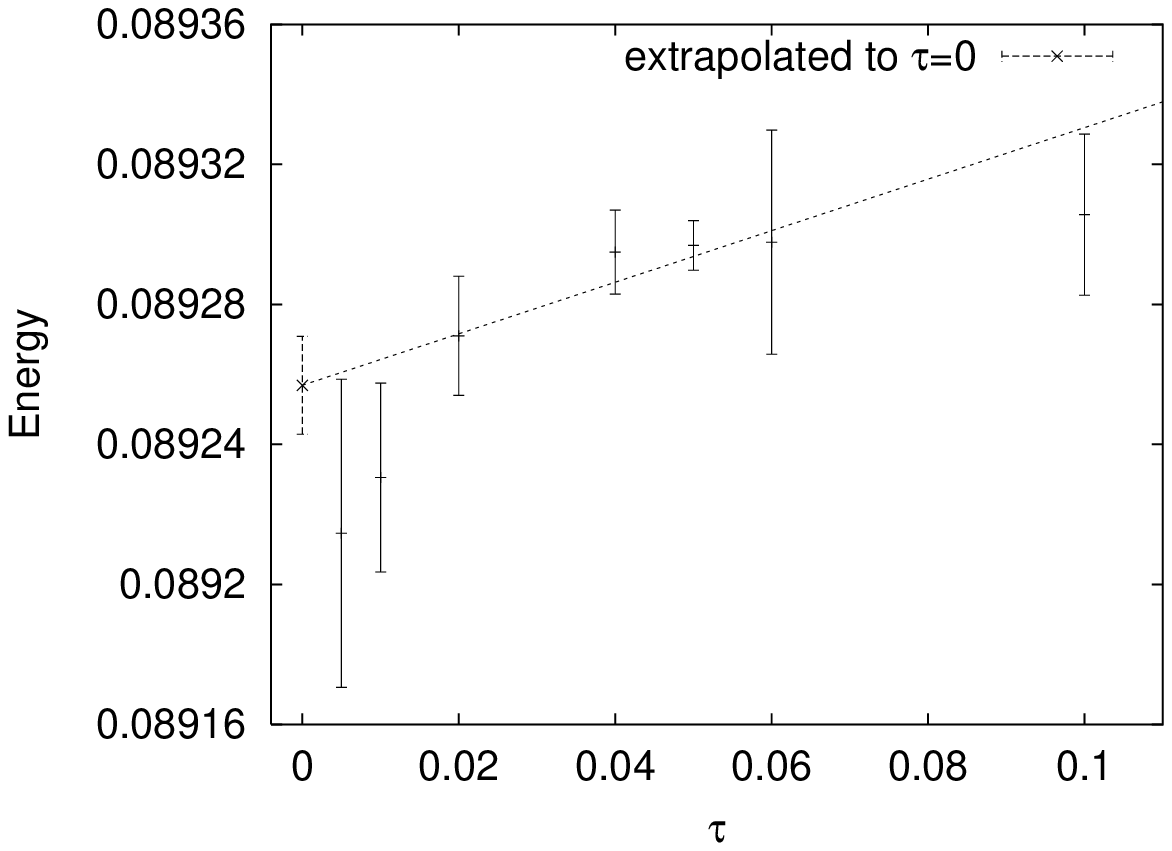} &
  \includegraphics[height=2.0in,trim=0 0 0 0]{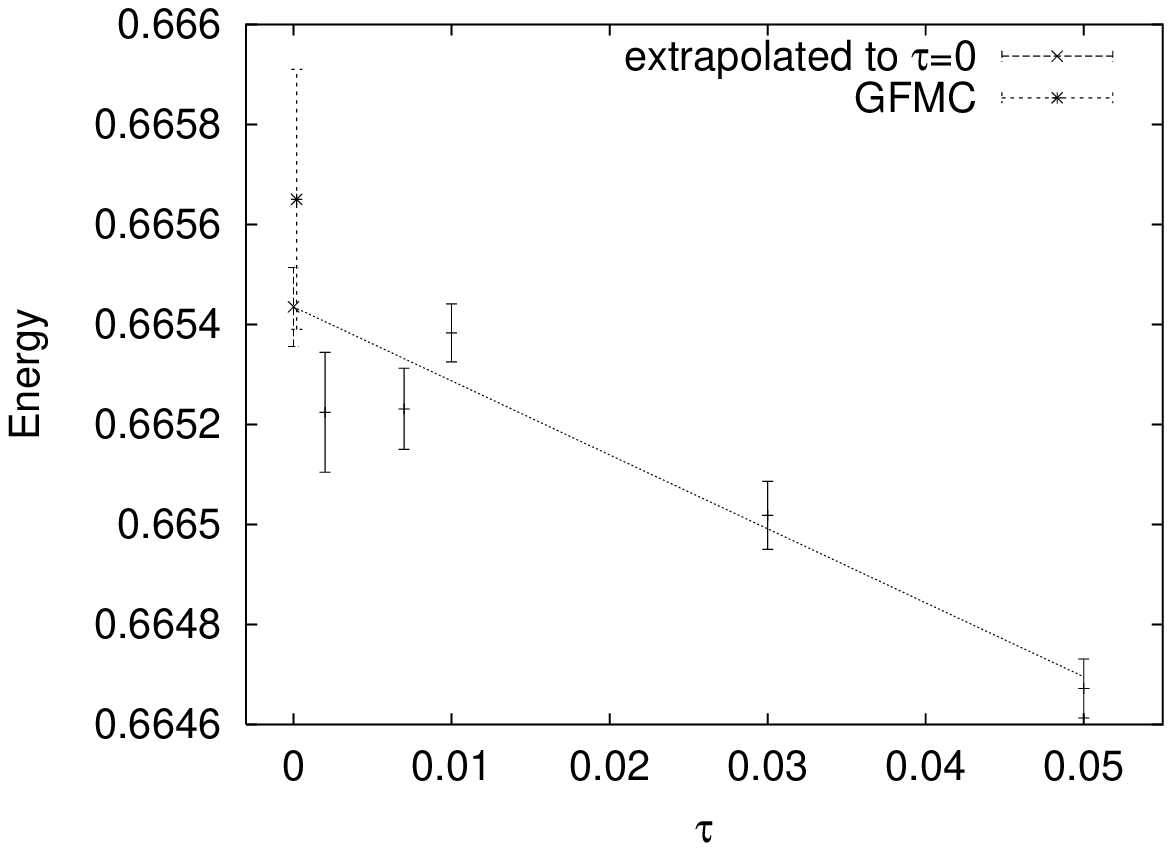} \\
  (a) & (b) \\
     & \\
  \includegraphics[height=2.0in,trim=0 0 0 0]{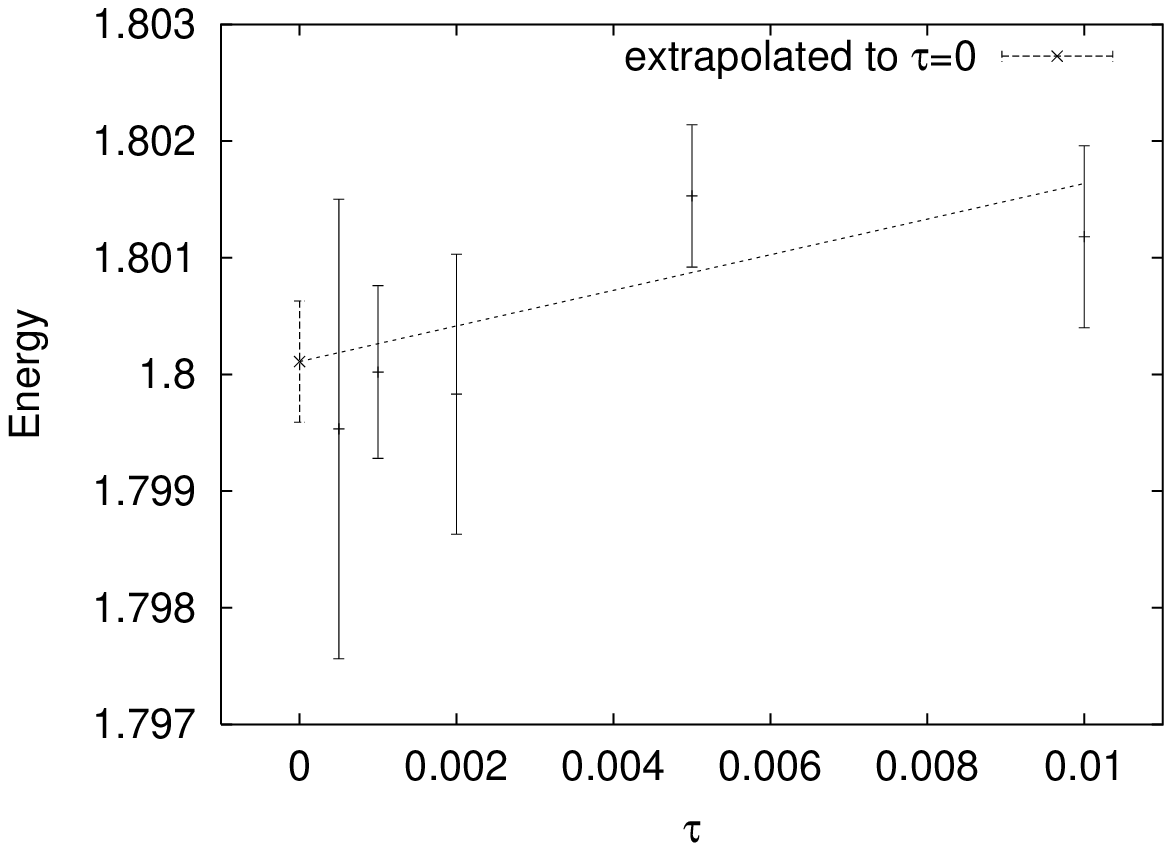} &
  \includegraphics[height=2.0in,trim=0 0 0 0]{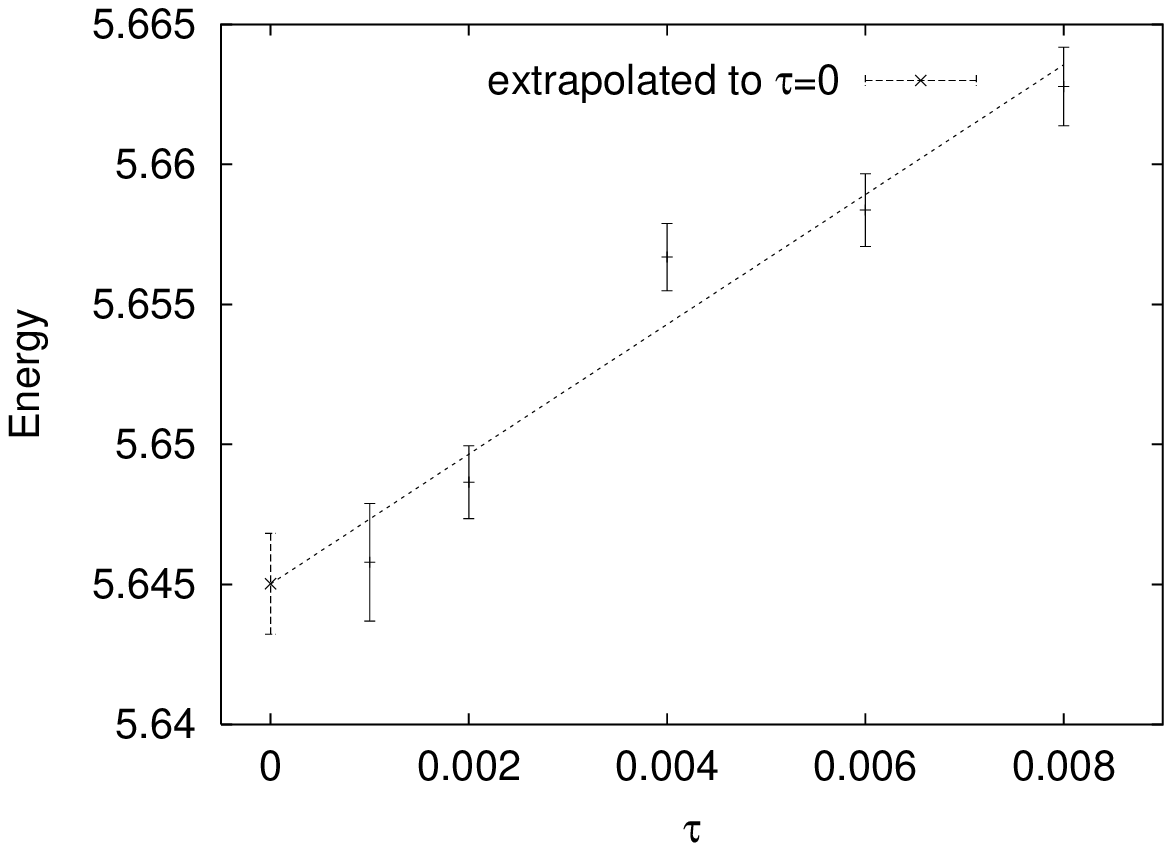} \\
  (c) & (d) \\
     & \\
 \end{tabular}
  \caption{Time step error for (a)\ $\rho=0.01$ \ (b)\ $\rho=0.05$ \ 
        (c)\ $\rho=0.1$ \ (d)\ $\rho=0.2$ \ 
   \label{tstep_err}}
\end{figure}

The DMC timestep errors should be local, and hence the same for 
all system sizes.
We did timestep extrapolation on systems with $N=40$ particles.
The timestep errors were found to be linear in $\tau$.
The extrapolations to $\tau=0$ are shown in Figure \ref{tstep_err}.


Green's Function Monte Carlo (GFMC) uses a different decomposition of the
Green's function that DMC.
The principle advantage of GFMC is that is has no time step error.
We ran GFMC at $\rho=.05$ and $N=40$ to verify the timestep errors.  The GFMC
data point is shown at $\tau=0$ in Figure \ref{tstep_err}.  
The importance sampling in GFMC
is not as effective as in DMC, hence the variance is larger, and GFMC is
less efficient than DMC.
GFMC has an efficiency of about 7, whereas DMC has an efficiency of 240
at $\tau=.002$ and an efficiency of $570$ at $\tau=.007$.
Even with computing at several timesteps to extrapolate to zero,
DMC is more efficient that GFMC.

\section{Finite Size Effects}

The main contribution to finite size effects in the energy 
is the long wavelength phonons.
The functional form for their contribution depends on the small
$k$ behavior of the structure factor, $S(k)$.
The energy can be written as 
\be
E = 4 \pi \int_{k_b} k^2 dk\  \epsilon(k)
\label{fs_integral}
\ee
where $k_b = 2 \pi/L$ is the small $k$ cutoff due to the finite box size.
The energy of the phonon excitations at small $k$ is
proportional to $S(k)$
\citep{feynman56}.
For a classical liquid, $S(k) \propto 1 + \calO(k^2)$.
For a Bose fluid, $S(k)$ should be proportional to $k$ and $S(k\ra0)=0$.

The VMC wave function has no long range part, and so we expect it
to behave like a classical fluid at small $k$.
Integrating Eq (\ref{fs_integral}), we get 
$E \propto k_b^3$, which is the same as scaling by $1/N$ for a fixed density.
A more rigorous derivation of this scaling is given by \cite{lebowitz61}.
The DMC algorithm should pick up the correct long wavelength behavior,
leading to 
an $S(k)$ that is linear in $k$.
Integrating Eq. (\ref{fs_integral}) we get $E \propto k_b^4$.
This then gives us $1/N^{4/3}$ scaling.

\begin{figure}
 \begin{tabular}{cc}
  \includegraphics[height=2.0in,trim=0 0 0 0]{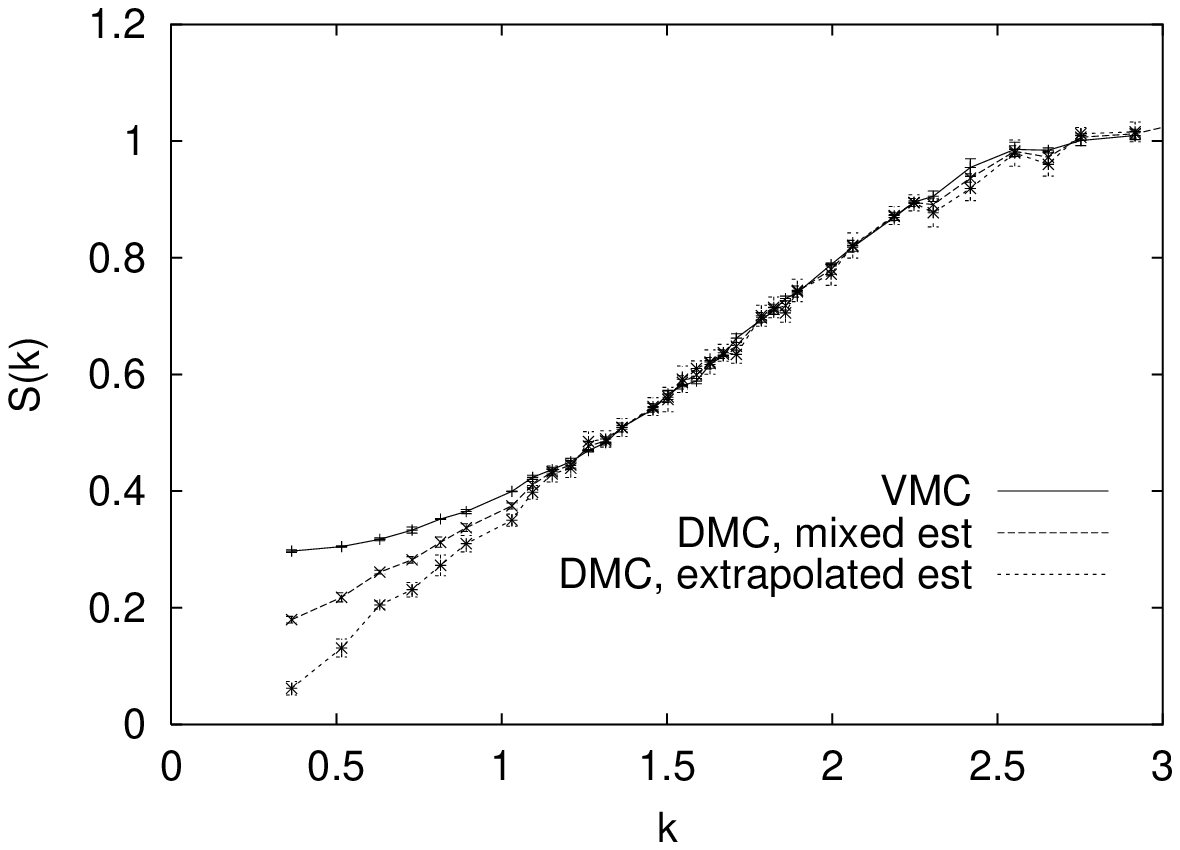} &
  \includegraphics[height=2.0in,trim=0 0 0 0]{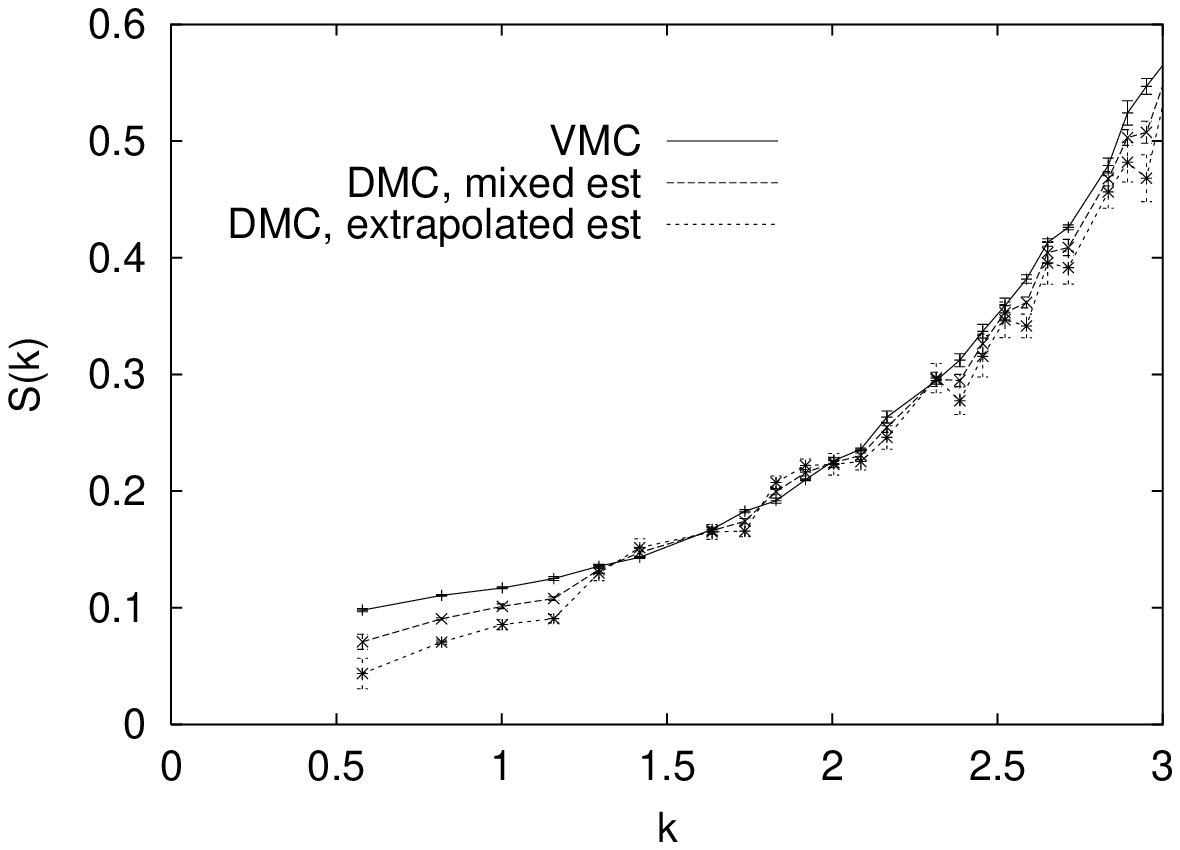}\\
  (a) & (b) \\
     & \\
 \end{tabular}
  \caption{S(k) for (a) $\rho=0.05$ (b) $\rho=0.2$ 
   \label{skfig}}
\end{figure}

The small $k$ behavior for S(k) can be seen nicely for
$\rho=0.05$, shown
in Figure \ref{skfig}a.  
The graph shows that the VMC structure factor appears quadratic as expected. 
The DMC mixed estimator shows the $S(k)$ behaving linearly, but still
not headed to zero.   The extrapolated estimator looks like it over corrects
and lowers S(k) too much.

\begin{figure}
 \begin{tabular}{cc}
  \includegraphics[height=2.0in,trim=0 0 0 0]{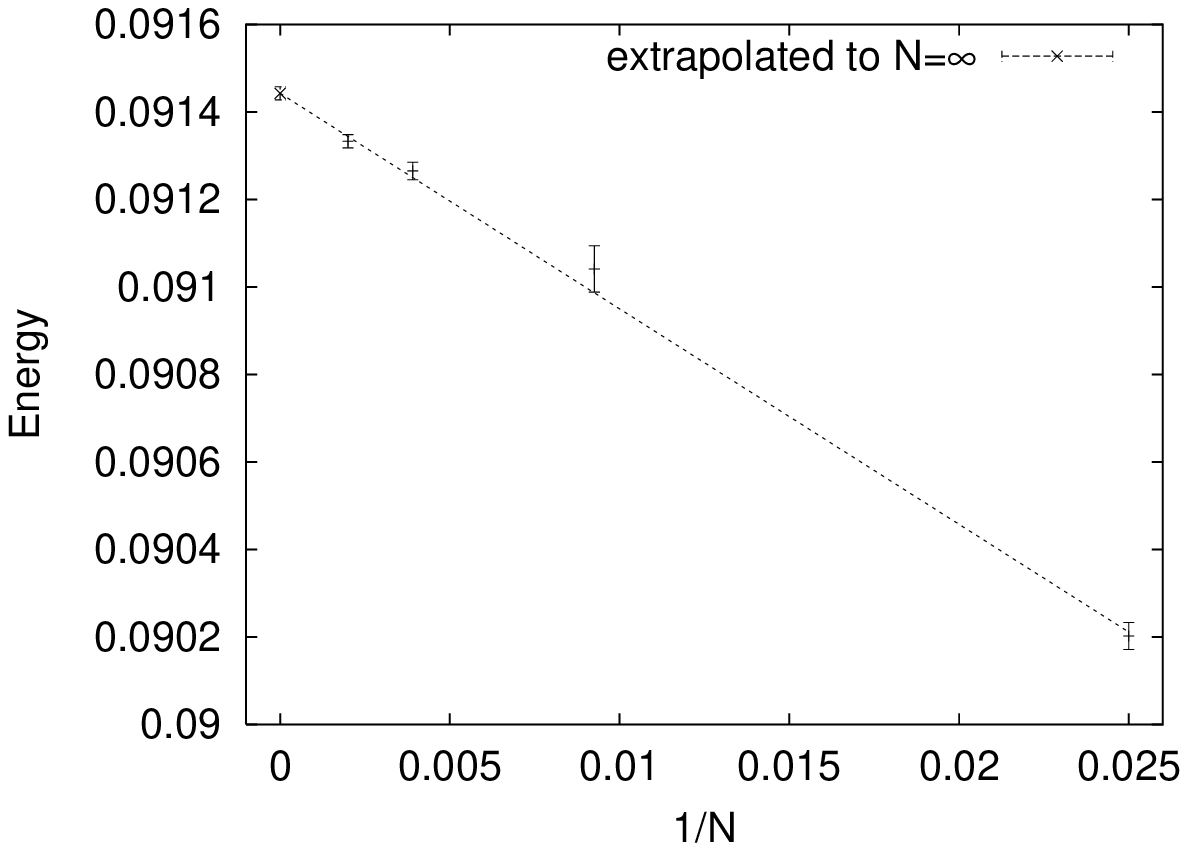} &
  \includegraphics[height=2.0in,trim=0 0 0 0]{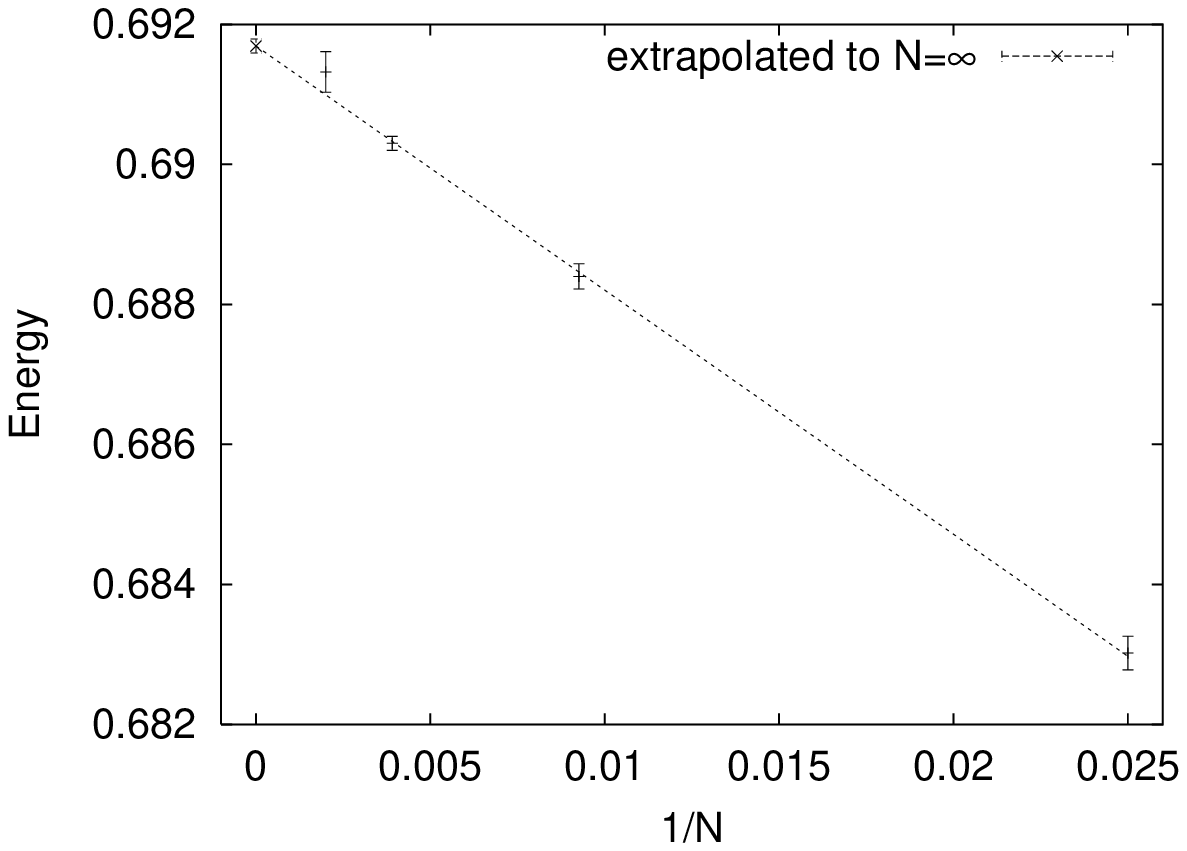} \\
  (a) & (b) \\
     & \\
  \includegraphics[height=2.0in,trim=0 0 0 0]{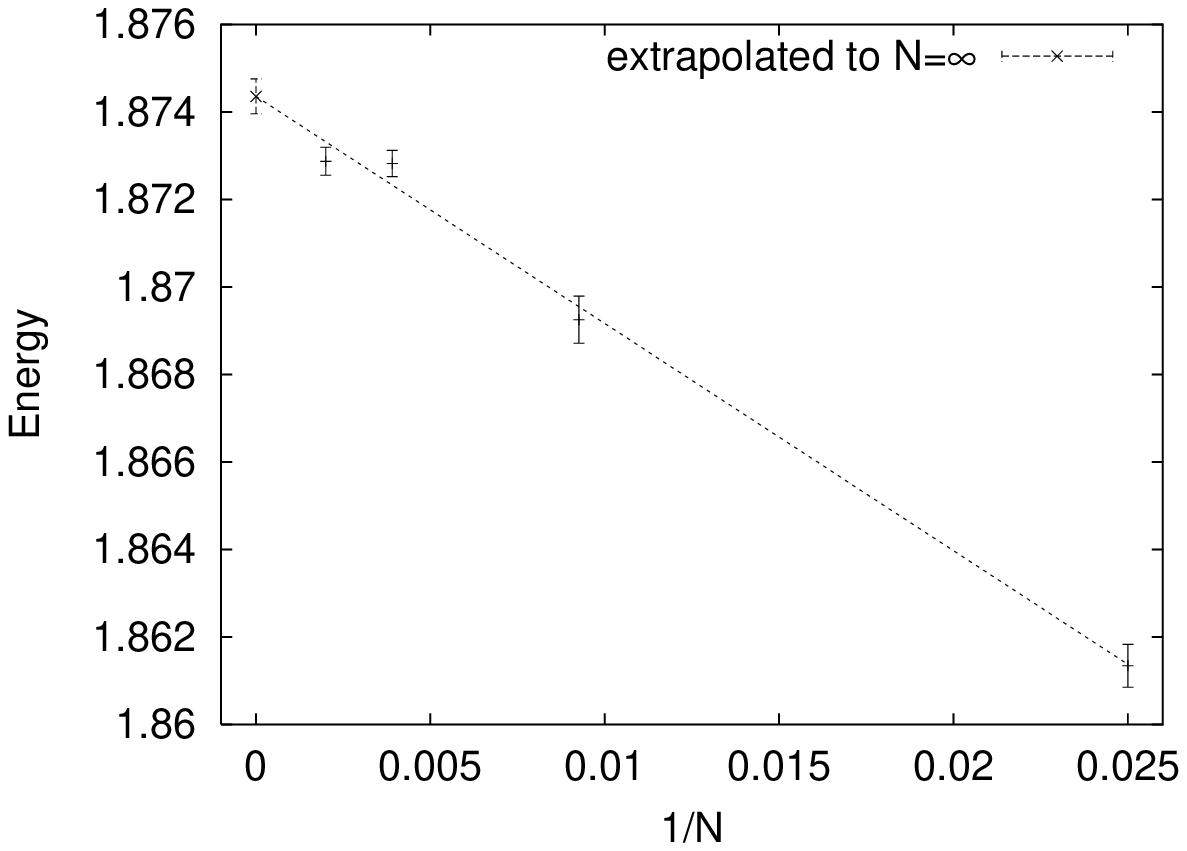} & 
  \includegraphics[height=2.0in,trim=0 0 0 0]{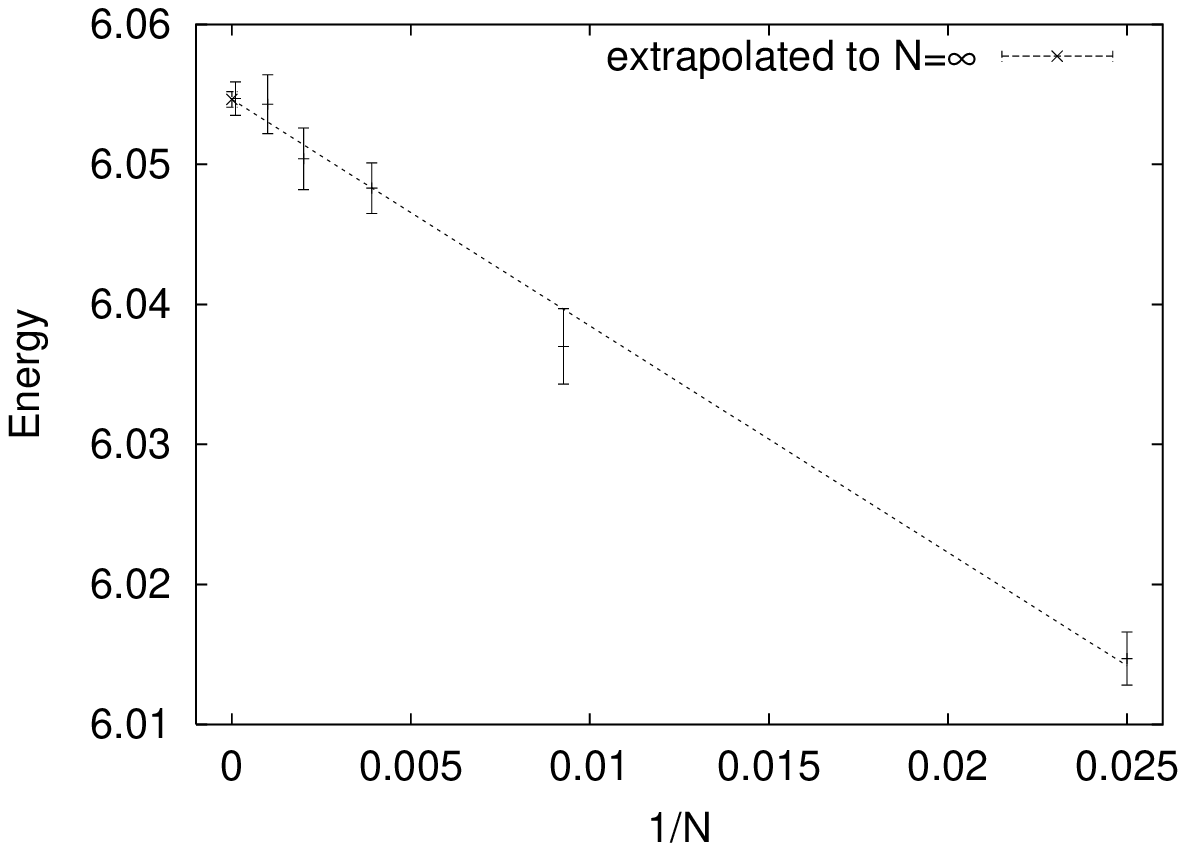} \\
  (c) & (d) \\
     & \\
 \end{tabular}
  \caption{VMC finite size effects for (a) $\rho=0.01$ (b) $\rho=0.05$
        (c)  $\rho=0.1$  (d) $\rho=0.2$
   \label{vmc_fs_err}}
\end{figure}

\begin{figure}
 \begin{tabular}{cc}
  \includegraphics[height=2.0in,trim=0 0 0 0]{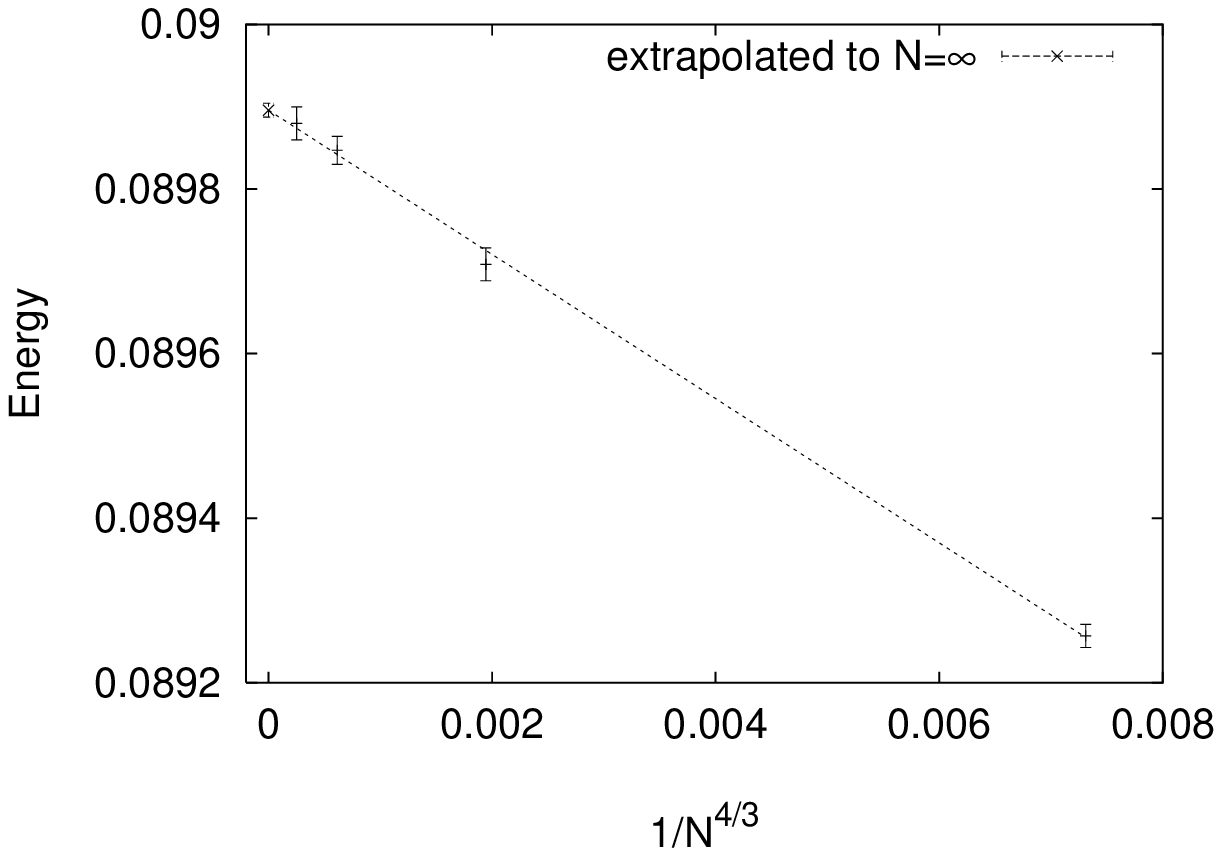} &
  \includegraphics[height=2.0in,trim=0 0 0 0]{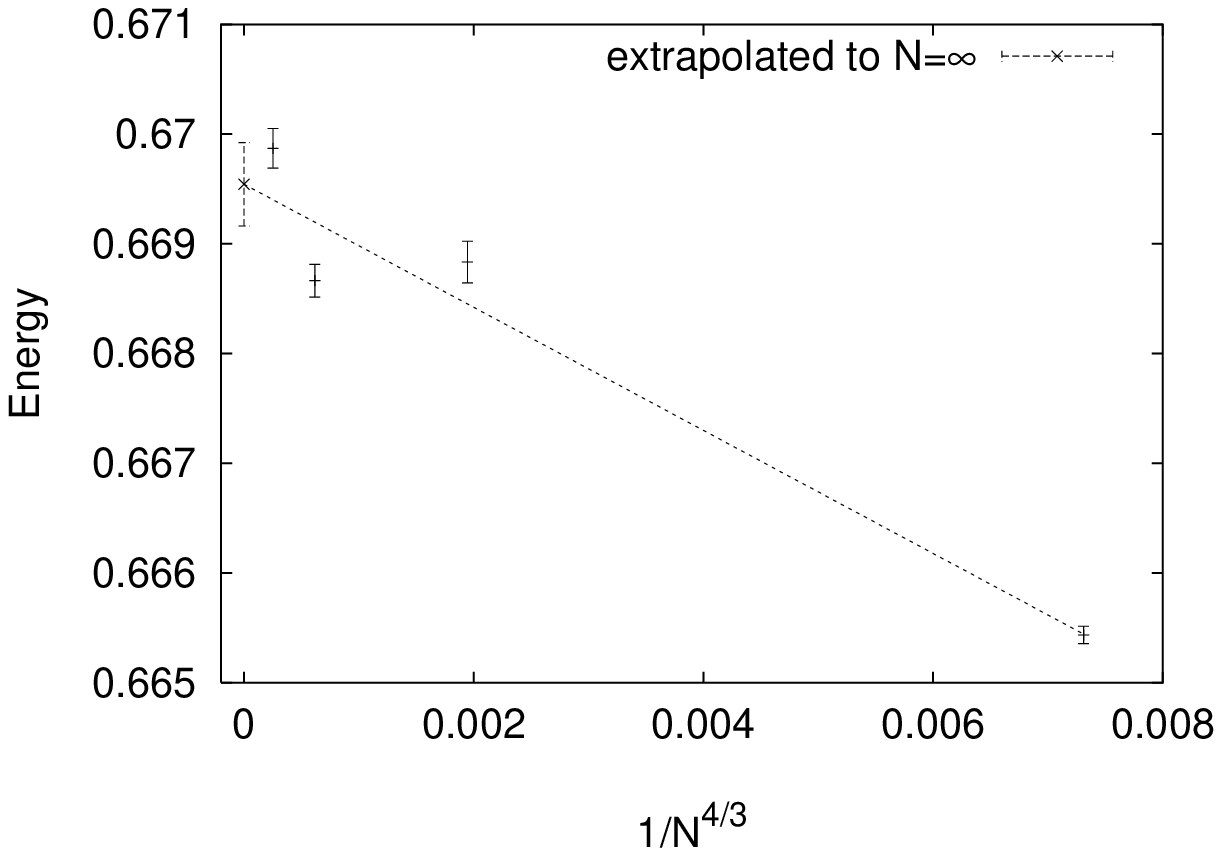}\\
  (a) & (b) \\
     & \\
  \includegraphics[height=2.0in,trim=0 0 0 0]{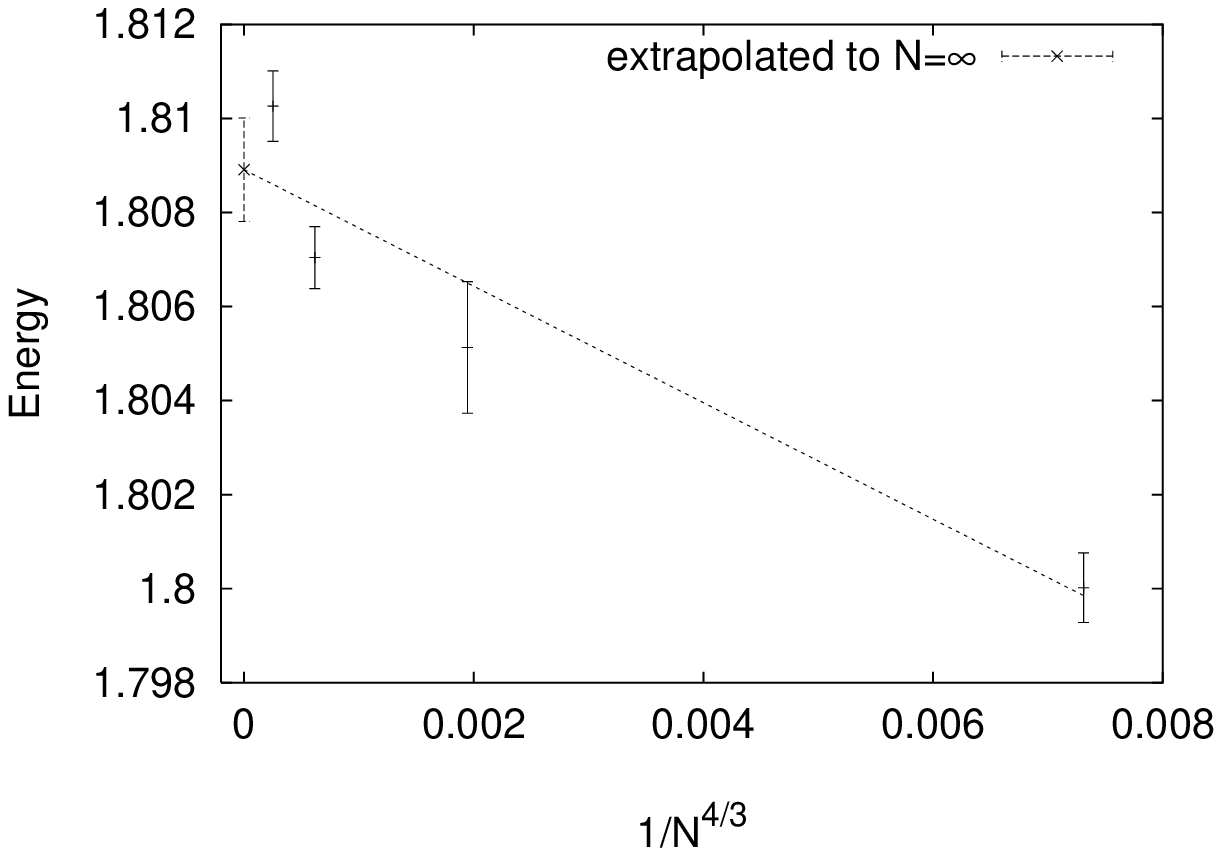} &
  \includegraphics[height=2.0in,trim=0 0 0 0]{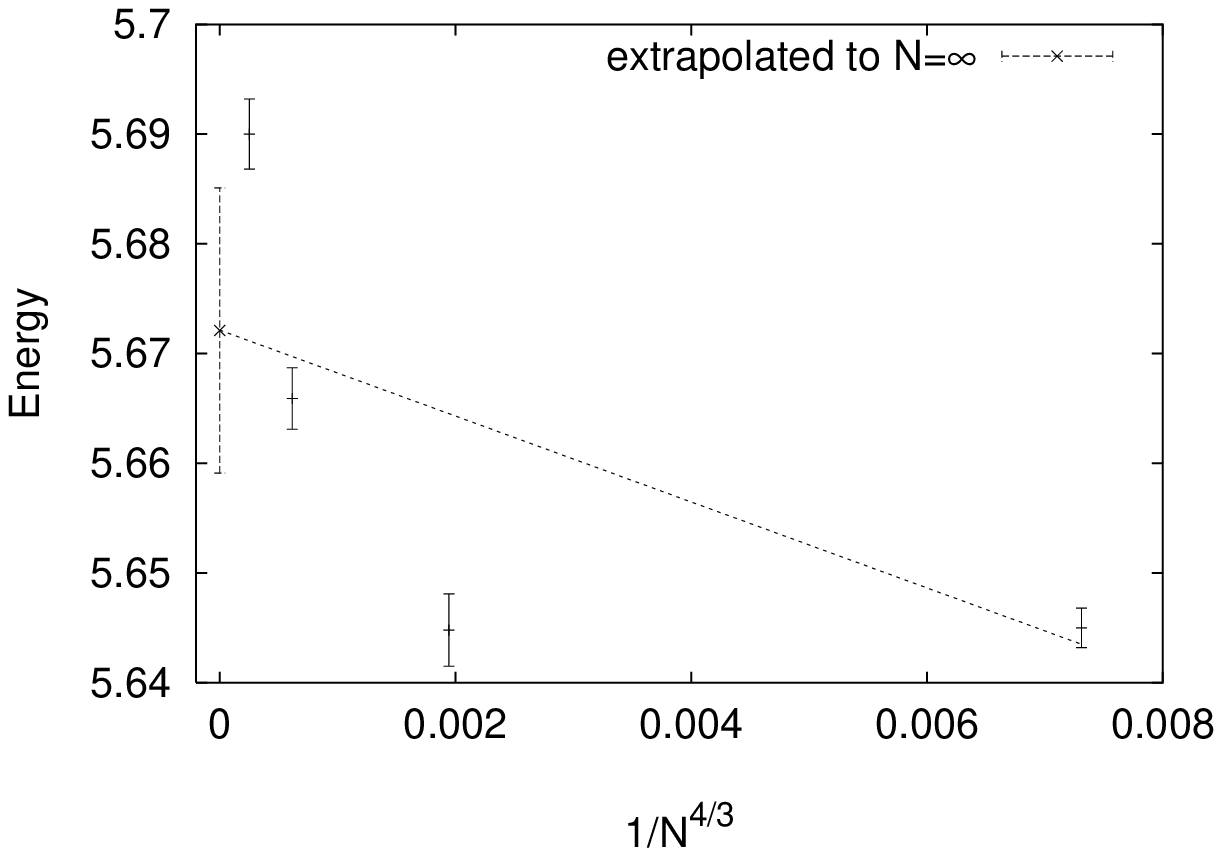}  \\
  (c) & (d) \\
     & \\
 \end{tabular}
  \caption{DMC finite size effects for (a) $\rho=0.01$ (b) $\rho=0.05$ 
        (c)  $\rho=0.1$  (d) $\rho=0.2$
   \label{dmc_fs_err}}
\end{figure}

We did calculations for systems with 40, 108, 256, and 500 particles.
For $\rho=.2$, additional VMC runs with $N=10^3$ and 
$N=10^4$ particles were done, and they are shown on the graph.
The VMC energy nicely fits the  $1/N$ behavior, as shown in 
Figure \ref{vmc_fs_err}.

The DMC finite size extrapolation is shown in Figure \ref{dmc_fs_err}.
At higher densities, the DMC energy does not appear to have a $1/N^{4/3}$
dependence (or even $1/N$ dependence).  However, the data is too sparse
and noisy to make a good determination as to what the functional form should
be, so we fit it to $1/N^{4/3}$.
The final infinite system results are given in Table \ref{infinite}.



The long wavelength excitations will take a long time to sample,
and their effect on the energy  (and $S(k)$) may only be apparent with very
long runs.  And the time needed to sample them will increase with box size.  The
larger box sizes may be insufficiently converged, causing the energy to be
too high.  This may explain the apparent curvature in the wrong direction.

There are several approaches for resolving the problem at
larger box sizes.  
The first is simply to perform even longer runs to see if the
energy drops.
Similarly, data for more system sizes would be helpful in outlining
the functional form of the finite size dependence.
Finally, explicit long
range correlations could be added to the wave function, 
of the form proposed by \cite{chester66}.

\begin{table}
\begin{center}
\caption{Energy extrapolated to infinite system size (in units of
$\frac{\hbar^2}{m\sigma^2}$)
  \label{infinite}}
\begin{tabular}{c|ccc}
$\rho$ & VMC & DMC & \cite{giorgini99} \\ \hline
.2  & $6.0546(6)$ & $5.67(1)$ \\
.1  & $1.8744(4)$ & $1.809(1)$ & 1.8130(35)\\
.05 & $0.6917(1)$ & $0.6690(4)$& 0.6690(5)\\
.01 & $9.144(2)\times 10^{-2}$ & $8.9896(8) \times 10^{-2}$ & 
    $8.980(5) \times 10^{-2}$ \\
\end{tabular}
\end{center}
\end{table}

\begin{figure}
 \begin{center}
  \includegraphics[height=3.0in,trim=0 0 0 0]{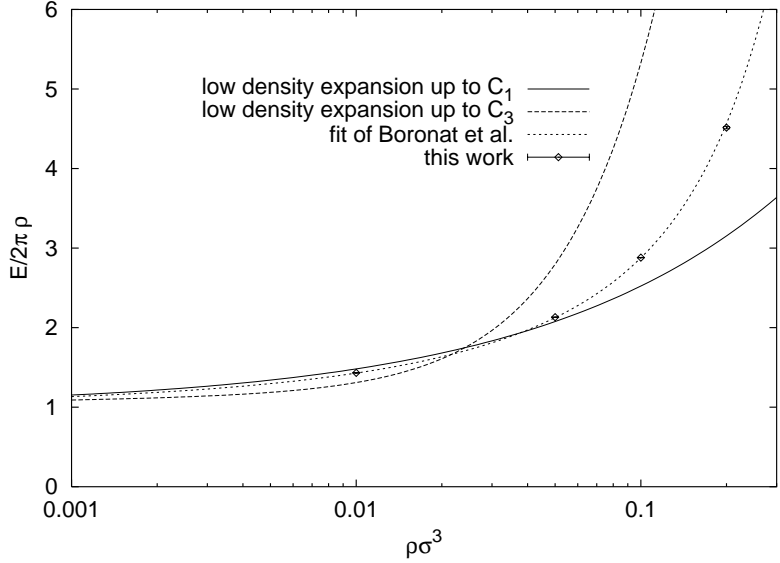}
  \caption{Energy vs. density
   \label{en_fig}}
 \end{center}
\end{figure}

The energy versus density is shown in Figure \ref{en_fig}, relative to the first
order term , which is linear in the density.
The low density exansion up to the $C_1$ term is also shown, as well
as up to $C_3$, using the fitted value of $73.296$ \citep{keller96}.
It is clear from the graph, and was noted by \cite{hugenholtz59},
that these additional terms by themselves do not help the expansion.

\cite{boronat00} treated $C_2$ and $C_3$ as adjustable
parameters and added two additional terms, $\rho^{5/2} \log(x)$ and
 $\rho^{5/2}$.
They get a very good fit to the high density data, as seen in 
Figure \ref{en_fig}.

The results of \cite{giorgini99} are also given in Table \ref{infinite}
for the three common densities computed.
Their data and ours agree within the error bars. 

\section{Distribution Functions and Condensate Fraction}
\begin{figure}
 \begin{center}
  \includegraphics[height=3.0in,trim=0 0 0 0]{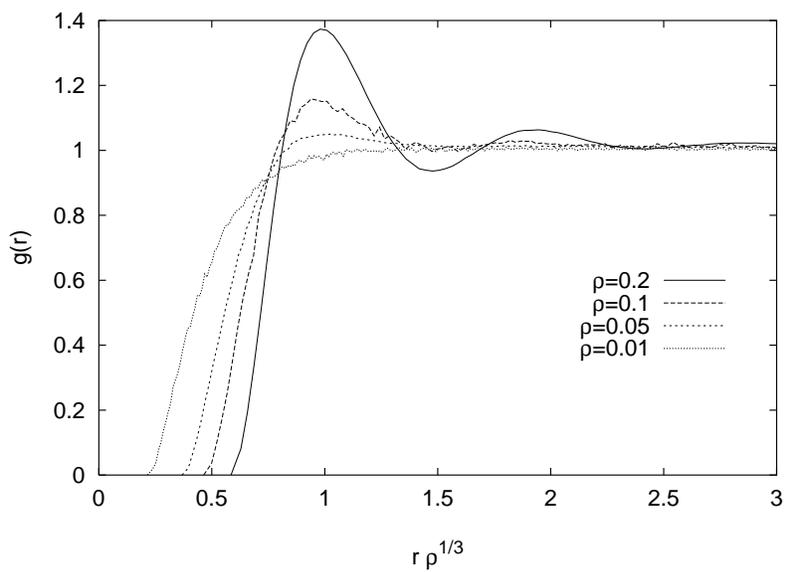}
  \caption{Pair distribution function for several densities
   \label{grfig}}
 \end{center}
\end{figure}

The two particle correlation function for all densities was calculated with a
system size of 256 particles.
The results for $g(r)$ using the extrapolated estimator are shown in 
Figure \ref{grfig}.
We see the liquid shell
structure developing as the density increases.

\begin{figure}
 \begin{center}
  \includegraphics[height=3.0in,trim=0 0 0 0]{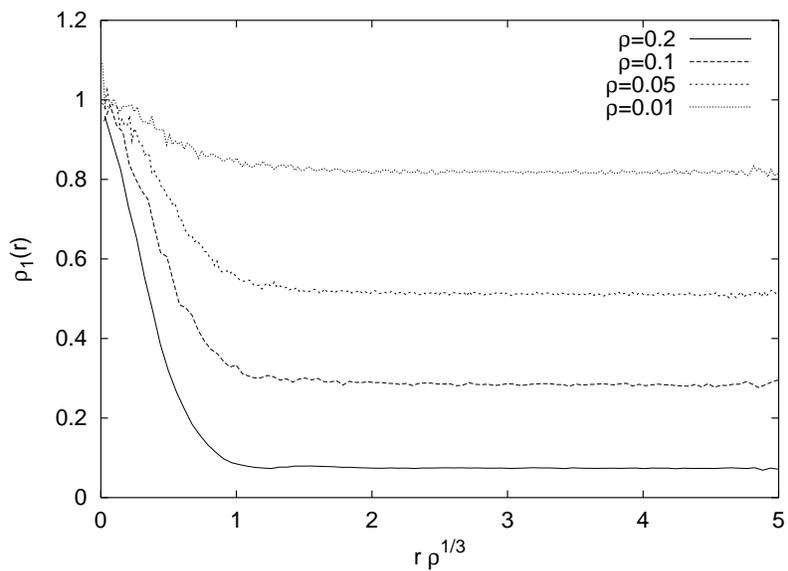}
  \caption{Single particle density matrix for several densities
   \label{prfig}}
 \end{center}
\end{figure}


The single particle density matrix is the projection of 
a many-body wave function
on to a single particle space.  
It is defined by
\be
\rho_1(r,r') = \int dr_2 ... dr_N \ \psi(r,r_2,...,r_N) \psi^*(r',r_2,...,r_N)
\ee
In the homogeneous case, $\rho_1$ only depends on the distance between 
$r$ and $r'$.
The large $r$ behavior of $\rho_1$ (or the $k=0$ behavior
of its Fourier transform, $n(k)$) is the condensate fraction.
At zero temperature, the many body wave function is in the ground state.
Because of interations,
not all the particles are in the single body zero momentum state
($k=0$ plane wave state in this case).

We used a method for sampling $\rho_1(r)$ which was given
by \cite{mcmillan65}.
The condensate fraction was obtained by integrating the single particle 
density matrix for distances greater than some cutoff, $r_c$, chosen to
be where $\rho_1(r)$ had reached a plateau.

\begin{figure}
 \begin{center}
  \includegraphics[height=3.0in,trim=0 0 0 0]{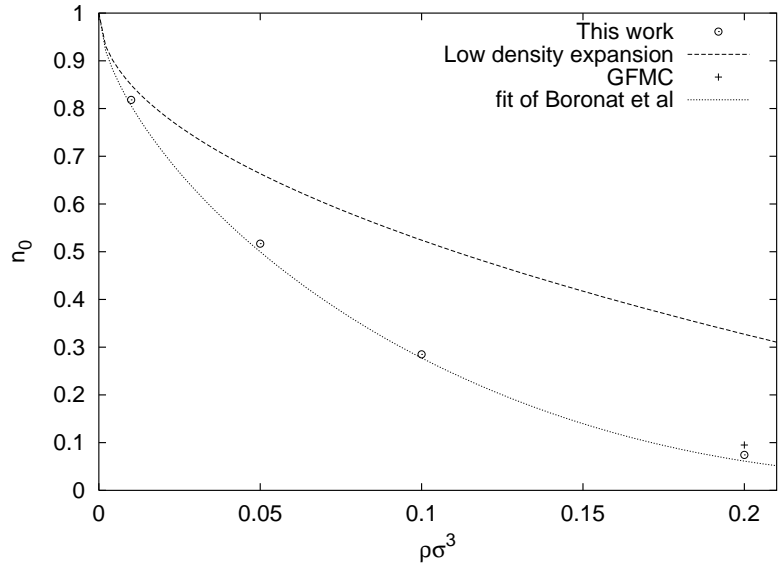}
  \caption{Condensate fraction vs. density
   \label{conds_fig}}
 \end{center}
\end{figure}

The condensate fraction is given in Table \ref{cond-frac} and shown
in Figure \ref{conds_fig}.  Also shown in the figure is the GFMC result
of $n_0 = 0.095(1)$ at $\rho=0.2$.
The low density expansion is given by
\be
n_0 = 1 - \frac{8}{3 \sqrt{\pi}} \rho^{1/2}
\ee
Similar to their treatment of the energy, \cite{boronat00} 
added two additional terms, $\rho$ and $\rho^{3/2}$.
Their fit does a good job at higher densities, where, 
as expected, the low density expansion misses the full extent of the depletion.


\begin{table}
\begin{center}
\caption{Condensate fraction
 \label{cond-frac}}
\begin{tabular}{c|lll}
$\rho$ & VMC & DMC (mixed) & Extrapolated  \\ \hline
.2  & $0.1009(5)$ & $0.0876(3)$ & $0.0743(8)$ \\
.1  & $0.307(2)$  & $0.2960(5)$  & $0.285(2)$ \\
.05 & $0.563(2)$  & $0.5401(4)$  & $0.517(8)$ \\
.01 & $0.834(1)$  & $0.826(1)$  & $0.818(2)$ \\
\end{tabular}
\end{center}
\end{table}


%% file: chap7.tex
\chapter{Hydrogen}




Hydrogen has been the subject of many experimental and theoretical studies.
Theoretically, its simple electronic structure make it a favorable
first target for various methods.
Experimentally, hydrogen has been compressed 
by shock waves and also with a diamond anvil cell.
We will present some CEIMC simulations and compare the results
with those from one of the gas gun shock wave experiments.

\section{Experiment}
The high pressure experiments fall into two categories - transient compression
from a shock wave or static compression from a diamond anvil cell.
The shock wave experiments reach higher temperatures and pressures,
but obtain more limited data.  
A high-velocity projectile hits a stationary target, inducing a
shock wave in the target.
The target is analyzed by the Hugoniot relations, derived by 
 treating the shock wave as an ideal discontinuity and applying
conservation of mass, momentum, and energy across it \citep{zeldovich66}.
The relations are then
\bea
P - P_0 &=& \rho_o u_s u_p \\
\rho &=& \rho_0 u_s/(u_s-u_p) \\
E - E_0 &=&  \half(V_0-V)(P+P_0)
\eea 
where $E_0$, $P_0$, $V_0$, and $\rho_0$ are the initial energy, pressure,
volume, and density, respectively.
The velocity of the shock wave is $u_s$ and $u_p$ is the velocity
of the projectile driving the shock.


There are a number of methods for accelerating a projectile
\citep{cable70}, but the two most prominent methods for hydrogen targets are
the two stage light gas gun \citep{nellis83,holmes95,weir96,nellis99} 
and a large laser \citep{dasilva97,collins98,celliers00}.

Recent advances make it possible to measure the temperature
by light emission from the samples during compression \citep{holmes95}.
Measurement of the conductivity is also possible,
used in recent experiments to detect metallic hydrogen \citep{weir96}.

The diamond anvil cell (DAC) is used to generate large static pressures.
It has been used to study the fluid phase and several solid phases
\citep{mao94}.  
It has also been used to determine the melting curve for hydrogen
up to 500K
\citep{diatschenko85,datchi00}.


\section{Theory}

Free energy models 
are typically based on the chemical picture, where molecules, atoms
and various types of ions are all treated as different species of 
particles.
Solving the physical picture, where the only fundamental particles are
electrons and protons, is much more difficult (see Path Integral Monte
Carlo below).
The free energy of the various phases is constructed from a variety
of fits to experimental data, equation of state data from reference systems
(Lennard-Jones and hard sphere),
and empirical and theoretical interaction potentials.

One of the best known models is that of Saumon and Chabrier
\citep{saumon91,saumon92}.
Extensive tables for astrophysical use were
published by \cite{saumon95}.
Another model was developed by \cite{kitamura98}, to study the
plasma (metal-insulator) transition.



The Path Integral Monte Carlo (PIMC) method is in principle the best method 
for simulations, since it treats both the electrons and protons quantum
mechanically at non-zero temperature \citep{pierleoni94,magro96,militzer00b,
militzer-thesis00}. 
The only major uncontrolled approximation is the location of the electron nodes,
with problems and a solution similar to the fixed node method in DMC.
\cite{militzer00a} have made progress in improving the nodal structure
used in these calculations.
PIMC is based on breaking up a thermal density matrix into a product of high
temperature components, and consequently it works well at high temperature
and becomes less efficient as the temperature decreases.
About 5000K is currently the lower limit for PIMC calculations.
Our CEIMC simulation technique should make a nice complement to the
PIMC method.

There have also been path integral studies using empirical potentials,
in order to examine the quantum effects of the nuclei on the system
\citep{wang96,wang97,cui97,chakravarty99}.

The Car-Parrinello method has been used to simulate this system
\citep{hohl93,kohanoff97,pfaffenzeller97,galli00}.
At low temperature, it it necessary to treat the  nuclei with 
path integrals \citep{biermann98,kitamura00}.
Some studies used LDA with the $\Gamma$ point approximation (using
only one $k$-point for the integral over the Brillouin zone),
which is not sufficient to converge the anisotropic behavior of the
potential \citep{mazin95}, and gives rise to unphysical planar structures 
\citep{kohanoff97}.


\section{Pressure and Kinetic Energy}
 The pressure is computed by a virial estimator based on the potential and
kinetic energies
\be
  P = \frac{1}{3V} \left[ 2 \expect{K} + \expect{\calV} \right]
\ee
where $V$ is the volume and $\calV$ is the potential energy.
In these MC simulations, only the kinetic energy of the electrons is
explicitly computed.  
The kinetic energy of the nuclei must also be added.

We are only considering
hydrogen in the molecular state, and further assume that 
rotational and vibrational motion can be separated.
The characteristic temperature for
quantum effects for rotational motion is about 85 K for H$_2$ \citep{landau80}.
At our simulation temperatures, we can use the classical expression for 
the rotational kinetic energy, $E_{\mathrm{rot}} = 2kT$.

The characteristic vibrational temperature for $H_2$ is $\theta_v = 6100$ K, 
so it is necessary 
to use the quantum expression for the vibrational kinetic energy.
It is 
\be
E_{\mathrm{vib}} = \frac{\theta_v}{e^{-\theta_v/T}  - 1}
\ee
For $D_2$, the characteristic temperature should be a factor 
of $\sqrt{2}$ lower.

Of course, these expressions are only valid for a free molecule.
To truly treat the kinetic energy of the nuclei correctly in the 
interacting system, path integrals should be used for the nuclei.




\section{Individual Configurations}

We took several configurations from PIMC simulations at 5000K at
two densities ($r_s = 1.86$ and $r_s = 2.0$), 
and compared the electronic energy using VMC, DMC, DFT-LDA, and 
some empirical potentials.
The DFT-LDA results  were obtained from a plane wave code using an energy
cutoff of 60 Rydbergs, and using the $\Gamma$ point approximation \citep{ogitsu00}.

The empirical potentials are the Silvera-Goldman \citep{silvera78}
and the Diep-Johnson \citep{diep00a,diep00b}.
To these we added the energy from the Kolos \citep{kolos64} intramolecular 
potential to get the energy as a function of bond length variations.
The Silvera-Goldman potential was obtained by fitting to 
low temperature experimental data,
with pressures up to 20 kbar, and is isotropic.
The Diep-Johnson potential
is the most recent in a number of potentials for the isolated H$_2$-H$_2$
system.
It was fit to the results of accurate quantum chemistry calculations 
for a number of 
H$_2$-H$_2$ configurations.
It is an anisotropic potential.

\begin{figure}
 \begin{center}
  \includegraphics[height=3.0in,trim=0 0 0 0]{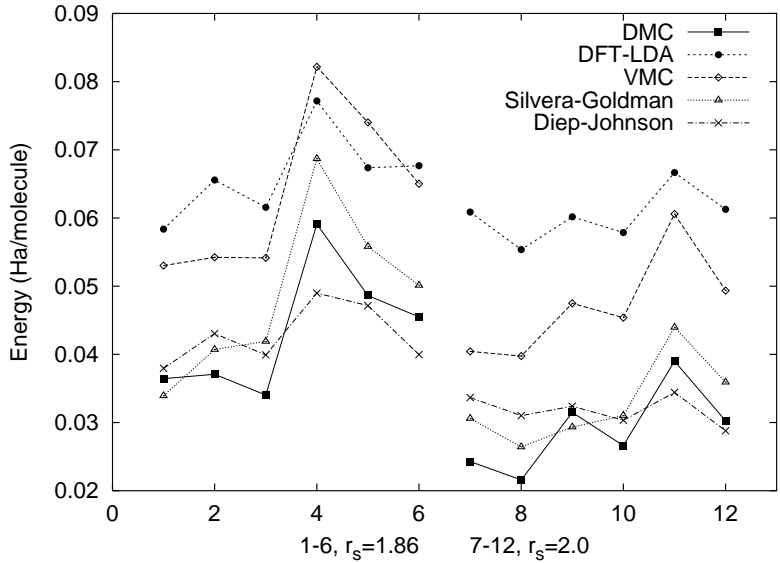}
  \caption{Electronic energy for several configurations computed by several
methods.  The energy is relative to an isolated H$_2$ molecule.
   \label{pimc-cfgs}}
 \end{center}
\end{figure}

The energies relative to an isolated H$_2$ molecule 
are shown in Figure \ref{pimc-cfgs}.  The first thing we notice is that 
the classical potentials are more accurate than VMC or DFT.
The Silvera-Goldman mostly does a good job of reproducing the DMC results.
\footnote{It should be noted that we are taking the SG potential far
from the temperature range it was fit to.}
Some of the failures of the SG potential can be attributed to the lack
of anisotropy.
The isolated H$_2$-H$_2$ potential (Diep-Johnson) has much weaker 
interactions, compared with interactions in a denser system.

The PIMC method itself gives an average energy of about $0.07(3)$ Ha
for both densities.
Improvements in the fermion nodes appear to lower the energy 
\citep{militzer00a,militzer00b}, 
although the error bars are still quite large.
There are also corrections to some internal approximations that lower the
energy by an additional 0.02 Ha \citep{militzer-thesis00}.  
These effects combined 
seem to bring the PIMC energy in rough agreement with the DMC energy.

We used the Silvera-Goldman potential for pre-rejection.  As seen in
the Figure \ref{pimc-cfgs}, it resembles the DMC potential even though
it lacks anisotropy.
A hybrid potential was created by \cite{cui97},
taking the isotropic part from a potential that was fit to high density,
and combining that  with the anisotropic part from one of the 
isolated H$_2$-H$_2$ 
potentials.   We did not pursue this approach for
constructing a better potential for pre-rejection.


\section{Results}

We obtained results from simulations at three state points, two of
which can 
be compared with the gas gun data of \cite{holmes95}.
The pressure is given in Table \ref{gasgundata}, with results
from the gas gun experiments,
the Saumon-Chabrier model,
from simulations using the Silvera-Goldman potential, 
and from our CEIMC simulations.  These state points are in the fluid
molecular H$_2$ phase.
For the gas gun experiments, the uncertainties in the mearsured temperatures
are around 100-200K.
The experimental uncertainties in the volume and pressure were not given, but
previous work indicates that they are about 1-2\% \citep{nellis83}.

We did CEIMC calculations using VMC or DMC for computing the underlying
electronic energy, which are the first such QMC calculations in this range.
The simulations at $r_s=2.1$ and $r_s=1.8$ were done with 32 molecules, and 
the simulations at $r_s=2.202$ were done with 16 molecules. 
We see that the pressures from VMC and DMC are very similar,
and that for $r_s=2.1$ we get good agreement with experiment.

There is a larger discrepancy with experiment at $r_s=2.202$.
The finite size effects are fairly large, especially with DMC.
We also did simulations at $r_s=2.1$ with 16 molecules and obtained
pressures of $0.264(3)$ Mbar for CEIMC-VMC and $0.129(4)$ Mbar for
CEIMC-DMC.
The Silvera-Goldman potential showed 
much smaller finite size effects than the CEIMC simulations, so we
that the electronic part of the simulation is largely responsible for the
observed finite size effects.

\begin{figure}
 \begin{tabular}{cc}
  \includegraphics[height=2.0in,trim=0 0 0 0]{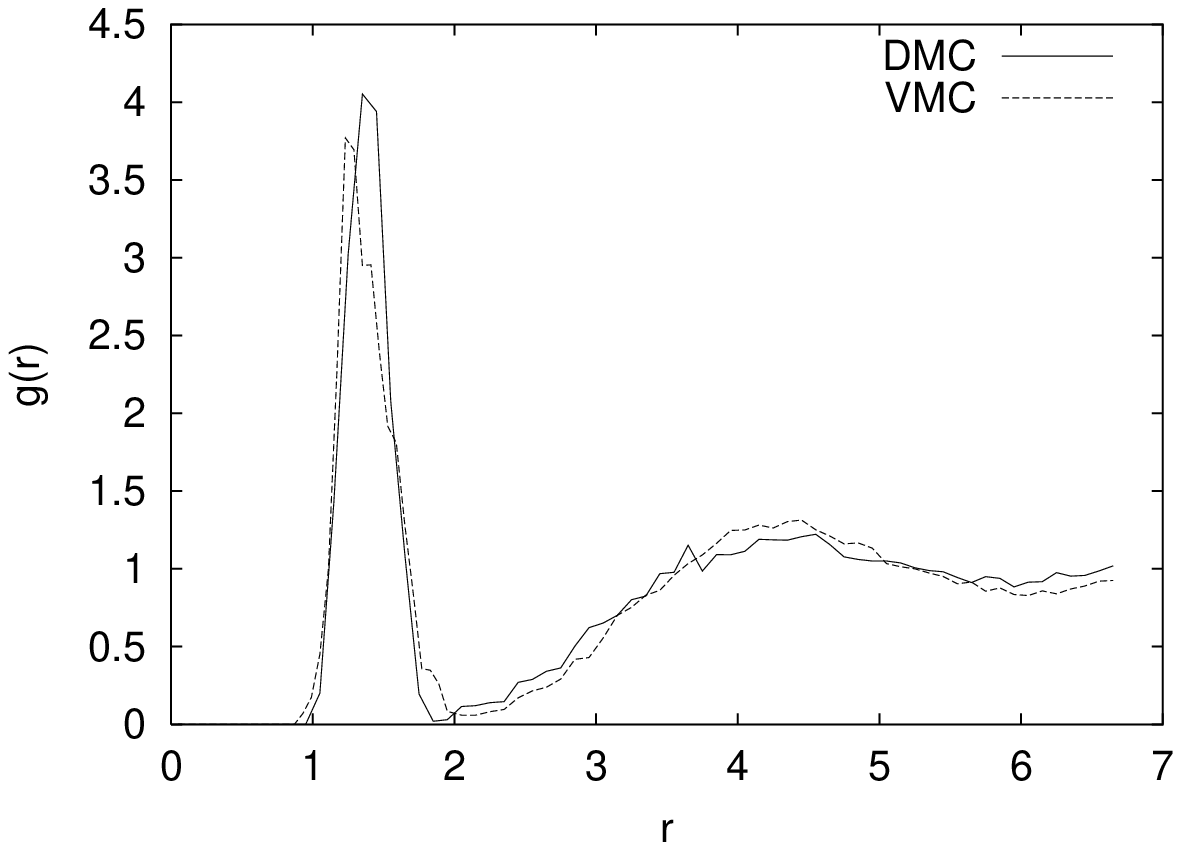} &
  \includegraphics[height=2.0in,trim=0 0 0 0]{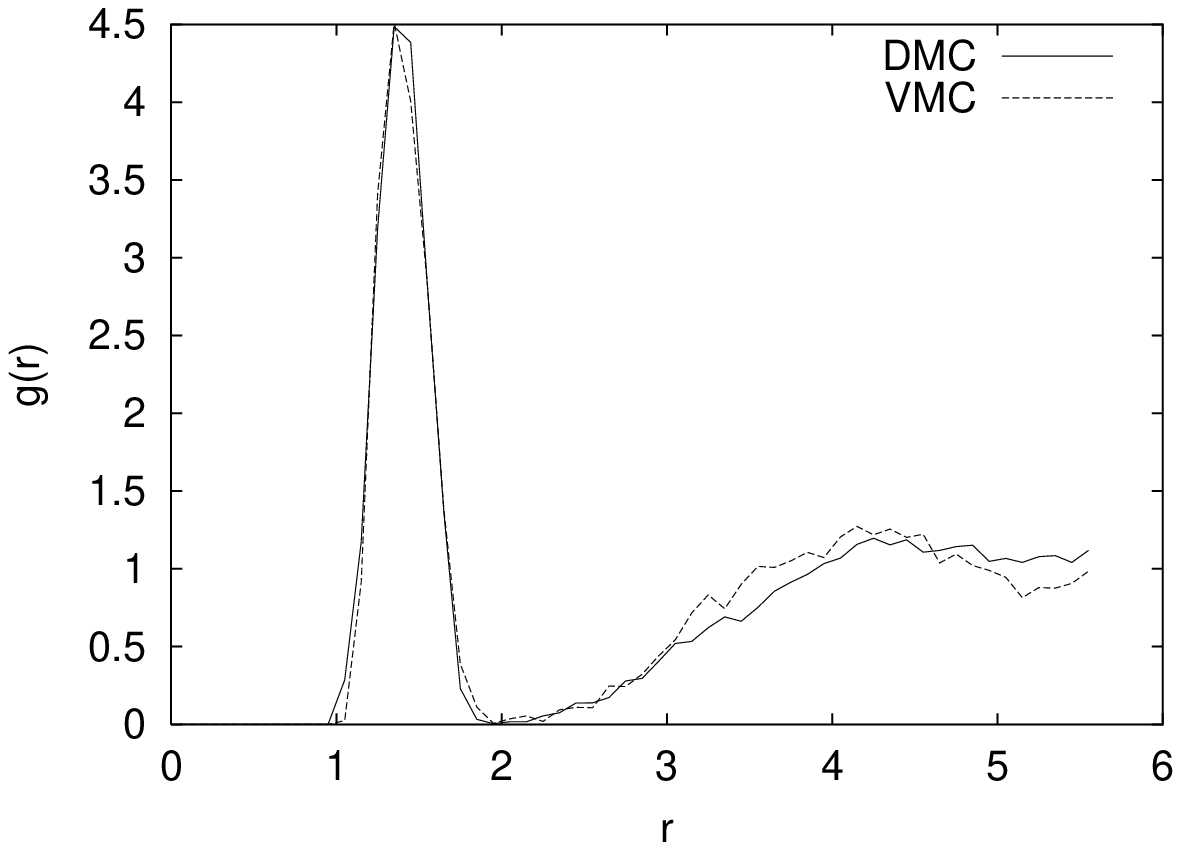} \\
  (a) & (b) \\
     & \\
 \end{tabular}
  \caption{Proton pair distribution function $g(r)$ for (a) $r_s = 2.1$ and
T=4530 K  (b) $r_s=2.202$ and T=2820 K
   \label{gr-results}}
\end{figure}

The energies for all these systems are given in Table \ref{h_energy}.
The energy at $r_s=2.1$ with 16 molecules for CEIMC-VMC 
is $0.0711(4)$ Ha and for CEIMC-DMC is $0.0721(8)$ Ha.
The average molecular bond length is given in Table \ref{h_avebond},
and we see the bond length is compressed relative to the free molecule.
The proton-proton distribution functions comparing CEIMC-VMC and
CEIMC-DMC are shown in Figure \ref{gr-results}.
The VMC and DMC distribution functions look similar,
with the first large intramolecular peak around $r=1.4$ and 
the intermolecular peak around $r=4.5$.


\begin{table}
\begin{center}
\caption{Pressure from simulations and shock wave experiments
 \label{gasgundata}}
 \begin{tabular}{cccccccc}
r$_s$  &    V(cc/mol)  &    T(K)   & \multicolumn{5}{c}{Pressure (Mbar)} \\
     &      &      &   Gasgun  &  S-C &  S-G & CEIMC-VMC & CEIMC-DMC \\\hline
 2.100     &    6.92      &   4530    &    0.234  
    &  0.213  & 0.201  & 0.226(4)  & 0.225(3)  \\
 2.202     &    7.98      &   2820    &    0.120   
    & 0.125   &  0.116      & 0.105(6)   &  0.10(5)  \\
 1.800      & 4.36  & 3000 & - & - & 0.528 & - & 0.433(4)  \\
 \end{tabular}
\end{center}
\end{table}



\begin{table}
\begin{center}
\caption{Energy from simulations and models,
 relative to the ground state of an
isolated H$_2$ molecule. The H$_2$ column is a single 
thermally excited molecule plus the quantum vibrational KE.
 \label{h_energy}}
 \begin{tabular}{cccccccc}
r$_s$  &    V(cc/mol)  &    T(K)   & \multicolumn{5}{c}{Energy (Ha/molecule)} \\
     &      &      &   H$_2$  &  S-C &  S-G & CEIMC-VMC & CEIMC-DMC \\\hline
 2.100     &    6.92      &   4530    &    0.0493
    &  0.0643  & 0.0689  & 0.0663(8)  & 0.0617(2)  \\
 2.202     &    7.98      &   2820    &    0.0290 
    & 0.0367   &  0.0408      & 0.0305(8)   &  0.0334(9)  \\
 1.800     &   4.36  & 3000 & 0.0311   & - & 0.0722   & - & 0.055(1)
 \end{tabular}
\end{center}
\end{table}


\begin{table}
\begin{center}
\caption{Average molecular H$_2$ bond length.
 The H$_2$ column is a single
thermally excited molecule in free space.
 \label{h_avebond}}
 \begin{tabular}{ccccc}
r$_s$  &  T(K)   & \multicolumn{3}{c}{Average bond length (Bohr)} \\
       &         &   H$_2$  & CEIMC-VMC & CEIMC-DMC \\\hline
 2.100  & 4530 &   1.550  & 1.431(1)  & 1.413(3) \\
 2.202  & 2820 &   1.486  & 1.443(1)  & 1.429(6) \\
 1.800  & 3000 &   1.492  &  -      & 1.410(1)
 \end{tabular}
\end{center}
\end{table}

The CEIMC-VMC simulations at $r_s=1.8$ and 3000 K never converged. Starting
from a liquid state, the energy decreased the entire simulation.
Looking at the configurations revealed they were forming a plane.
It is not clear whether it was trying to freeze, or forming structures similar
to those found in DFT-LDA calculations with insufficient Brillouin zone
sampling
\citep{hohl93,kohanoff97}.
The CEIMC-DMC simulations did not appear to have any difficulty, so
is seems the VMC behavior was due to inadequacies of the wave function.

\begin{figure}
 \begin{center}
  \includegraphics[height=3.0in,trim=0 0 0 0]{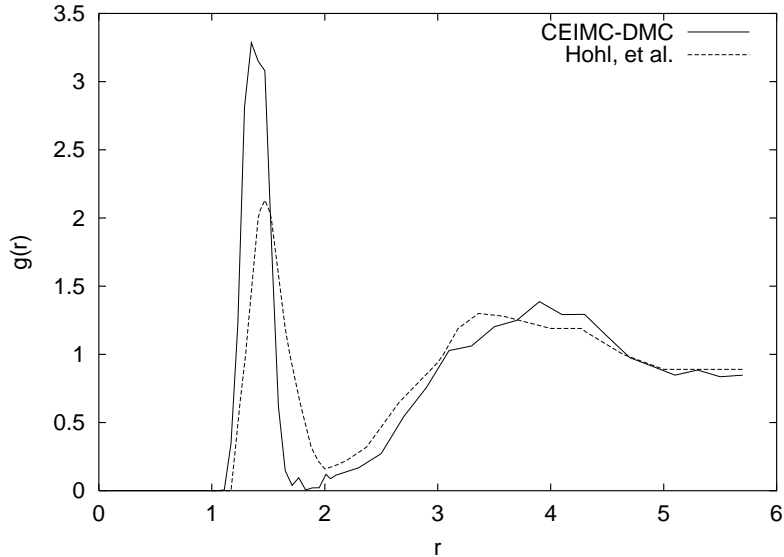}
  \caption{The proton pair distribution function, $g(r)$, 
 close to $r_s = 1.8$ and T = 3000K. 
   \label{rs18}}
 \end{center}
\end{figure}

\cite{hohl93} did DFT-LDA simulations at $r_s=1.78$ and T=3000K, which is
very close to our simulations at $r_s=1.8$.  The resulting proton-proton
distribution functions are compared in Figure \ref{rs18}.
The discrepancy between CEIMC and LDA in the intramolecular
portion of the curve has several possible causes.
On the CEIMC side, it may be due to an insufficiently long run or 
due to the molecular nature of the wave function, which does not
allow dissociation.  
The deficiencies of LDA may account for it preferring fewer and less
tightly bound molecules. LDA is known to overestimate the bond length
of a free hydrogen molecule \citep{hohl93}, which would account for the 
shifted location of the bond length peak.

\section{Simulation analysis}

We also recorded some diagnostic information about the workings of the 
simulation,
such as the average noise level, the relative noise parameter 
$f = (\beta \sigma)^2 t/t_0$,
and a quantity called the additional noise rejection ratio,
$\eta$.
When a move is rejected with the penalty method, it is useful
to  recompute the acceptance decision with the same random number and 
without the noise penalty.  If the move would have been accepted without
the noise penalty, it is considered a rejection due to noise (as opposed
to a rejection due to a trial move that increases the energy).
This can be used to monitor the effects of the noise on the simulation.
The additional noise rejection ratio is defined as
\be
\eta = \frac{N_\mathrm{noise\_rej}}{N_\mathrm{noise\_rej} + N_\mathrm{accept}}
\ee
If $\eta$ is small, the noise is causing few additional rejections.
If $\eta$ is 1/2, the noise is causing as many moves to be
rejected as accepted.  As $\eta \ra 1$, the noise is causing
many moves to be rejected.

Table \ref{sim-quantities} show the noise level $(\beta \sigma)$, 
the relative noise
parameter, $f$, the additional noise rejection ratio, a ratio of 
the error level for the direct method and the two-sided method, and the
time for a single quantum step.
Looking at $f$, we see it is small for VMC and large for DMC. This 
is because of VMC optimization takes a proportionately larger amount
of time in the VMC run than in the DMC run.

We used the two-sided method for computing energy differences of trial
moves with
VMC, but only used the direct method with DMC.  
The column headed $\sigma^2_{d}/\sigma^2_{ts}$
shows how the efficiency of computing the energy difference is
improved by using the two-sided method.
This improvement is only in the energy difference part of the total time, 
the optimization time is unaffected (and is
a large part of the run time, since the $f$ parameter is small).
In Chapter 3, there was an example showing that the two-sided method was not as
effective for DMC as for VMC.  But in these simulations the DMC
runs have a much larger $f$ parameter, so even small reductions in 
the noise level would have an impact on the run time.

Some of DMC energy differences had values of noise many times greater
than the average, which
may be due to an instability in the DMC algorithm.
We removed these outliers in computing the average noise
level.

\begin{table}
\begin{center}
 \caption{ Simulation quantities ordered according to average noise level, 
       $\beta \sigma$. 
 The time column is the time
for a single quantum step in minutes on an SGI Origin 2000. N is the number
of molecules in the simulation.
  \label{sim-quantities}}
 \begin{tabular}{cccc|cccccc} 
  $r_s$ & T(K) & N & QMC & $\beta \sigma$ & $f$  & 
      $\eta$ & $\sigma^2_{d}/\sigma^2_{ts}$ & time (min)\\ \hline
 2.100 & 4530 & 16 & VMC &  0.68  & 0.17  & 0.11 & 2.2 & 18 \\
 2.202 & 2820 & 16 & VMC &  0.70  & 0.27  & 0.13 & 3.2 & 21 \\
 2.100 & 4530 & 32 & VMC &  0.90  & 0.29  & 0.16 & 2.3 & 70 \\
 1.800 & 3000 & 32 & VMC &  0.91  & 0.30  & 0.15 & 7.7 & 89 \\
 2.100 & 4530 & 16 & DMC &  1.62  & 2.28  & 0.28 & - & 76 \\
 2.100 & 4530 & 32 & DMC &  1.74  & 5.30  & 0.29 & - & 440 \\
 2.202 & 2820 & 16 & DMC &  2.02  & 5.33  & 0.40 & - & 92 \\
 1.800 & 3000 & 32 & DMC &  2.42  & 13.1  & 0.42 & - & 510 \\
 \end{tabular}
\end{center}
\end{table}

We tried the method for even lower temperatures with a simulation 
at T=800 K and $r_s=1.8$ and it had a promising
start, but after a while
the acceptance ratio dropped and we were unable to get
any usable data.

%
%
%

\section{Future Work}

The finite size effects in DMC need to be resolved.
Using Ewald sums for computing the Coulomb interaction might help
alleviate some of the finite size effects.
Extending the wave function to allow for dissociated molecules and
to provide for ionization would help make the simulations more accurate,
particularly at higher temperatures and pressures.


%% file: chap8.tex
\chapter{Conclusions}

In this work we have developed new methods for increasing the 
scope of QMC calculations, and for increasing their efficiency.
Variational Monte Carlo depends on optimizing parameters, but
the presence of noise makes it difficult.  We have examined 
several different kinds of optimization approaches and compared them.
Further work should improve these methods even more.





The boson hard sphere model is an important theoretical model.
We have performed ``computational experiments'' to obtain the ground
state energy of this model.
The effects of long range correlation on the energy are masked
by the current uncertainty in the infinite system size results.
However, if more accurate results are desired, the nature of the
long range correlations and their effect on the energy 
will need to be more clearly resolved.

 
As a method for including increasingly more detailed
and accurate physical effects in our simulations, we
have developed the Coupled Electronic-Ionic Monte Carlo method.
The central idea is simple, but several supporting developments were
needed to make it computationally feasible.
The penalty method enables use of energy differences 
with a noise level of approximately 
$k_B T$, rather than needing noise smaller than some 
fraction of $k_B T$ to avoid bias.
The two sided energy difference method can stably compute these
energy differences.


The CEIMC method was applied to a system of molecular hydrogen
at a few state points.  It shows promise for generating accurate
simulation results.


%% file: appen1.tex
\chapter{Determinant Properties}

The elements of the Slater matrix are $D_{ij} = \phi_j(r_i)$.
The Slater determinant looks like
\be
\Det{
 \fcmatrix{\phi_1(r_1)}{\phi_n(r_1)}{\phi_1(r_n)}{\phi_n(r_n)}
}
\ee
We assume that the single particle orbitals depend only on a single
coordinate (ie, no backflow).

The determinant of a matrix can be computed using the expansion by cofactors.
This expands the determinant of an $N \times N$ matrix into a sum
of $N$ determinants of $(N-1)\times (N-1)$ matrices.
As a recursive algorithm for computing the determinant, it is not very
efficient, but for theoretical analysis, it is very useful for
isolating the influence of a single row or column.

Define the cofactors of a matrix $M$ to be
\be
c_{ij} = (-1)^{i+j} \Det{M_{ij}}
\ee
where the matrix formed by $c_{ij}$ is called the cofactor matrix.
The matrix $M_{ij}$ is an $(N-1) \times (N-1)$ matrix formed by removing
row $i$ and column $j$ from $A$.
The determinant of $A$ can then be written as
\be
\Det{A} = \sum_j a_{kj} c_{kj} = \sum_i a_{ik} c_{ik}
\ee
for $k=1 \ldots N$.
The transpose of the cofactor matrix is called the adjoint of $A$.
Now the adjoint is related to the inverse by
\be
{\mathrm{adj}\ } A = \Det{A} A^{-1}
\ee

To compute the ratio of determinants, 
expand the determinant of $D(r'_k)$ in cofactors about the $k$th row.
Note that then the cofactors have no dependence on $r'_k$.

\bea \nonumber
 \Det{D(r'_k)} &=& \sum_i \phi_i(r'_k) c_{ki} \\ \nonumber
         &=& \sum_i \phi_i(r'_k) \Det{D(r_k)} (D^{-1}(r_k))_{ik} \\\nonumber
 \frac{\Det{D(r'_k)} }{ \Det{D(r_k)} } &=& \sum_i \phi_i(r'_k)
     (D^{-1}(r_k))_{ik}
\eea

If a move is accepted, the inverse matrix can be updated in $O(N^2)$ time
(rather than $O(N^3)$ for recomputing the inverse).
The formula for updating an inverse if only a single row (or column) changes
was given by 
\cite{sherman50}.
Let $q$ be the ratio of
determinants given above.  Row $k$ merely needs
to be updated to reflect the new determinant, $D^{-1}_{kj} = D^{-1}_{kj}/q$ . 
The other rows are updated as
\be
D^{-1}_{ij} = D^{-1}_{ij} - \frac{D^{-1}_{ik}}{q}
              \sum_l D^{-1}_{lj} \phi_l(r'_k)  \quad i \neq k
\ee

%% file: appen2.tex
\chapter{Elements of the Local Energy}

The wave function has the form
\be
\psi_T =D  \exp\left[-U\right]
\ee
where $D$ is the product of a spin up and a spin down Slater determinant
and 
\be
U = \sum_{i<j} u(r_{ij}).
\ee
The local energy is then
\be
E_{\mathrm L} =  \half\del^2 U - \half\del U \cdot \del U
               - \half \left(\frac{\del^2 D}{D}\right)
               +  \left(\frac{\del D}{D}\right) \cdot \del U
               + V
\ee
In Diffusion Monte Carlo, we need the quantum force, 
$F_{\mathrm Q} = \del \ln \abs{\psi}^2$.
\be
F_{\mathrm Q} = 2 \left(\frac{\del D}{D}\right) - 2 \del U
\ee
The derivatives of the Jastrow factors are
\bea
\del_k U_{ee} &=& \sum_{i \neq k} u'_{ee}(r_{ik})
               \frac{\vc{r_i}-\vc{r_k}}{r_{ik}} \\
\del_k U_{ne} &=& \sum_{\alpha=1}^{M} u'_{ne}(r_{k\alpha})
                 \frac{\vc{r_k} - \vc{R_\alpha}}{r_{k\alpha}} \\
\del^2_k U_{ee} &=& \sum_{i \neq k} \frac{2}{r_{ik}} u_{ee}'(r_{ik})
                  + u''_{ee}(r_{ik}) \\
\del^2_k U_{ne} &=& \sum_{\alpha=1}^M \frac{2}{r_{k\alpha}} u_{ne}'
              (r_{k\alpha}) + u''_{ne}(r_{k\alpha})
\eea
For the gradient with respect to particle $k$, expand the determinant by
cofactors about row $k$.  Then the cofactors have no $r_k$ dependence.
\bea
\frac{\del_k \Det{d}}{\Det{d}}
     &=& \sum_i \left[\del_k \phi_i(r_k)\right] d^{-1}_{ik} \\
\frac{\del^2_k \Det{d}}{\Det{d}}
     &=& \sum_i \left[\del^2_k \phi_i(r_k)\right] d^{-1}_{ik}
\eea

%% file: appen3.tex
\chapter{Cusp Condition}

When two Coulomb particles get close, the potential has a $1/r$ singularity.
The wave function must have the correct form to cancel this singularity.
First, consider an electron and a nucleus.  The relevant part of the
Schr\"odinger equation is
\be
\left[-\frac{1}{2M} \del_n^2 - \half \del_e^2 - \frac{Ze^2}{r}\right] \psi
 = E \psi
\ee
where $M$ is the nuclear mass and $Z$ is the nuclear charge.  Assume that
$M \gg m_e$, so the first term can be ignored.  Write the second term
in spherical coordinates and we get
\be
-\half \psi'' -\frac{1}{r}\left( Ze^2 \psi + \psi' \right) = E\psi
\ee
In order for the singularity to cancel at small $r$,
the term multiplying $1/r$ must vanish.  So we have
\be
\frac{1}{\psi} \psi' = -Z e^2
\ee
If $\psi= e^{-cr}$ we must have $c=Ze^2$.

For the case of two electrons, the Schr\"odinger equation is
\be
\left[-\frac{1}{2} \del_1^2 - \half \del_2^2 + \frac{e^2}{r_{12}}\right] \psi
 = E \psi
\ee
Switching to relative coordinates $r_{12} = r_1 - r_2$ gives us
\be
\left[- \del_{12}^2  + \frac{e^2}{r_{12}}\right] \psi
 = E \psi
\ee
Electrons with unlike spins (no antisymmetry requirement) have
an extra factor of $1/2$ in the cusp condition compared with
the electron-nucleus case. So we have $c= -e^2/2$.

In the antisymmetric case, the electrons will be in a relative $p$ state,
reducing the cusp condition by $1/2$, so $c=-e^2/4$.
Having the correct cusp for like spin electrons gains very little
in the energy or the variance, since the antisymmetry requirement keeps
them apart anyway.

%% file: vita.tex
\chapter*{Vita}
\addtocontents{toc}
{%
\protect\contentsline {chapter}{\protect {Vita}}{\arabic{page}}%
}

Mark Dewing was born on May 18, 1971 in San Diego, California.
He received a B. S. in physics from Michigan Technological University
in 1993.  He received an M. S. in physics from the University of
Illinois at Urbana-Champaign in 1995.